\pdfoutput=1
\documentclass[10pt,twocolumn,letterpaper]{article}

\usepackage[letterpaper,margin=0.75in]{geometry}
\usepackage{times}
\usepackage{xspace}
\usepackage[dvipsnames]{xcolor}
\usepackage{graphicx}
\usepackage{amsmath}
\usepackage{amssymb}
\usepackage{booktabs}
\usepackage{longtable}
\usepackage[numbers,sort&compress]{natbib}
\usepackage[format=plain,labelformat=simple,labelsep=period,font=small]{caption}
\usepackage[font=footnotesize,skip=3pt,subrefformat=parens]{subcaption}
\usepackage[shortlabels,inline]{enumitem}
\setlist[itemize]{noitemsep,leftmargin=*,topsep=0em}
\setlist[enumerate]{noitemsep,leftmargin=*,topsep=0em}

\newcommand{\rev}[1]{#1}
\newenvironment{revblock}{}{}

\DeclareRobustCommand{\addressedpoint}[2]{}
\DeclareRobustCommand{\addressedfiglink}[1]{}
\DeclareRobustCommand{\addressedtablelink}[1]{}
\DeclareRobustCommand{\addressedbacklink}[2]{}

\usepackage[export]{adjustbox}
\usepackage{float}
\usepackage{siunitx}
\usepackage{array}
\newcolumntype{N}{S[table-format=+2.1]}

\makeatletter
\DeclareRobustCommand\onedot{\futurelet\@let@token\@onedot}
\def\@onedot{\ifx\@let@token.\else.\null\fi\xspace}

\makeatother

\definecolor{linkblue}{rgb}{0.15,0.35,0.60}
\usepackage[breaklinks,colorlinks,allcolors=linkblue]{hyperref}
\usepackage[capitalize]{cleveref}
\hypersetup{
  pdftitle={JUMP-lite: Compact, reproducible benchmarking of cell representations},
  pdfauthor={Alán F. Muñoz, Johan Fredin Haslum, Runxi Shen, Anne E. Carpenter, Shantanu Singh}
}

\title{JUMP-lite: Compact, reproducible benchmarking of cell representations}

\author{
Al\'an F.\ Mu\~noz\thanks{Equal contribution.},
Johan Fredin Haslum\footnotemark[1],
Runxi Shen,
Anne E.\ Carpenter,
Shantanu Singh\thanks{Corresponding author.}\\
Broad Institute of MIT and Harvard, Cambridge, MA 02142, USA\\
{\tt\small \{amunozgo,jfredinh,shenrunxi,anne,shsingh\}@broadinstitute.org}
}
\date{}

\begin{document}
\maketitle

\begin{abstract}

Image-based profiling captures rich phenotypic signatures for drug discovery and functional genomics. Large public datasets like JUMP Cell Painting now provide millions of images for systematic study. \rev{However, JUMP alone occupies 115 TB, and fragmented evaluation practices make systematic comparisons of representation methods impractical for many researchers.} Here we present \textit{Nahual}, an open-source framework for reproducible model deployment, and JUMP-lite, \rev{a 92.0 GB subset of JUMP that is approximately 1,250-fold smaller, selected to cover genetic modalities and compound annotations and reduced via lossy JPEG XL compression.} \rev{Using these resources}, we benchmark five representation methods, including classical features (CellProfiler) and deep learning models (MorphEM, OpenPhenom, SubCell, DINOv2). \rev{Moderate compression broadly retains signal relative to uncompressed images.} Standardized phenotypic activity and consistency metrics reveal meaningful performance differences across methods. Together, JUMP-lite and \textit{Nahual} provide a foundation for accessible, reproducible benchmarking of image-based cell representations.

\end{abstract}
\section{Introduction}

Image-based profiling captures high-dimensional readouts of cell state from microscopy images, enabling applications in drug discovery and functional genomics \cite{chandrasekaranImagebasedProfilingDrug2020,wayEvolutionImpactHigh2023}.
Cell Painting, a standardized fluorescence microscopy assay that images cells across multiple organelle channels \cite{gustafsdottirMultiplexCytologicalProfiling2013,brayCellPaintingHighContent2016,ciminiOptimizingCellPainting2023}, has become the dominant platform for such profiling.
Profiling reduces each image to a numerical representation of cell state, and the goal is a representation general enough to transfer across laboratories, assays, and perturbation types.
The field has long relied on classical analysis pipelines that extract thousands of morphological features describing cell shape, texture, and intensity, and these handcrafted representations remain competitive baselines.
Deep learning is the natural candidate for such a general representation, having transformed many areas of biological image analysis, from medical imaging to cell segmentation \cite{stringerCellposeGeneralistAlgorithm2021}.
\rev{However, its advantages for cell representation learning remain inconsistent because improvements over classical features are often marginal or task-dependent \cite{chandrasekaranImagebasedProfilingDrug2020}. The resulting uncertainty can lead researchers to miss drug-target associations or fail to prioritize useful compounds.}
\rev{In contrast to most other computer vision subdomains, there is currently no conclusive evidence that deep learning outperforms engineered features \cite{chandrasekaranImagebasedProfilingDrug2020}. The lack of evidence remains an open challenge for the computer vision community.}

\begin{figure*}[ht]
    \centering
    \begin{minipage}{0.52\textwidth}
 \centerline{\small (a) \rev{JUMP-lite selection and perturbations}}
          \centering   \includegraphics[width=0.8\textwidth]{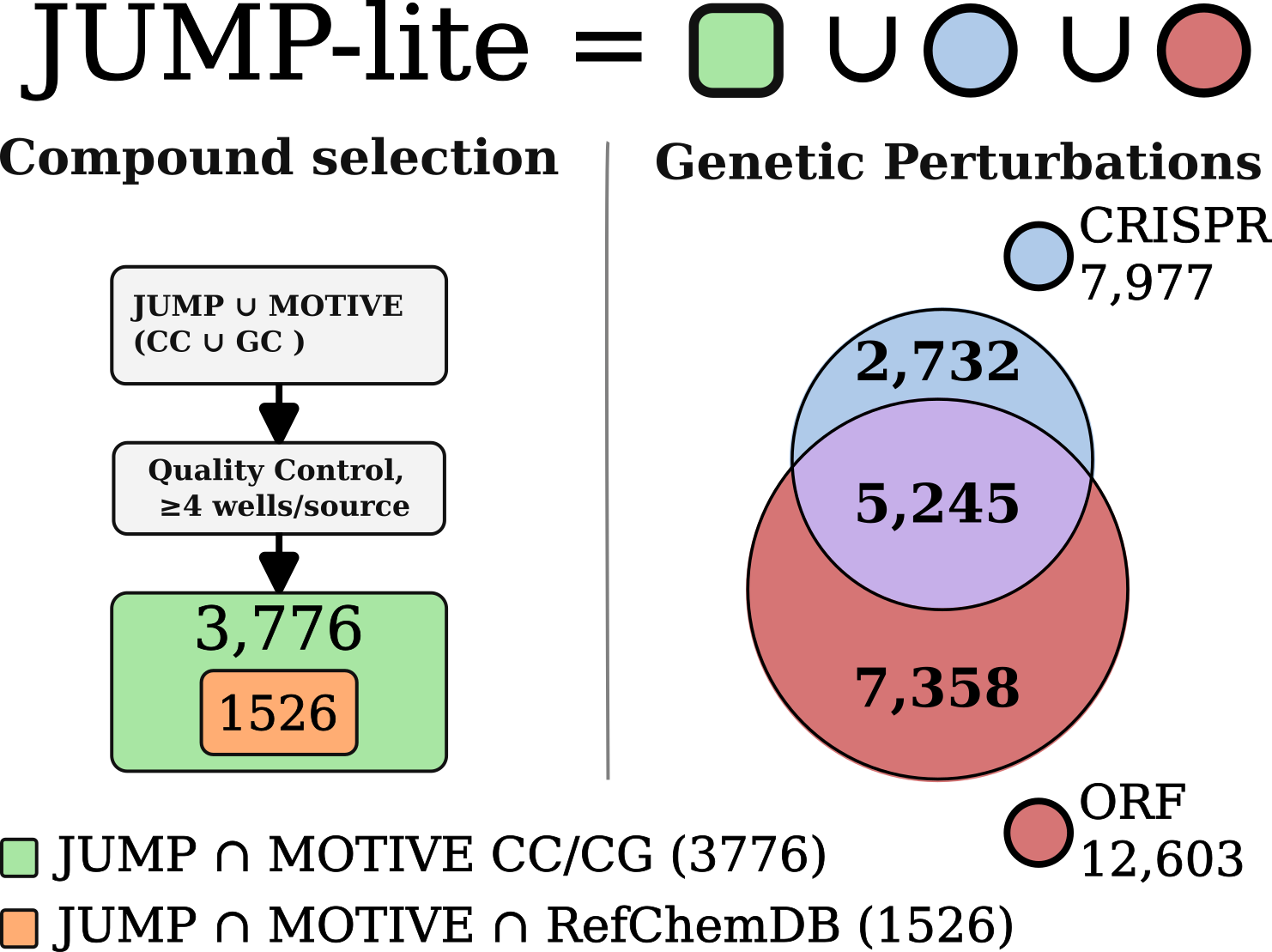}
        \centering
 
    \end{minipage}
    \begin{minipage}{0.35\textwidth}
        \centering
        \centerline{\small (b) JUMP-lite size.}
        \includegraphics[width=0.7\textwidth]{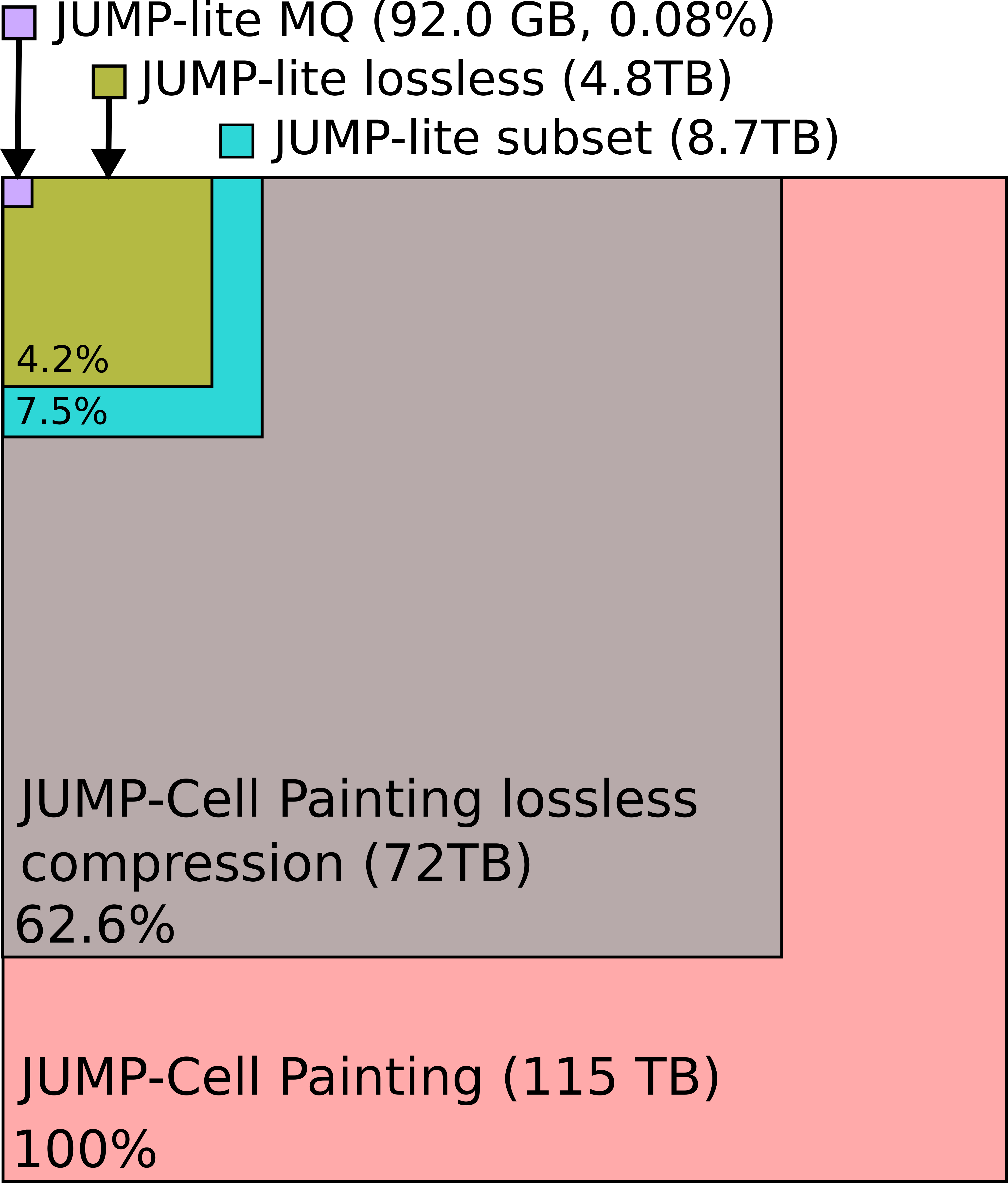}
    \end{minipage}
    \\[1.5ex]  
    \begin{minipage}{1.0\textwidth}
        \centering
        \includegraphics[width=0.9\textwidth]{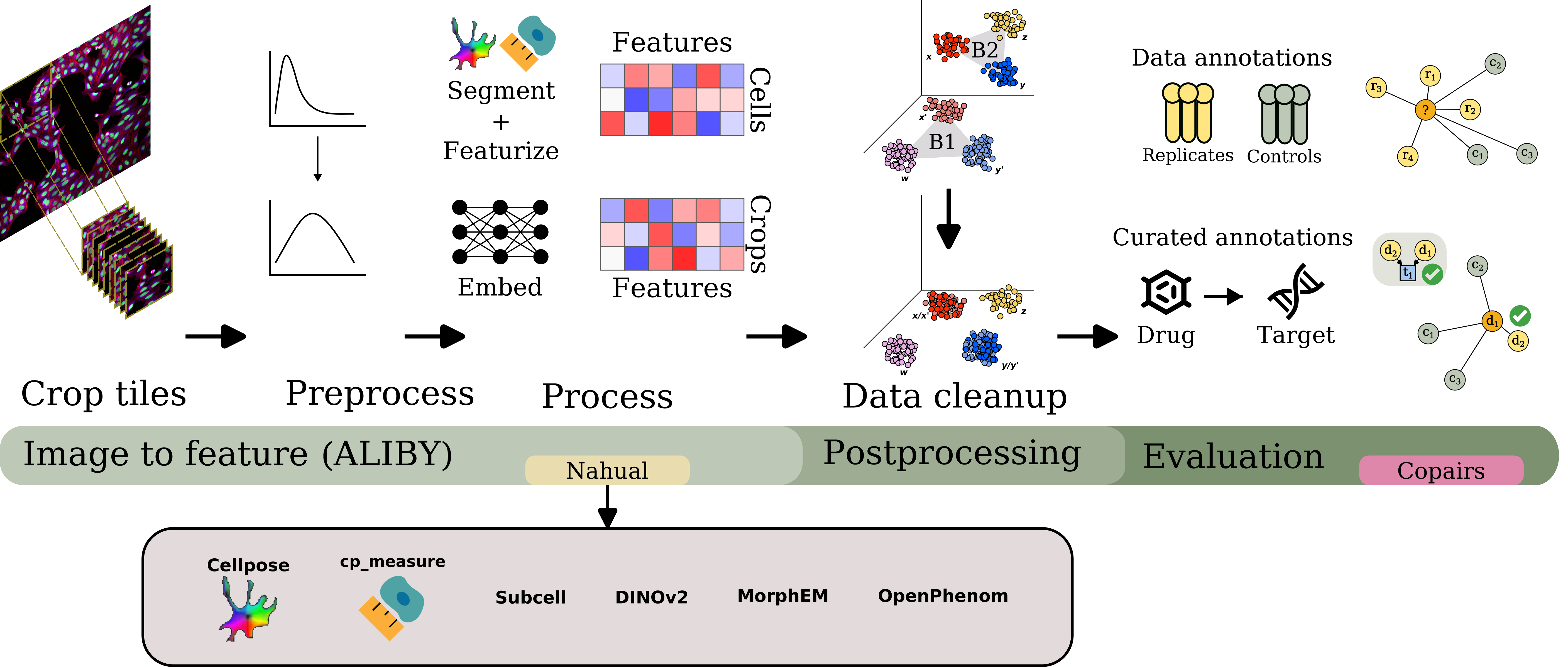}
        \centerline{\small (c) Overview of processing and evaluation pipeline.}
    \end{minipage}
    \caption{\rev{\textbf{JUMP-lite substantially reduces storage while focusing on annotated perturbations for benchmarking.} \textbf{(a)} Compound candidates were mapped to the union of MOTIVE's compound--compound (CC) and compound--gene (CG) graphs, then passed through filters for plate quality, profile quality and replicate count. Supplementary Section~\ref{sec:target2-identifier-accounting} provides detailed component and annotation accounting, and Supplementary Figure~\ref{fig:overlap_per_source} shows target overlap across the compound, CRISPR, and ORF components. \textbf{(b)} Disk-size comparison of JUMP Cell Painting with lossless and lossy compression options. Each square's area is directly proportional to its disk size. \textbf{(c)} Standardized pipeline from images to evaluation. ALIBY \cite{munozPhenotypingSingleCells2023} orchestrates segmentation (Cellpose) and feature extraction (cp\_measure), while \textit{Nahual} connects isolated environments for deep learning models. A normalization sweep processes the profiles for comparison across methods.}\addressedbacklink{perturbations}{A8}\addressedbacklink{accounting}{A9}\addressedbacklink{codec-names}{A10}\addressedbacklink{terminology}{A15}}
    \label{fig:overview}
\end{figure*}

The resources needed to answer this question now exist.
Large public datasets provide millions of images: JUMP Cell Painting alone spans 115 TB across chemical and genetic perturbations \cite{chandrasekaranJUMPCellPainting2023}.
Multiple representation models now exist, from generalist vision transformers like DINOv2 \cite{oquabDINOv2LearningRobust2024} to microscopy-specific methods such as \rev{MorphEM \cite{agrawalCHAMMI75PreTrainingMultiChannel2025,caicedolabMorphEmModelCard2026}, OpenPhenom \cite{krausMaskedAutoencodersMicroscopy2024,recursionpharmaOpenPhenomModelCard2026}}, and SubCell \cite{gupta2025subcell}.
Despite the availability of datasets such as JUMP, their scale makes comparisons challenging: storage cost, compute requirements, and preprocessing effort create prohibitive barriers, both for academic laboratories with limited local storage and for cloud-based workflows where I/O and bandwidth dominate costs.
Crucially, studies are not directly comparable because each project adopts different data selection, preprocessing pipelines, and evaluation metrics. It is impossible to determine whether an apparent improvement reflects the representation model or the choices surrounding its evaluation.

For image-based profiling, a small number of datasets have been adopted as shared test beds, among them BBBC021 \cite{caie2010high}, RxRx1 \cite{sypetkowskiRxRx1DatasetEvaluating2023}, and CP-JUMP1 \cite{chandrasekaran2024three}, but these suffer from limited biologically relevant tasks and an insufficient scale to represent the complexity of real large-scale imaging screens.

Although large cell-imaging datasets are readily available, resources such as the Cell Painting Gallery \cite{weisbartCellPaintingGallery2024}, IDR \cite{williams2017image}, and EU-OPEN \cite{wolff2024morphological} contain tens of millions of images with heterogeneous formats, varied metadata schemas, prohibitive storage requirements, and unclear biological tasks and ground truth for evaluation, making them practically inaccessible for systematic benchmarking.
Recent work, RxRx3-core, was designed to address this problem \cite{krausRxRx3coreBenchmarkingDrugtarget2025} and represents meaningful progress: it scaled up the number of samples relative to earlier benchmarks, capturing greater complexity, while using data sub-sampling and lossy compression to reduce the dataset footprint to 18 GB.

A complementary need remains for a Cell Painting benchmark that reaches the scale of modern high-content screens, typically 10,000+ perturbations \cite{zhangTahoe100MGigaScaleSingleCell2025,replogle2022mapping}, while remaining accessible, drawing on multiple experimental sources to enable cross-batch and cross-lab corrections and more generalizable evaluation, providing a principled assessment of the compression-fidelity trade-off, and supporting reproducible model deployment.
We address this with three contributions.
First, we present \textbf{JUMP-lite}, a \rev{92 GB} curated subset of JUMP Cell Painting spanning more than 24,000 perturbations across CRISPR (gene knockouts), ORF (gene overexpression), and chemical compounds drawn from multiple laboratories. Storage is reduced by \rev{approximately 92\% through annotation-linked compound selection and plate/replicate filtering while retaining broad coverage}, and by a further \rev{98.9\%} through lossy compression \cite{zhouDeepLearningBasedImageCompression2024}.
\rev{Second, we provide the first systematic characterization of lossy compression artifacts in Cell Painting data. JPEG XL at high quality (HQ) and medium quality (MQ) preserves biological signal. HQ reduces storage by 97.3\%, with a 1.2\% decrease in downstream performance, whereas MQ reduces storage by 98.9\% (61.3\% relative to HQ), with a 6.5\% decrease in performance.}
Third, we provide \textit{Nahual}, a framework for reproducible extraction of representations from models with otherwise incompatible dependencies.
Together, JUMP-lite and \textit{Nahual} provide an accessible, reproducible, and systematic foundation for benchmarking image-based cell representations.
% Curated accessible datasets, diverse representation methods, and principled evaluation metrics existed as independent components until now.

\section{Background}

\subsection{Public Cell Painting Datasets}
Cell Painting is a multiplexed fluorescence microscopy assay designed to capture morphological signatures of cellular state \cite{ciminiOptimizingCellPainting2023,sealCellPaintingDecade2025}.
\rev{It images eight cellular compartments in five channels using six fluorescent dyes.}
Cells treated with unique perturbations, as well as unperturbed cells, are spread across multi-well plates.
Unperturbed cells make perturbations comparable across plates and experiments. 
This standardization makes Cell Painting the dominant assay for image-based profiling by enabling consistent profiling across laboratories.

Beyond JUMP (115 TB; 116k compound and genetic perturbations over 15k genes \cite{chandrasekaranJUMPCellPainting2023}), RxRx3 is the largest public Cell Painting dataset by image count (17k CRISPR knockouts and 1,674 compounds at eight doses \cite{fayRxRx3PhenomicsMap2023}).
Its curated subset RxRx3-core (18 GB) achieves compactness through sample selection, cropping, bit-depth reduction, and JPEG 2000 compression \cite{krausRxRx3coreBenchmarkingDrugtarget2025}, but the effect of compression on downstream performance was never assessed.
Other resources (e.g., CHAMMI-75, IDRCell100k \cite{agrawalCHAMMI75PreTrainingMultiChannel2025,bourriezChAdaViTChannelAdaptive2024}) prioritize cross-modality generalization but include limited Cell Painting data, annotations, and evaluation tasks.

\subsection{Data Compression}
\rev{Storing raw microscopy images can incur substantial disk, bandwidth, and input/output costs. Compression can reduce file size. Lossy compression can reduce footprints by orders of magnitude more than lossless compression, but risks losing important information. As datasets grow, the trade-off between file-size reduction and information loss is being explored across scientific domains \cite{underwood2022understanding, zhao2020sdrbench, cappello2025lossy}.}
To our knowledge, this has not been systematically evaluated for morphological profiling, leaving practitioners uncertain about the impact of lossy compression on biological signal.

\subsection{Representation Methods}
Two paradigms dominate feature representation of cell morphologies.
\rev{Classical pipelines, exemplified by CellProfiler \cite{stirling2021cellprofiler}, segment individual cells and extract thousands of handcrafted features that describe the morphology, texture, and intensity distribution of each cell \cite{caicedoDataAnalysisStrategiesImageBased2017}.}
A recent reimplementation, cp\_measure \cite{munozCp_measureAPIfirstFeature2025}, exposes \rev{and speeds up the calculation of} these features through a programmatic API.
Deep learning approaches instead learn representations directly from images.
\rev{These range from generalist self-supervised models trained on natural images, such as DINOv2 \cite{oquabDINOv2LearningRobust2024}, to models pretrained specifically on microscopy data, including MorphEM \cite{agrawalCHAMMI75PreTrainingMultiChannel2025,caicedolabMorphEmModelCard2026}, OpenPhenom \cite{krausMaskedAutoencodersMicroscopy2024,recursionpharmaOpenPhenomModelCard2026}, SubCell \cite{gupta2025subcell}, scDINO \cite{pfaendlerSelfSupervisedVisionTransformers2023}, and CytoSelf \cite{kobayashiSelfSupervisedDeepLearning2022}.}
% Many other models are also available; resources such as the BioImage Model Zoo provide standardized formats for sharing bioimaging models for a multitude of tasks \cite{ouyangBioImageModelZoo2022}.}
\rev{We selected released representations that Nahual could deploy reproducibly on five-channel Cell Painting sites and that span natural-image and microscopy pretraining. The set is deliberately non-exhaustive. We did not include all existing microscopy models: scDINO, because it is trained on immune cell images whose channel meanings differ from Cell Painting; CytoSelf, because it specializes on protein and nucleus-derived inputs rather than whole-site Cell Painting images.}
SubCell was trained on Human Protein Atlas data using a masked autoencoder (MAE) setup.
\rev{The documented training corpora for OpenPhenom and MorphEM include JUMP images \cite{caicedolabMorphEmModelCard2026,recursionpharmaOpenPhenomModelCard2026}; the Supplementary Section \ref{sec:model-audit} offers further details.}

\subsection{Evaluation Metrics}
A range of metrics has been used to assess image-based profiles, including replicate-vs-non-replicate correlation, distribution-based distances such as energy distance and MMD, classification or clustering accuracy against known labels, and retrieval-based mean average precision (mAP) \cite{kalininVersatileInformationRetrieval2025,celikBuildingBenchmarkingExploring2024}.
Recent work argues that retrieval-based metrics are well suited to comparing high-dimensional embeddings across perturbation types and modalities \cite{kalininVersatileInformationRetrieval2025}.
A separate line of evaluation grounds representations in external biological knowledge through cross-modality retrieval against curated drug-target annotations, as in MOTIVE \cite{arevaloMOTIVEDrugTargetInteraction2024}.

Curated drug-target resources \cite{konigsHeterogeneousPharmacologicalMedical2022,freshourIntegrationDrugGene2021,corsello2017drug} take different approaches to assembling annotations, from manual curation, as in the Drug Repurposing Hub \cite{corsello2017drug}, to literature- and database-integration pipelines.
We draw on two complementary resources for our evaluations: \rev{RefChemDB} \cite{judson2018workflow}, which reports the number of independent assay records supporting each compound-target annotation, enabling explicit selection of high-confidence links, and MOTIVE \cite{arevaloMOTIVEDrugTargetInteraction2024}, which aggregates drug-target relationships across seven public knowledge bases and covers cross-modality retrieval.
We adopt Phenotypic Activity (PA, perturbation vs.\ negative controls), Phenotypic Consistency (PC, same-target vs.\ different-target perturbations grounded in \rev{RefChemDB}), and cross-modality recall@K against MOTIVE annotations as our primary evaluation tasks.

\begin{figure}[ht]
    \centering
    % --- Row 1: 7 compression levels ---
    \centerline{\small (a) Example images at different compression levels.}
    \begin{minipage}[t]{0.135\columnwidth}\centering
        \adjustbox{trim={0.40\width} {0.38\height} {0.5\width} {0.52\height}, clip, width=\linewidth}{%
            \includegraphics{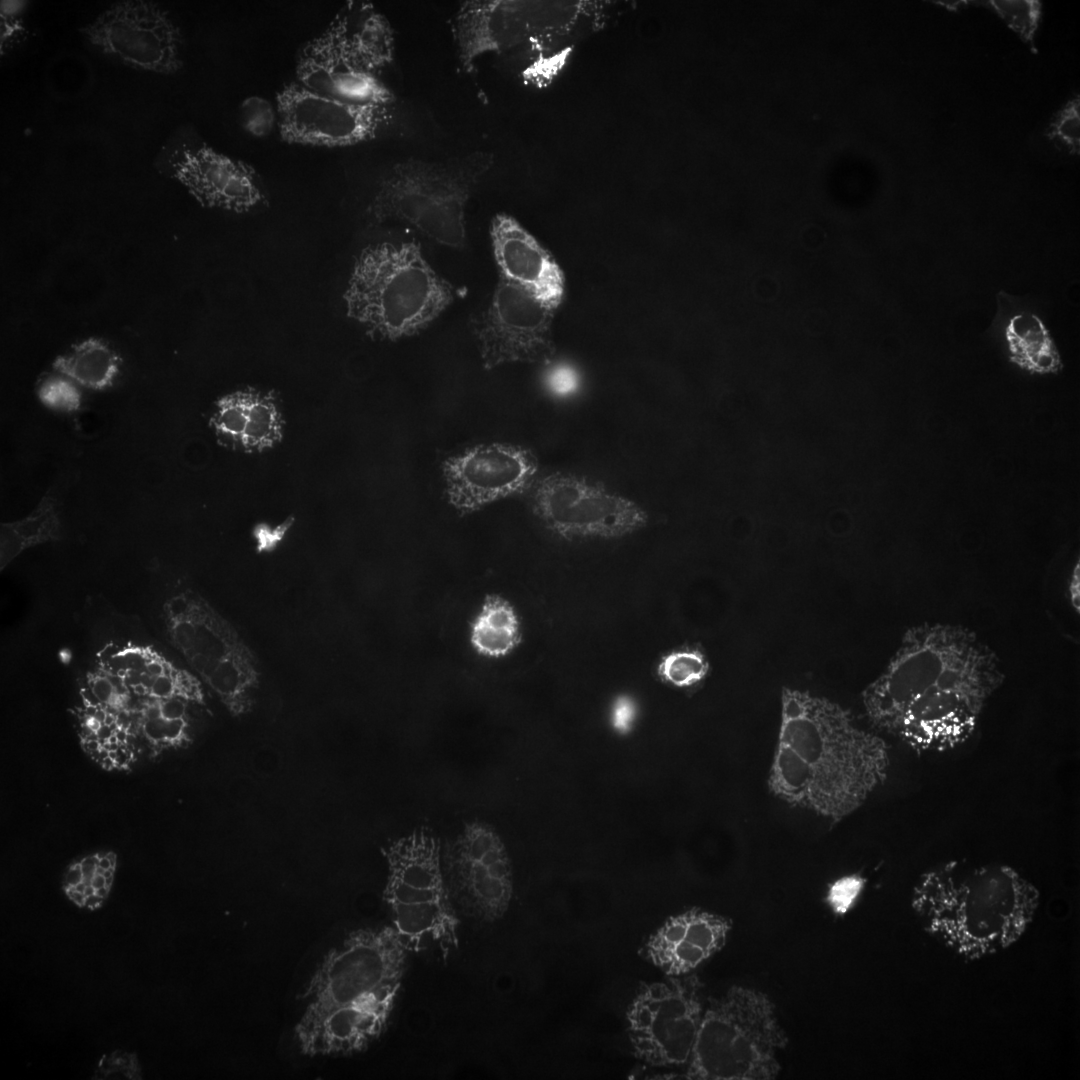}}
        \centerline{\scriptsize \rev{Raw}}
    \end{minipage}%
    \begin{minipage}[t]{0.135\columnwidth}\centering
        \adjustbox{trim={0.40\width} {0.38\height} {0.5\width} {0.52\height}, clip, width=\linewidth}{%
            \includegraphics{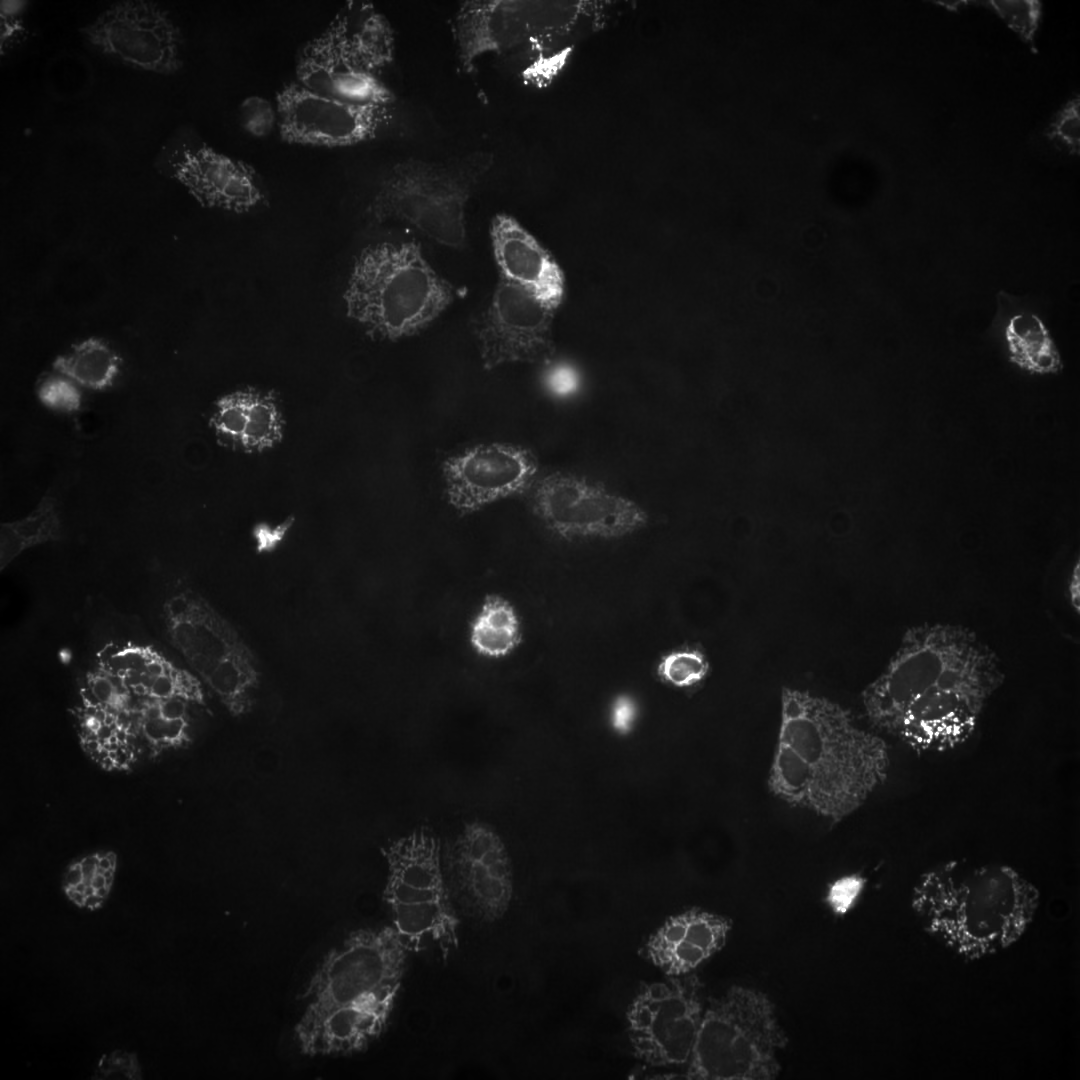}}
        \centerline{\scriptsize \rev{HQ}}
    \end{minipage}%
    \begin{minipage}[t]{0.135\columnwidth}\centering
        \adjustbox{trim={0.40\width} {0.38\height} {0.5\width} {0.52\height}, clip, width=\linewidth}{%
            \includegraphics{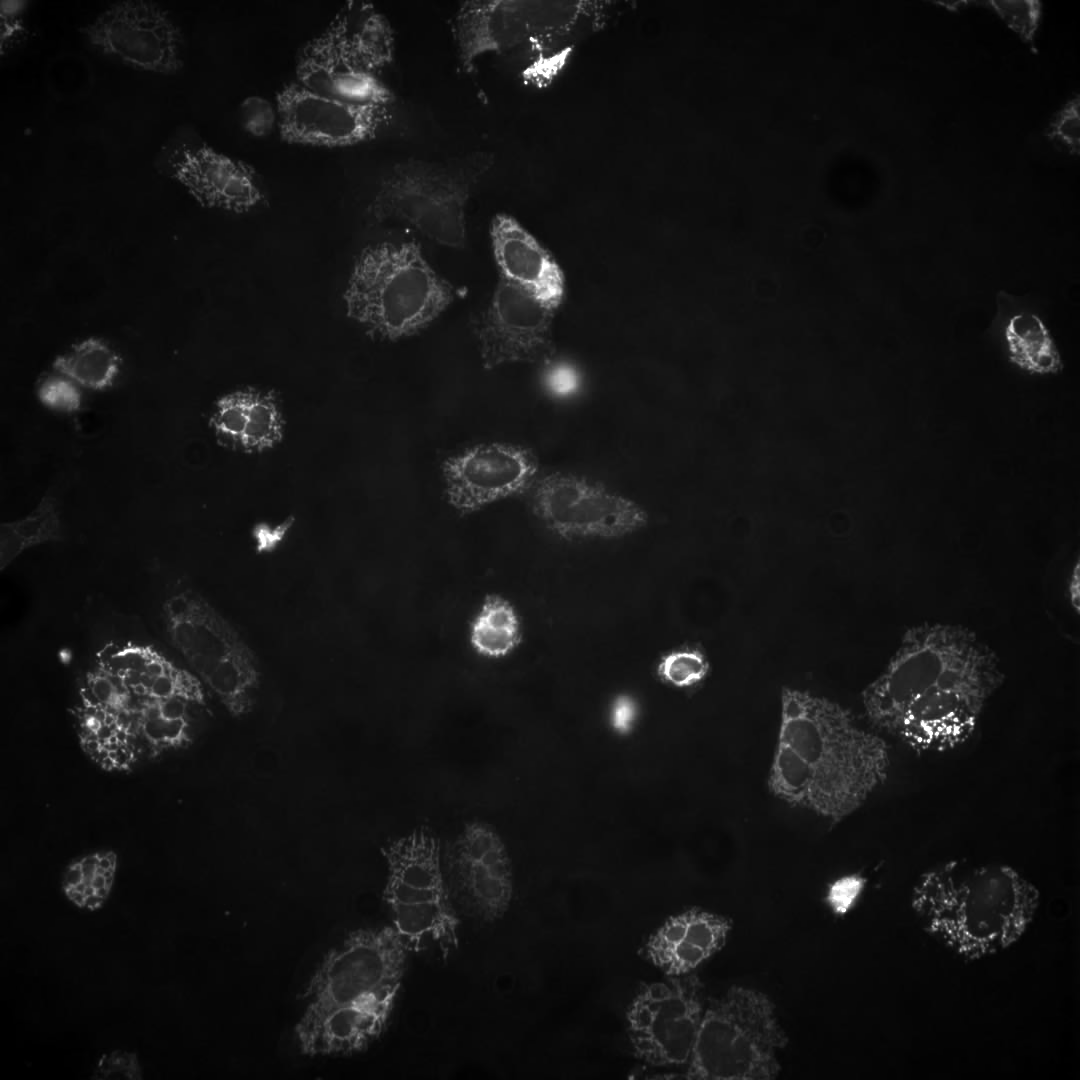}}
        \centerline{\scriptsize \rev{E3}}
    \end{minipage}%
    \begin{minipage}[t]{0.135\columnwidth}\centering
        \adjustbox{trim={0.40\width} {0.38\height} {0.5\width} {0.52\height}, clip, width=\linewidth}{%
            \includegraphics{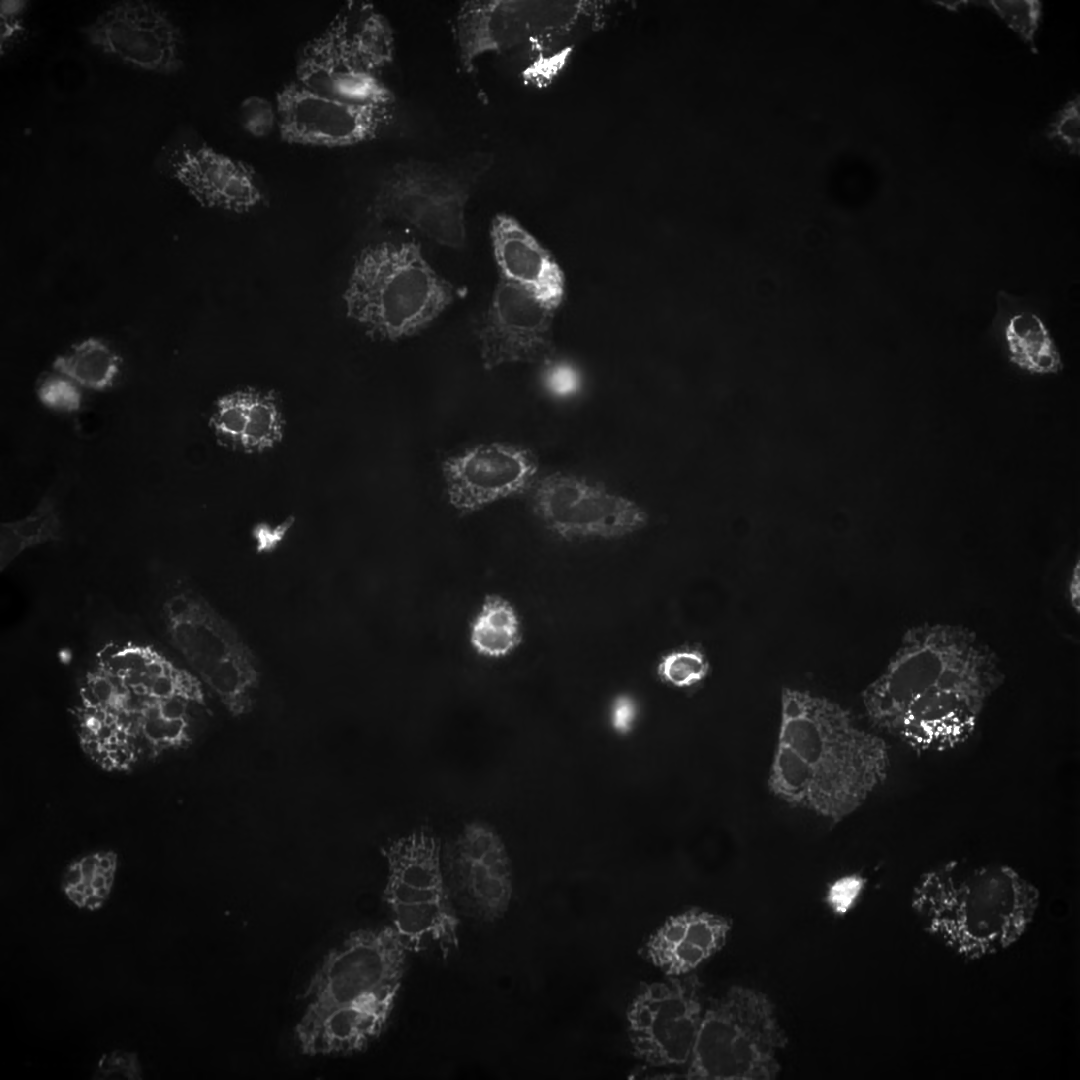}}
        \centerline{\scriptsize \rev{D2-E8}}
    \end{minipage}%
    \begin{minipage}[t]{0.135\columnwidth}\centering
        \adjustbox{trim={0.40\width} {0.38\height} {0.5\width} {0.52\height}, clip, width=\linewidth}{%
            \includegraphics{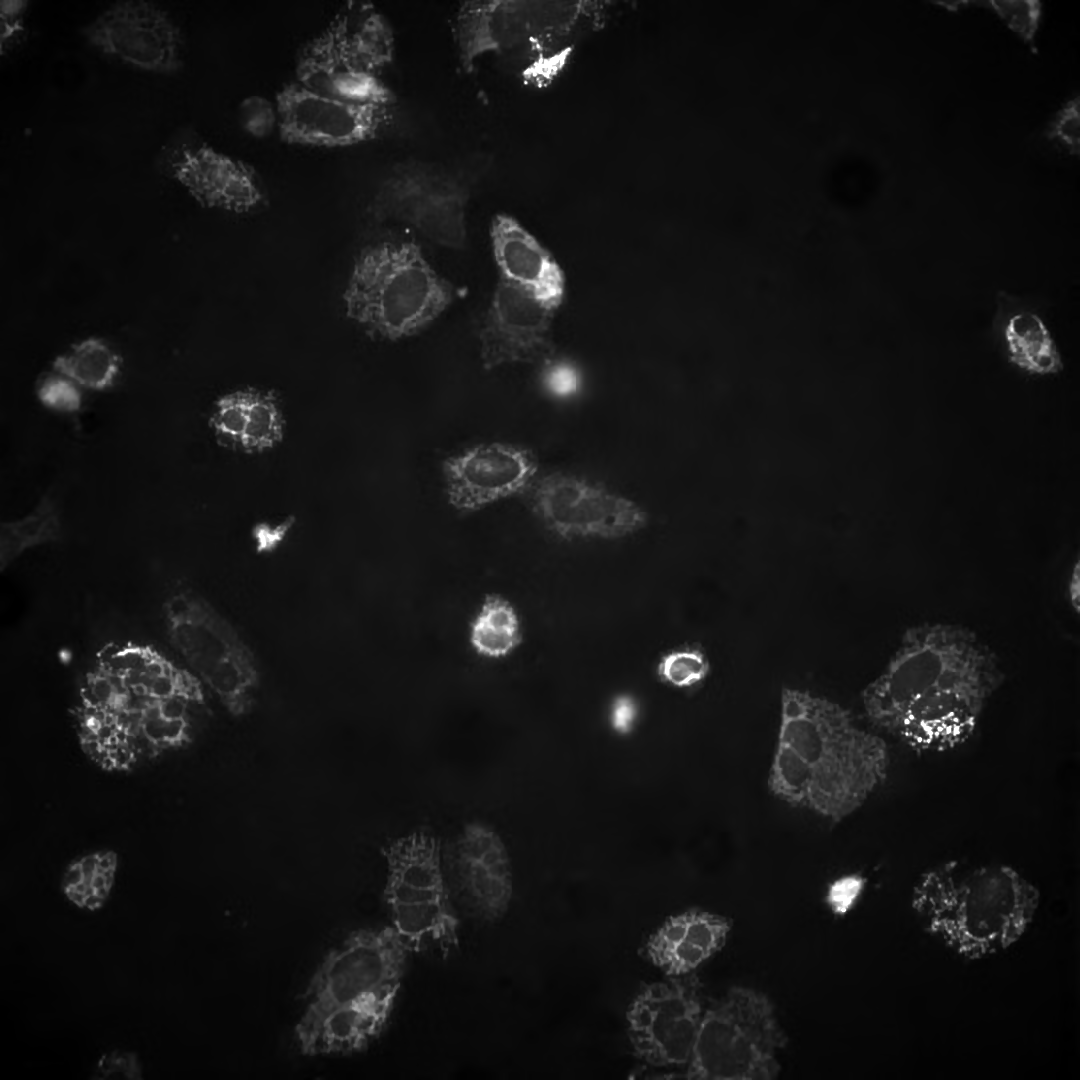}}
        \centerline{\scriptsize \rev{MQ}}
    \end{minipage}%
    \begin{minipage}[t]{0.135\columnwidth}\centering
        \adjustbox{trim={0.40\width} {0.38\height} {0.5\width} {0.52\height}, clip, width=\linewidth}{%
            \includegraphics{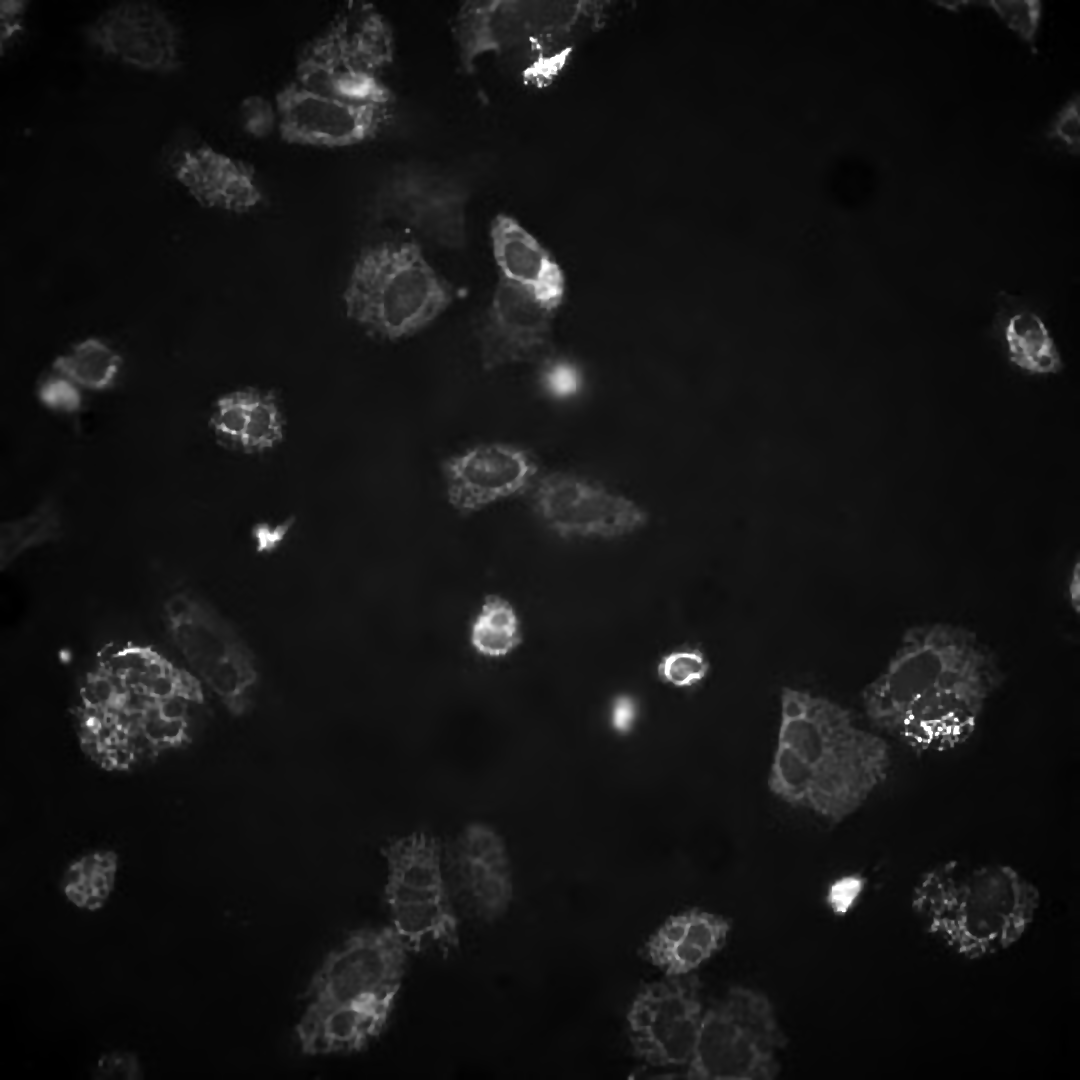}}
        \centerline{\scriptsize \rev{LQ}}
    \end{minipage}%
    \begin{minipage}[t]{0.135\columnwidth}\centering
        \adjustbox{trim={0.40\width} {0.38\height} {0.5\width} {0.52\height}, clip, width=\linewidth}{%
            \includegraphics{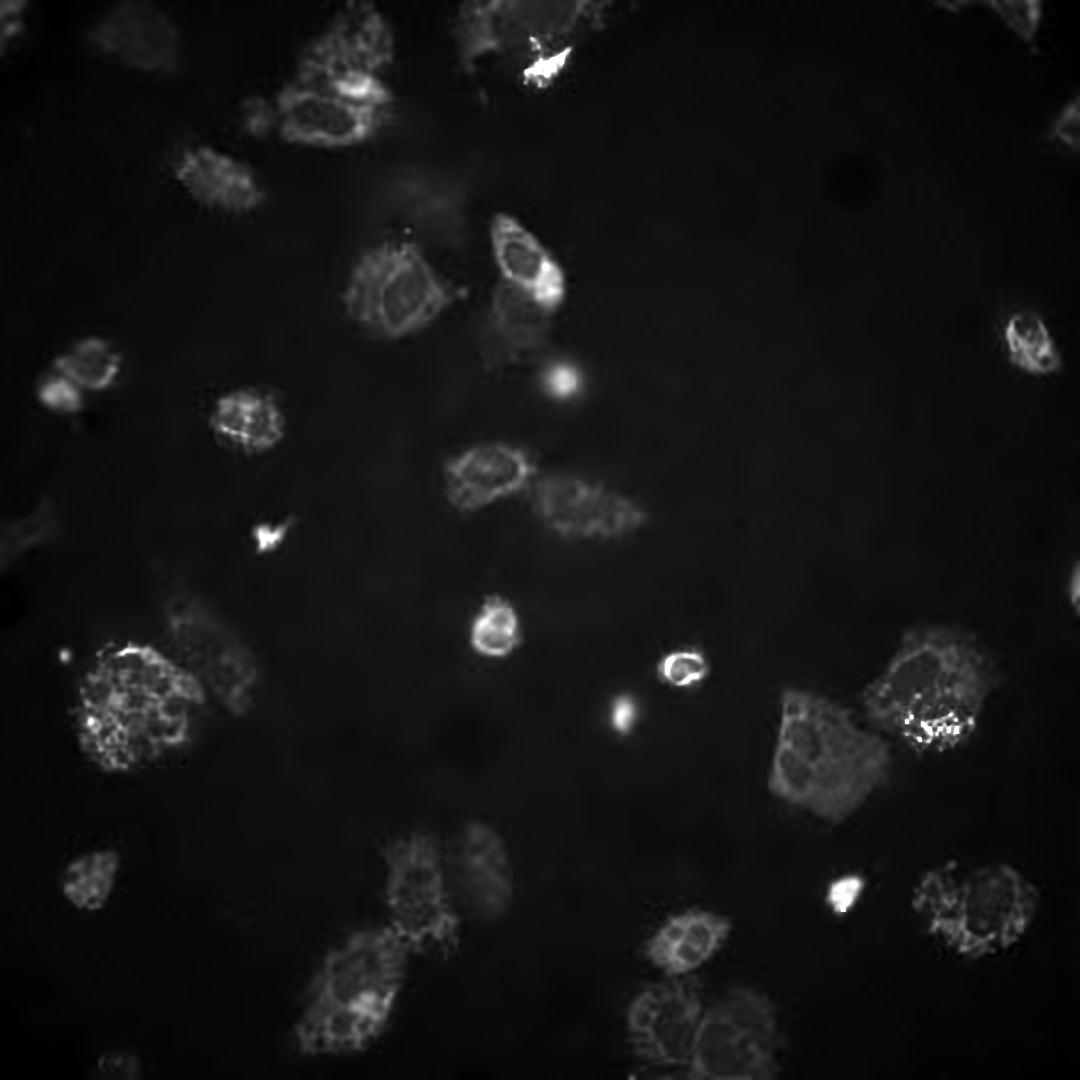}}
        \centerline{\scriptsize \rev{D10}}
    \end{minipage}
    \\[1ex]
    % --- Row 2: phenotype comparison ---
    \begin{minipage}[t]{\columnwidth}\centering
        \begin{minipage}[t]{0.4\columnwidth}
            \centerline{\small Unperturbed}
            \adjustbox{trim={0.10\width} {0.5\height} {0.1\width} {0.1\height}, clip, width=\linewidth}{%
                \includegraphics[width=0.9\columnwidth]{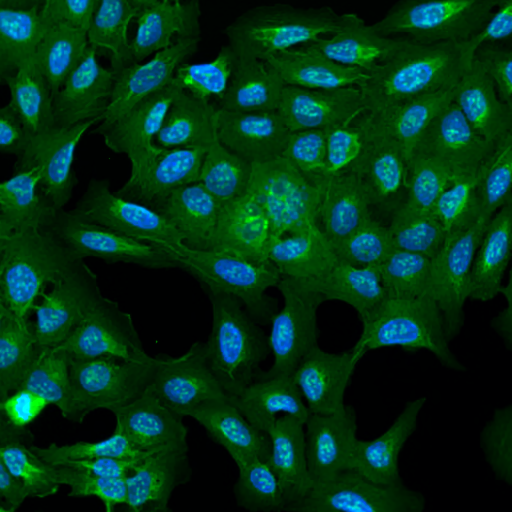}}
        \end{minipage}\hspace{2em}%
        \begin{minipage}[t]{0.4\columnwidth}
            \centerline{\small With perturbation}
            \adjustbox{trim={0.10\width} {0.5\height} {0.1\width} {0.1\height}, clip, width=\linewidth}{%
                \includegraphics[width=0.9\columnwidth]{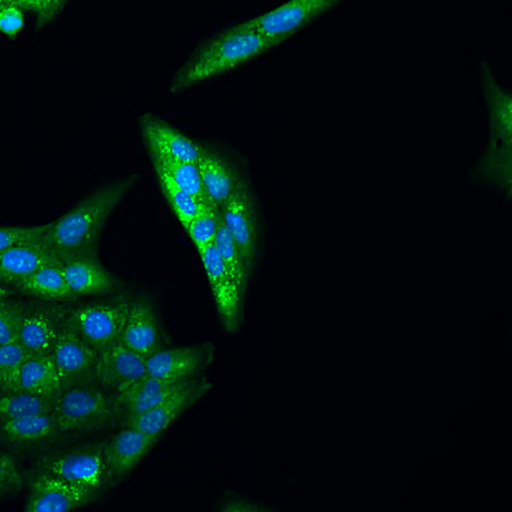}}
        \end{minipage}
    \end{minipage}
    \\
    \centerline{\small (b) Visibly distinct phenotypes.}
    \caption{\rev{\textbf{Image quality across compression levels.} \textbf{(a)} Example images at compression levels ranging from Raw (leftmost) to the more heavily compressed D10 setting (rightmost). We calculated the Structural Similarity Index Measure (SSIM) at all compression levels (Figure~\ref{fig:compression_impact_plot}).
    \textbf{(b)} Control cells (no active chemical) vs. cells perturbed with a chemical. 
     Figure~\ref{fig:segmentation_example_sup} shows an example of compression-related segmentation differences.}}
    \label{fig:compression_impact_image}
\end{figure}

\section{Results}

\subsection{JUMP-lite: a curated, compressed subset}
\label{sec:jump_lite_overview}

\textbf{Dataset curation}

\rev{We constructed JUMP-lite as a compact benchmark spanning genetic and chemical perturbations in JUMP (Figure~\ref{fig:overview}a). We included the available CRISPR (knockout) and ORF (overexpression) profiles and selected compounds through MOTIVE's compound--compound and compound--gene graphs \cite{arevaloMOTIVEDrugTargetInteraction2024}. RefChemDB annotations \cite{judson2018workflow} are used for downstream target-based evaluation.}

\rev{After plate and profile quality control \cite{chandrasekaranMorphologicalMapOverexpression2025}, the retained compounds were grouped into bioactive-library (0.625~$\mu$M, single source) and diversity (10~$\mu$M, multiple sources) subsets according to their origin and assay conditions. Within each source group, compounds with fewer than four replicate wells were excluded. Target-2 was retained as a separate cohort for complementary analyses. Section~\ref{sec:sources-dose} details the construction criteria.}

\rev{CRISPR knockout and ORF overexpression of the same gene are treated as distinct perturbations because they are biologically different interventions. The final dataset comprises 24,356 such perturbations across 163,776 wells and 655,101 five-channel image sites, totaling 8.7~TB of raw TIFFs; almost all wells have four imaging sites. The release also provides independently labeled Cellpose cell and nuclei instance masks for the 632,672-site JUMP-lite dataset under Raw (lossless Zstd) and MQ.}

\rev{JUMP-lite was not designed to reproduce the broader-JUMP distribution. Section~\ref{sec:broader-jump-coverage} details the overlap.}

\textbf{Lossy compression.} We chose JPEG XL compression because it is an open standard, supports both lossless and lossy modes, and facilitates the long-term storage of imaging data.
\rev{In our Target-2 experiment, we used six compression levels: High Quality (HQ), Medium Quality (MQ), Low Quality (LQ), aggressive compression (D10), and the intermediate settings E3 and D2-E8. The labels use the JPEG XL compression parameters \textit{Distance} (D) and \textit{Effort} (E). For tractability, we reduced this set to four levels for the full JUMP-lite release: Raw, HQ, MQ, and D20. D20 replaced D10 as an even more aggressive setting at which we expected biological signal to deteriorate (Figure~\ref{fig:compression_impact_image}a). Table~\ref{tab:zarr_benchmark} maps the pilot labels to their settings; Section~\ref{sec:compression} distinguishes the pilot and full-release D20 settings.}

At \rev{MQ} compression, JUMP-lite occupies \rev{92.0 GB: approximately 94-fold smaller than the curated 8.7 TB image set and 1,250-fold smaller than the original 115 TB JUMP release} (Figure \ref{fig:overview}b). See Section \ref{sec:compression} for the full-release settings.
This compact footprint made systematic, multi-model benchmarking tractable, using standardized evaluation via the ALIBY framework \cite{munozPhenotypingSingleCells2023} and our \textit{Nahual} library for reproducible model deployment (Figure \ref{fig:overview}c).

\textbf{Annotations.} JUMP-lite couples two complementary annotation sources to ground its evaluations. RefChemDB \cite{judson2018workflow} provides high-confidence compound--target labels, while MOTIVE \cite{arevaloMOTIVEDrugTargetInteraction2024} aggregates drug-target interactions from seven public databases and additionally exposes compound-compound and gene-gene relationships our retrieval tasks rely on. See Section~\ref{sec:annotations} for details on curation.

\rev{\textbf{Evaluation tasks for biological signal.} Using \textit{copairs} \cite{kalininVersatileInformationRetrieval2025}, phenotypic activity (PA) retrieves replicates against same-plate negative controls, whereas phenotypic consistency (PC) retrieves perturbations sharing a RefChemDB target against different-target perturbations. Both use Normalized Average Precision (NAP), for which random retrieval maps to 0 and perfect separation to 1. We combine them using the PA-PC product, $\mathrm{PA}\times\mathrm{PC}$. For post-processing selection only, we first min--max rescale PA and PC across the swept configurations and then compute their product. MOTIVE recall@K tests within-modality compound--compound and gene--gene retrieval and cross-modality compound--gene retrieval. ORF and CRISPR are treated as separate modalities.}

\textbf{Profile processing.} \rev{Following standardized image-based profiling workflows, we filtered out low-quality and redundant features, normalized each plate against its negative controls, and corrected plate-to-plate technical variation.
For each model--compression pair, we varied the post-processing parameters to optimize each representation. We report the configuration with the highest rescaled PA-PC product. See Section~\ref{sec:profile-processing} for the parameter ranges.}

\subsection{Nahual: reproducible model deployment}

\rev{We developed \textit{Nahual} to evaluate models with incompatible software dependencies in one consistent framework. This Python library uses inter-process communication to connect models running in isolated environments to the ALIBY orchestration framework \cite{munozPhenotypingSingleCells2023}.}
Each model runs in a dedicated Nix environment \cite{dolstraNixSafePolicyFree2004}, ensuring reproducibility while avoiding dependency conflicts. See \ref{sec:methods_nahual} for details on the design of Nahual.

\rev{In the 9,216-site pilot, individual deep learning models processed 13,104 to 22,752 sites per hour. Cellpose \cite{pachitariu2025cellpose} plus cp\_measure \cite{munozCp_measureAPIfirstFeature2025} processed 332 sites per hour with the newest cp\_measure version across the MQ and lossless queues (Table~\ref{table:speed_benchmark}).}
% While these pipelines differ substantially (yielding tile-based embeddings versus per-cell features), the speed differential highlights how deep learning approaches can reduce iteration time for large-scale benchmarking.

\subsection{Compression preserves phenotypic signal}

\begin{figure*}[!ht]
    \centering
    \begin{minipage}{0.3\textwidth}
    \includegraphics[width=\textwidth]{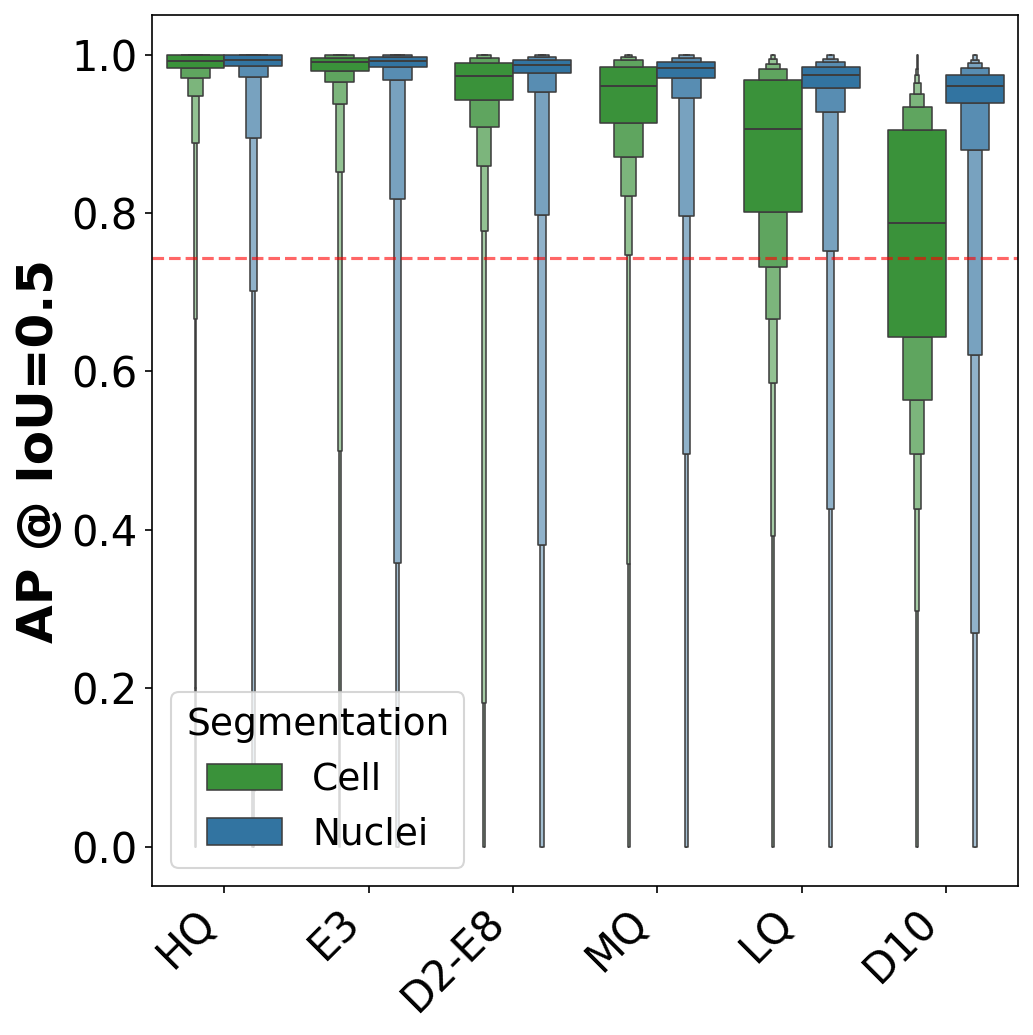}      
        \centerline{\small \textbf{(a)} Segmentation AP.}
    \end{minipage}
    \centering
    \begin{minipage}{0.3\textwidth}
    \includegraphics[width=\textwidth]{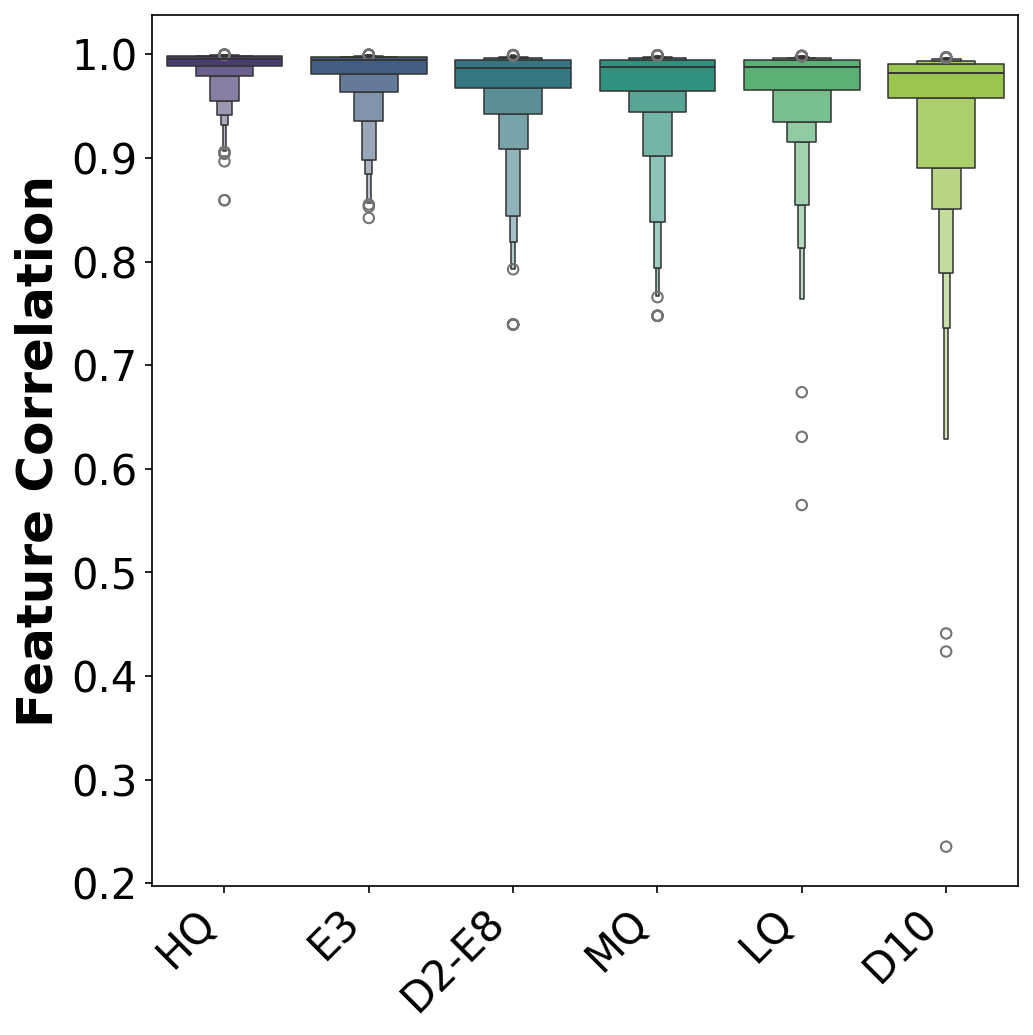}
        \centerline{\small \textbf{(b)} Feature correlation.}
    \end{minipage}
    \centering
    \begin{minipage}{0.3\textwidth}
    \includegraphics[width=\textwidth]{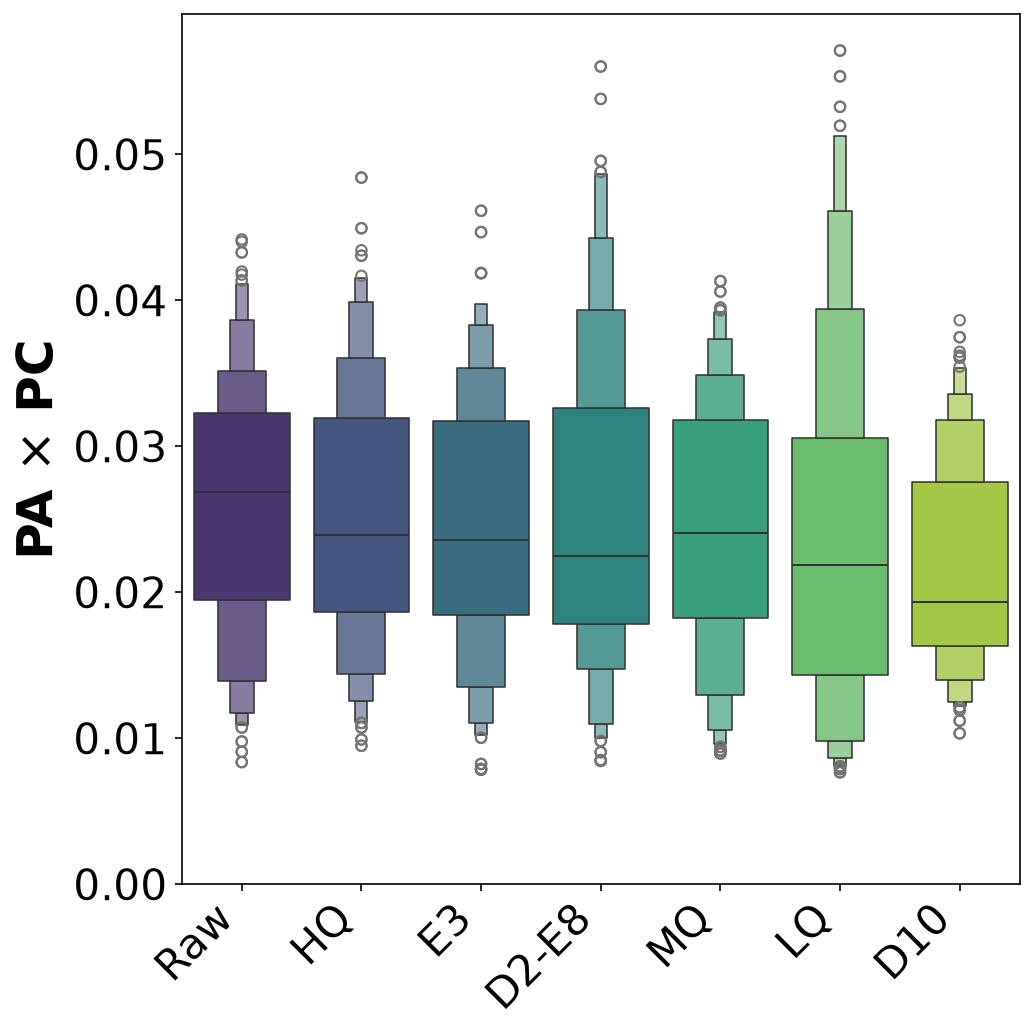}
    \centerline{\small \textbf{(c)} Biological metrics}
    \end{minipage}

    \centering
    \caption{\rev{\textbf{Target-2 segmentation stability and feature consistency across compression levels.}
    \textbf{(a)} Segmentation-mask agreement across compression levels, measured as average precision (AP) at an intersection-over-union (IoU) threshold of 0.5. Masks from Raw images are the reference. For comparison with agreement between human annotators, the red line shows the AP achieved when experts segmented the same images multiple times \cite{pachitariu2025cellpose}. Figure~\ref{fig:sup_seg_f1} shows additional IoU thresholds.
    \textbf{(b)} Pearson correlation between features from compressed and uncompressed images. Only features that passed conventional CellProfiler feature selection are included; Figure~\ref{fig:sup_feature_correlation} describes the selection.
    \textbf{(c)} The PA-PC product, $\mathrm{PA}\times\mathrm{PC}$. For each compression setting, results are pooled over five model families (MorphEM, OpenPhenom, SubCell, DINOv2, and cp\_measure), with approximately 48 normalization configurations per family. The median decreases as compression becomes more aggressive but does not collapse. The distribution is wide because both the model and the feature-normalization recipe can influence the score. Supplementary Figure~\ref{fig:mq-d2e8-explanation} expands on the non-monotonic ordering of D2-E8 and MQ.}\addressedbacklink{segmentation}{A13}}
    \label{fig:compression_iou_feature_correlation}
\end{figure*}

We investigated how much image fidelity can be sacrificed before biological signal degrades, evaluating compression on two complementary subsets of JUMP: a smaller pilot subset for detailed sweeps across compression levels, and the full JUMP-lite dataset for downstream evaluation. \rev{The pilot comprises 306 sample identifiers from the JUMP Target-2 set, each associated with evidence-supported protein targets \cite{chandrasekaranJUMPCellPainting2023}. The corresponding samples were imaged across four 384-well plates (approximately 9,200 images in total), with one plate per laboratory.} Replicates of each treatment appear across plates, providing both biological diversity and technical variability in imaging hardware, sample preparation, and acquisition settings. \rev{Section~\ref{sec:target2-identifier-accounting} expands on the dataset composition.}

To illustrate how imaging can convey perturbation differences, Figure \ref{fig:compression_impact_image}b shows a paired example of an untreated control and a treated sample. 
As a check on the visual impact of compression, we computed the Structural Similarity Index (SSIM) between original and \rev{JPEG XL-compressed} images at each compression level (Figure \ref{fig:compression_impact_plot}). Even at the most aggressive setting, all images retained \textit{SSIM} $>$ 0.95.

\textbf{Segmentation.} \rev{A common task in image-analysis pipelines is the segmentation of cells or nuclei.}
In many classical image-based profiling pipelines, segmentation occurs early and influences all downstream processing. 
%\rev{We evaluated segmentation stability under compression by comparing masks from compressed images against Raw-derived masks using average precision at IoU = 0.5, following previous methods \cite{caicedoNucleusSegmentationAcross2019}.} 
\rev{Mask agreement was well maintained across all compression levels (Figure~\ref{fig:compression_iou_feature_correlation}a), with AP values above 0.73. For comparison, expert annotators achieved an AP of 0.73 when they segmented similar data multiple times \cite{pachitariu2025cellpose}.} For matched cells (those with IoU above 0.5), the per-cell IoU likewise remained high across compression levels (Supplementary Figure \ref{fig:segmentation_iou}).

\textbf{Feature correlation.} Many features extracted via cp\_measure \cite{munozCp_measureAPIfirstFeature2025} are directly interpretable and useful to biologists. \rev{Features from uncompressed images show strong Pearson correlation with those from compressed images (Figure~\ref{fig:compression_iou_feature_correlation}b).
Although feature correlations decrease as compression becomes more aggressive, the effect is modest. The low-correlation tail contains texture descriptors, such as Zernike features, that exhibit high variance even without compression (Figure~\ref{fig:sup_feature_correlation}).}

\textbf{Downstream biological tasks} The most convincing evaluation for image-based profiling is its performance on downstream tasks of interest such as phenotypic activity (PA) and phenotypic consistency (PC) \cite{kalininVersatileInformationRetrieval2025}. The PA-PC product shows that more aggressive compression levels do result in lower scores relative to the uncompressed images (Figure \ref{fig:compression_iou_feature_correlation}c). The median PA-PC product of the highest-fidelity setting (\rev{HQ}) is the highest, and that of the lowest-fidelity setting (\rev{D20}) sits at the bottom of the ranking. The ordering across intermediate levels is non-monotonic, reflecting sensitivity to model family and deterministic post-processing recipe as well as the limited statistical power of four plates; we addressed the latter by extending the analysis to the entirety of JUMP-lite. \rev{Supplementary Figure~\ref{fig:mq-d2e8-explanation} expands on the non-monotonicity of Figure~\ref{fig:compression_iou_feature_correlation}c.}

\subsection{\rev{Representations show three performance tiers}}

\rev{Having established the trend on the four-plate Target-2 subset, we scaled the evaluation to the entire JUMP-lite dataset. We evaluated multiple biological tasks using learned representations at Raw, HQ, MQ, and the aggressive D20 setting, alongside Raw-only CellProfiler and Cell Count references.
We report performance on PA, PC, and the MOTIVE retrieval tasks. The MOTIVE tasks use annotations for CRISPR knockout genes and two compound subsets (Diverse and Bioactive-library).
Absolute PA and PC differences for all groups relative to Raw are shown in Figure~\ref{fig:performance_delta_raw}. Equivalent MOTIVE cross-modality recall results are shown in Supplementary Figures~\ref{fig:main-results-combined-access-format}, \ref{fig:sub_motive_results_full}, and \ref{fig:sub_motive_delta_full}. Mean percentage changes across all tasks are summarized in Table~\ref{tab:codec_delta_pct_combined_summary}.

At HQ, the mean per-task change from Raw is $-1.2\%$. Per-task changes remain within $\pm 6\%$, and most remain within $\pm 3\%$, despite an approximately 97.3\% reduction in dataset size.
At MQ, the mean performance change across tasks is $-6.5\%$ at an approximately 98.9\% storage reduction. PA decreases by an average of 6--14\% across perturbation types, while PC and cross-modality recall decrease by up to 15\% in the most affected subsets.
Performance degrades substantially at D20, with PA decreasing by 20--34\% and PC by up to 31\%, while moderate settings (HQ and MQ) retain more signal. D20 is the maximum-information-loss setting provided by the JPEG XL implementation.}

% % Main task (RefChem) percent base performance degradation vs raw images
% UNREFERENCED TABLE - tab:codec_delta_pct
% \input{main/tables/may_8_codec_delta_pct_table_mean_only}

% % MOTIVE tasks percent base performance degradation vs raw images
% UNREFERENCED TABLE - tab:motive_codec_delta_pct_1pct
% \input{main/tables/motive_codec_delta_pct_table_1pct_noORF_mean_only}

% UNREFERENCED TABLE - tab:codec_delta_pct_combined_summary - not referenced in paper
\begin{table*}[htbp]
\small
\centering
\setlength{\tabcolsep}{2.5pt}
\caption{\rev{Mean percentage performance change relative to the Raw dataset. The models were MorphEM, SubCell, DINOv2, and OpenPhenom. RefChemDB reports NAP, and MOTIVE reports Recall@1\%. Per-tier aggregates are shown here; Supplementary Table~\ref{tab:codec_delta_pct_combined} shows per-model performance changes.}\addressedbacklink{operating-points}{A11}}
\label{tab:codec_delta_pct_combined_summary}
\begin{minipage}{0.7\linewidth}
\resizebox{\linewidth}{!}{%
\begin{tabular}{ll NNNN NN NN N NN N}
\toprule
 & & \multicolumn{4}{c}{} & \multicolumn{2}{c}{\textbf{\rev{RefChemDB}}} & \multicolumn{5}{c}{\textbf{MOTIVE}} & \multicolumn{1}{c}{\textbf{Mean}} \\
\cmidrule(lr){7-8} \cmidrule(lr){9-13} \cmidrule(lr){14-14}
 & & \multicolumn{6}{c}{\% NAP $\Delta$} & \multicolumn{5}{c}{\% Recall@1 $\Delta$} & \\
\cmidrule(lr){3-8} \cmidrule(lr){9-13}
 & & \multicolumn{4}{c}{PA} & \multicolumn{2}{c}{PC} & \multicolumn{2}{c}{CC} & \multicolumn{1}{c}{GG} & \multicolumn{2}{c}{CG $\to$ CRISPR} & \\
\cmidrule(lr){3-6} \cmidrule(lr){7-8} \cmidrule(lr){9-10} \cmidrule(lr){11-11} \cmidrule(lr){12-13}
\textbf{\rev{Compression setting}} & & {CRISPR} & {ORF} & {Divs} & {Bioact} & {Divs} & {Bioact} & {Divs} & {Bioact} & {CRISPR} & {Divs} & {Bioact} & {All} \\
\midrule
\rev{HQ} & & -4.1 & +1.1 & -2.7 & +1.4 & +1.3 & -6.0 & -1.4 & -2.8 & +0.1 & +1.1 & -1.8 & -1.2 \\
\midrule
\rev{MQ} & & -14.2 & -5.8 & -6.5 & -7.3 & -0.8 & -15.4 & -5.2 & -6.0 & -3.0 & +0.4 & -8.3 & -6.5 \\
\midrule
\rev{D20} & & -34.2 & -22.8 & -19.6 & -25.7 & -14.9 & -30.8 & -12.6 & -17.8 & -8.1 & -4.6 & -21.4 & -19.3 \\
\bottomrule
\end{tabular}%
}
\end{minipage}
\end{table*}

\begin{figure}[!ht]
    \centering
    \includegraphics[width=\columnwidth]{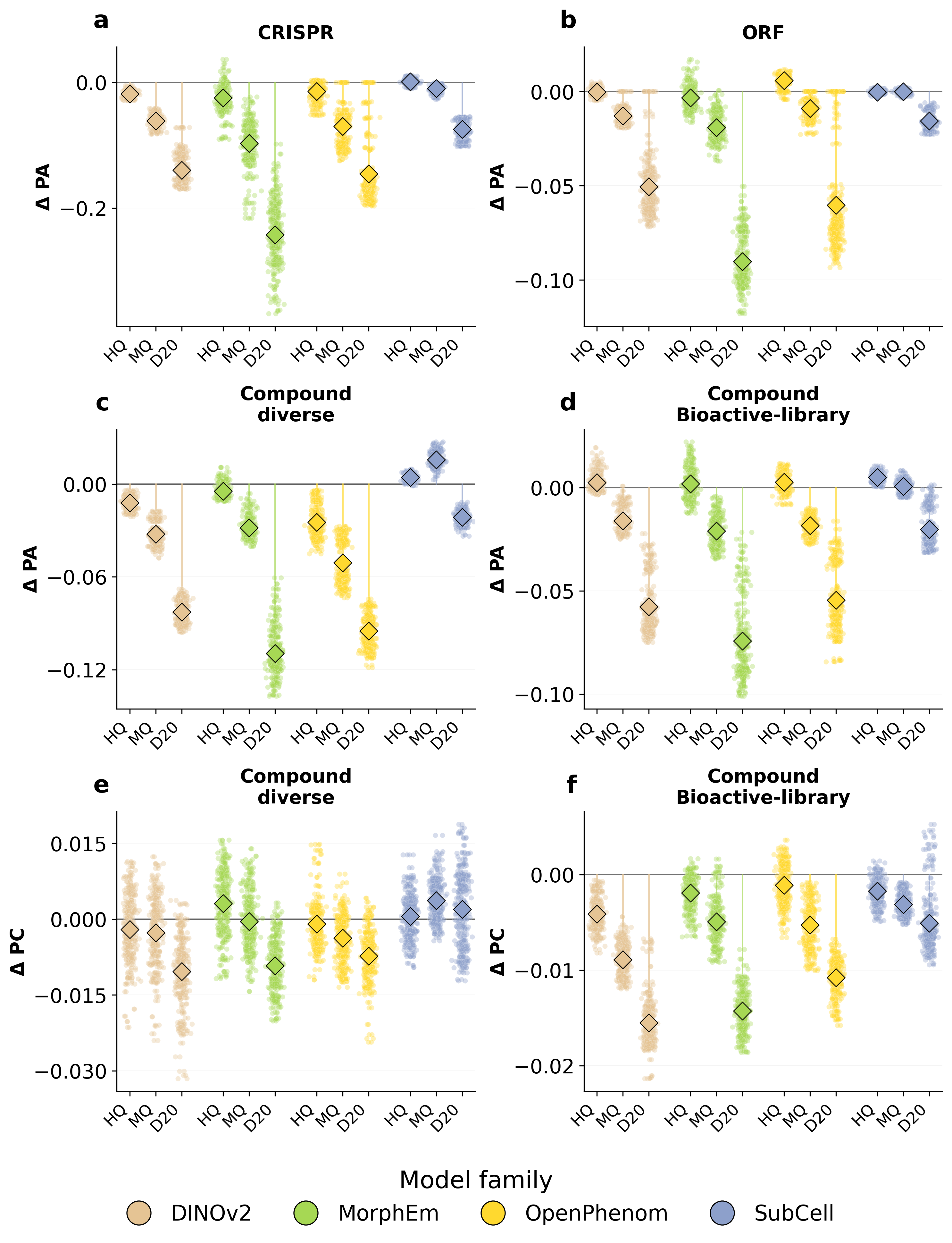}
    \caption{\rev{\textbf{JUMP-lite RefChemDB evaluation performance relative to Raw.} We evaluated the ability to retrieve known interactions annotated in RefChemDB. Each point represents one set of normalization parameters, and each diamond shows the mean across configurations. We use perturbation and negative-control annotations for PA and RefChemDB annotations for PC. \textbf{(a--d)} Phenotypic activity by perturbation type: CRISPR, ORF, Compound Diverse, and Compound Bioactive-library. \textbf{(e--f)} Phenotypic consistency for the Diverse and Bioactive-library compound subsets.}}
    \label{fig:performance_delta_raw}
\end{figure}

\begin{figure*}[!ht]
    \centering
    \includegraphics[width=\textwidth]{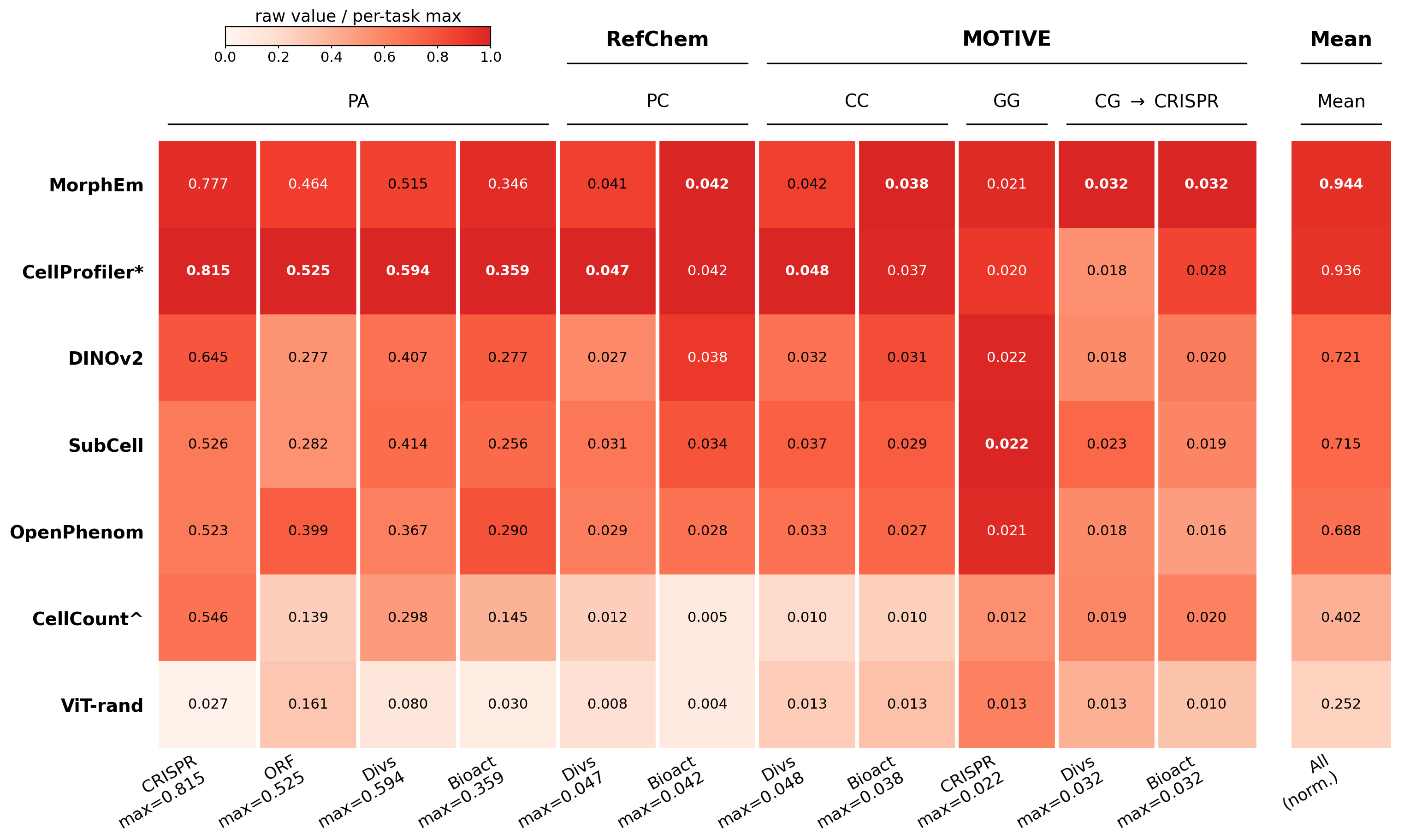}
    \caption{\rev{\textbf{Model performance per task on Raw images, normalized by the per-task maximum in each column.}
    Each cell shows a model's uncompressed performance on one task. Scores are NAP on six RefChemDB tasks and Recall@1\% on five MOTIVE tasks: compound--compound (CC) and compound--gene (CG) to CRISPR for the Diverse and Bioactive-library subsets, and gene--gene (GG) for CRISPR. Colors show the score divided by the maximum in each column (white is lowest and red is highest). Absolute maxima appear under the column labels.
    Representations are ordered vertically by the mean normalized value in the rightmost column. For each representation, we report the average across normalization configurations. Cell Count and CellProfiler are extracted and reported only at the uncompressed level.
    CellProfiler\textsuperscript{*} aggregates more than four sites per well, whereas the image-model representations use four. Cell Count\textsuperscript{\textasciicircum} was derived from those archived CellProfiler well profiles and therefore had access to the same additional sites. Bold values identify the highest-scoring displayed representation for each task and for ``All.''}\addressedbacklink{cellprofiler}{A5} }
    \label{fig:performance_ranking}
\end{figure*}

\rev{We next compared representations on Raw images to isolate representation performance. Figure~\ref{fig:performance_ranking} reports scores normalized by the task-specific maximum across four PA, two PC, and five MOTIVE retrieval tasks.}

\rev{The representations fall into three aggregate performance tiers. MorphEM (mean score 0.944) and CellProfiler features (0.936) form the leading aggregate tier ahead of DINOv2 (0.721), SubCell (0.715), and OpenPhenom (0.688), while individual-task leaders can differ. The Cell Count (0.402) and random-weight vision transformer (ViT; 0.252) baselines sit at the bottom.}
CellProfiler narrowly leads on phenotypic activity (e.g.\ 0.815 vs.\ 0.777 NAP on CRISPR, 0.594 vs.\ 0.515 on the Diverse subset), and the two are effectively tied on phenotypic consistency and on within-modality MOTIVE recalls (CC, GG). Their clearest separation appears in the cross-modality compound-to-CRISPR task (CG$\to$CRISPR), where MorphEM outperforms CellProfiler's recall on the Diverse subset (0.032 vs.\ 0.018). 
\rev{An audit of models whose training data includes part of JUMP is shown in Supplementary Section \ref{sec:model-audit}.}

\rev{To evaluate the stability of the observed ranking, we performed sensitivity tests (Figures~\ref{fig:sup_rank_stability} and~\ref{fig:main-results-combined-access-format}). Across Raw, HQ, and MQ, MorphEM remains the leading learned representation, and the overall performance tiers persist. However, exact middle-model ranks are not uniformly resolved. CellProfiler and Cell Count remain Raw-only descriptive references because their archived well profiles have no compression sweep and had access to more than four sites. Both are excluded from compressed-setting rankings. The random ViT baseline trails consistently and is insensitive to compression level because it carries no biological signal.}

\rev{The results in Figure~\ref{fig:performance_ranking} were calculated on uncompressed data. Figure~\ref{fig:performance_ranking_mq} reports MQ results for the four learned representations and ViT-rand. Raw-only CellProfiler and Cell Count are excluded before per-task normalization, winner identification, mean calculation, and row ordering. Subsampling supported the broad performance tiers, but no consistent ordering emerged within the middle tier (Section~\ref{sec:rank-subsampling}, Figure~\ref{fig:rank-subsampling}).}

\rev{In the exploratory separately optimized sweep, differences among learned models are generally larger than within-model variation across compression levels (Supplementary Figure \ref{fig:main-results-combined-access-format}).}
%The held-out fixed-recipe analysis provides the confirmatory learned-model contrasts and their conditional uncertainty (Supplementary Section \ref{sec:heldout-postprocessing}).
\rev{We provide a more detailed breakdown of performance across all tasks and model--compression combinations in Supplementary Figures~\ref{fig:main-results-combined-access-format}, \ref{fig:sub_pa_pc_results_full}, and \ref{fig:sub_motive_results_full}.}

Simple baselines remain informative in this comparison: cell count alone (mean 0.402) matches mid-tier deep-learning representations on CRISPR phenotypic activity, indicating that gross changes in cell number carry meaningful signal on genetic perturbations independent of finer morphology, and reinforcing the value of reporting such baselines alongside learned representations.

\rev{Taken together, these results show that JUMP-lite combined with \textit{Nahual} supports systematic benchmarking of learned cell image representations. Moderate JPEG XL compression broadly retains learned-representation performance tiers, with MorphEM leading learned models across compression settings, although exact middle-model ranks remain uncertain. CellProfiler remains competitive as a descriptive Raw reference, although its archived profiles use more sites per well.}

\section{Discussion}

% --- Step 1: Problem ---

\rev{Phenotypic profiling of large-scale Cell Painting datasets requires substantial storage and compute, which prevents many researchers from benchmarking cell representations.}

% --- Step 2: Core finding (compression preserves signal) ---

\rev{Our central finding is that lossy JPEG XL compression lowers this barrier substantially. Storage drops by 98.9\% while most of the biological signal is preserved. We first established this result on a small Target-2 pilot subset, where image similarity, segmentation agreement, and downstream phenotypic scores were retained under compression. We then confirmed the result on the full JUMP-lite dataset, which draws on six laboratories and includes annotation subsets for eleven biologically meaningful tasks.}

% --- Step 3: What this enables (shareable benchmarking resource) ---

\rev{Combined with our curation of a compact JUMP subset and the current aggregate $\sim$135$\times$ output rate advantage of the five deep learning models over the Cellpose plus cp\_measure MQ and lossless queues, this compression yields an easily shareable resource that supports benchmark runs on modest infrastructure. Users can follow the protocol we present here or use JUMP-lite as a starting point for more detailed evaluations.}

% --- Step 4: Benchmark findings ---

\rev{In the learned-representation comparison across compression settings, MorphEM leads across Raw, HQ, MQ, and D20, and broad performance tiers persist, although exact middle-model ranks are not uniformly resolved. CellProfiler is competitive only in the separate Raw comparison, which uses different numbers of sites per well, and has no compression sweep. A separate paired cp\_measure sensitivity analysis measures feature preservation and all-site object counts rather than benchmark scores, so we make no CellProfiler ranking claim across compression settings. Aggressive D20 compression causes appreciable degradation, whereas moderate settings retain substantially more signal.}

% --- Step 5: Adaptable framework ---

An additional contribution is the infrastructure required for rapid benchmarking. \rev{Researchers can adapt this infrastructure to more focused perturbation subsets, alternative annotations, or even single-cell evaluations.} We expect practitioners to adjust JUMP-lite to match their exact scientific or methodological question.

% --- Step 6: Caveats ---

Our reported rankings should be read with a few caveats. We evaluated each representation out-of-the-box, using its published normalization pipeline and default channel configuration; tuned normalization or task-specific adjustments could shift the ordering. \rev{MorphEM and OpenPhenom also have documented training exposure to JUMP-derived images. This exposure may give either model an edge over methods that have not encountered similar data. The full-scale CellProfiler evaluation is not fully comparable because it uses more sites per well than the learned-representation evaluation.} Our primary metrics (phenotypic activity and consistency) prioritize interpretability over completeness, and future work could complement them with batch-effect measures such as kBET and silhouette score. \rev{Finally, our compression evaluation is limited to five-channel Cell Painting and does not extend to bright-field or other immunofluorescence imaging, or to finer-grained downstream tasks such as subcellular protein localization.} We leave these directions to future work.

\rev{We release JUMP-lite images both as a lossless Zstd copy used as the Raw baseline and at the three evaluated JPEG XL levels (HQ, MQ, and D20).} Users can choose the fidelity-storage trade-off appropriate for their task. \rev{Our newly released library \textit{Nahual} provides straightforward deployment of a set of representation models and simplifies the addition of new methods.} We envision that these resources will lower the barrier for developing, deploying and evaluating new representation methods as the field continues to grow.

\section*{Data availability}
Code and release documentation are publicly available for JUMP-lite
(\url{https://github.com/afermg/JUMP_lite}) and Nahual
(\url{https://github.com/afermg/nahual}). The JUMP-lite software release is
archived on Zenodo (\url{https://doi.org/10.5281/zenodo.21779243}), and the
release indices and related tables are archived separately
(\url{https://doi.org/10.5281/zenodo.18705140}). JUMP-lite v1.0 has been
deposited in the Cell Painting Gallery and awaits gallery promotion.

{
    \small
    \bibliographystyle{ieeenat_fullname}
    \bibliography{bibliography}
}

\clearpage
\appendix
\renewcommand{\thesection}{S\arabic{section}}
\renewcommand{\thefigure}{S\arabic{figure}}
\renewcommand{\thetable}{S\arabic{table}}

\setcounter{figure}{0}
\setcounter{table}{0}

\newpage
\section{Supplementary Material}
\label{supplementary}

\subsection{\rev{Data accounting}}
\label{sec:target2-identifier-accounting}

\rev{The Target-2 analysis contains 306 non-negative-control \texttt{Metadata\_broad\_sample} identifiers but 302 \texttt{Metadata\_pert\_iname} labels because BVT-948, ME-0328, dexamethasone, and thiostrepton each map to two sample identifiers; 306 is the sample-level count reported in the main text. The release contains 163,776 wells and 655,101 image sites: 163,773 wells have four sites and three have three. Its 24,356 modality-specific identities comprise 24,347 treatments: 3,775 compound JCP identifiers, 7,974 CRISPR genes, and 12,598 ORF genes, plus nine controls (one compound negative; two CRISPR negative and one positive; four ORF negative and one positive).}

\rev{All 3,776 retained compound identities map to the union of MOTIVE's compound--compound and compound--gene graphs. RefChemDB was joined only downstream for evaluation: 1,526 compounds have a deposited RefChemDB row, while 2,250 do not. The bioactive-library and diversity compound groups were filtered separately by source group. JUMP-lite also contains 7,977 CRISPR-targeted genes and 12,603 ORF-overexpressed genes. The 5,245 genes shared by CRISPR and ORF are counted once per modality; together with the 3,776 compounds, this yields 24,356 modality-specific perturbations. Supplementary Figure~\ref{fig:overlap_per_source} visualizes the related downstream target-set overlaps across the compound, CRISPR, and ORF components; its target-level counts are not perturbation counts.}

% \subsection{\rev{Target-2 compression robustness}}

% \rev{The four-plate Target-2 curves are descriptive because each model--compression-setting pair was optimized separately. With one Zstd-selected recipe per learned representation fixed across Zstd, D10, and D15, 50,000 paired bootstrap resamples left D15--D10 PA--PC product differences unresolved for DINOv2, MorphEM, and OpenPhenom after Holm correction; only SubCell supported D15 over D10 ($+0.00879$, pointwise 95\% interval $[+0.00485,+0.01312]$, $p_{\mathrm{Holm}}=0.0008$). PA and PC were resampled independently, so the interval omits their covariance and is conditional rather than end-to-end. Separately, 87.6--94.3\% of 2,000 fixed-recipe, stratified 306-identifier subsamples contained an ordering reversal between adjacent compression settings. This is not a literal four-plate or D10/D15 replication, and the results support neither a universal monotonic ordering nor a denoising interpretation. The pilot set the levels for full-scale evaluation of four learned representations across compression settings; CellProfiler and Cell Count remained Raw-only references.}

\subsection{\rev{Figure 3c non-monotonicity}}
\label{sec:mq-d2e8-explanation}

\rev{With one Zstd-selected recipe per family fixed across compression settings, D2-E8 was supported over MQ only for MorphEM after Holm-Bonferroni correction; the other four contrasts were unresolved (Figure~\ref{fig:mq-d2e8-explanation}).}

\subsection{\rev{Broader-JUMP coverage and compound annotation records}}
\label{sec:broader-jump-coverage}

\rev{JUMP-lite's 3,775 retained compounds occupy 123 of 128 operational clusters. An unplotted sensitivity analysis at $K=64$ found 62 to 64 occupied clusters across five seeds and total variation distances of 0.386 to 0.425 from the eligible broader JUMP compound frequencies. Figure~\ref{fig:cluster-selection-coverage} also shows that existing MOTIVE records cover all 3,775 JUMP-lite compounds and 573 of 91,915 other eligible compounds.}

\subsection{\rev{Strict treatment-disjoint post-processing evaluation}}
\label{sec:heldout-postprocessing}

\rev{We split 26,877 unique \texttt{Metadata\_JCP2022} reagent identifiers before fitting any post-processing operation. This unit differs from the biological-treatment accounting because distinct ORF reagents can map to the same biological identity. In the primary split (seed 20260811), 5,375 identifiers were assigned to recipe selection and 21,502 were assigned to held-out evaluation, with no overlap. We selected one recipe per representation family using Raw recipe-selection PA and PC, then refit that fixed recipe separately for each supported compression setting using only that setting's recipe-selection treatments and shared controls. Figure~\ref{fig:heldout_fixed_recipe_codec} reports PA and PC across five independent end-to-end split replicates. Each replicate repeats the reagent-identifier split, Raw recipe selection, compression-setting-specific refitting, and held-out evaluation. Whiskers are 95\% Student-t intervals across seeds.}

\subsection{\rev{Model-ranking subsample sensitivity}}
\label{sec:rank-subsampling}

\rev{To assess whether representation tiers depend on the available treatment identities, we recomputed the 11-task Raw and MQ rankings across 20 deterministic 80\% subsamples with frozen transforms and recipes (Figure~\ref{fig:rank-subsampling}).}

\subsection{Methods}

\subsubsection{Data Curation}
\label{sec:sources-dose}
We selected an annotation-supported subset \rev{of JUMP. Our focus was to use annotated perturbations, necessary for most biologically-relevant benchmarks, not necessarily being representative of the compound-focused JUMP.}

\begin{figure*}[!ht]
    \centering
    \includegraphics[width=0.92\textwidth]{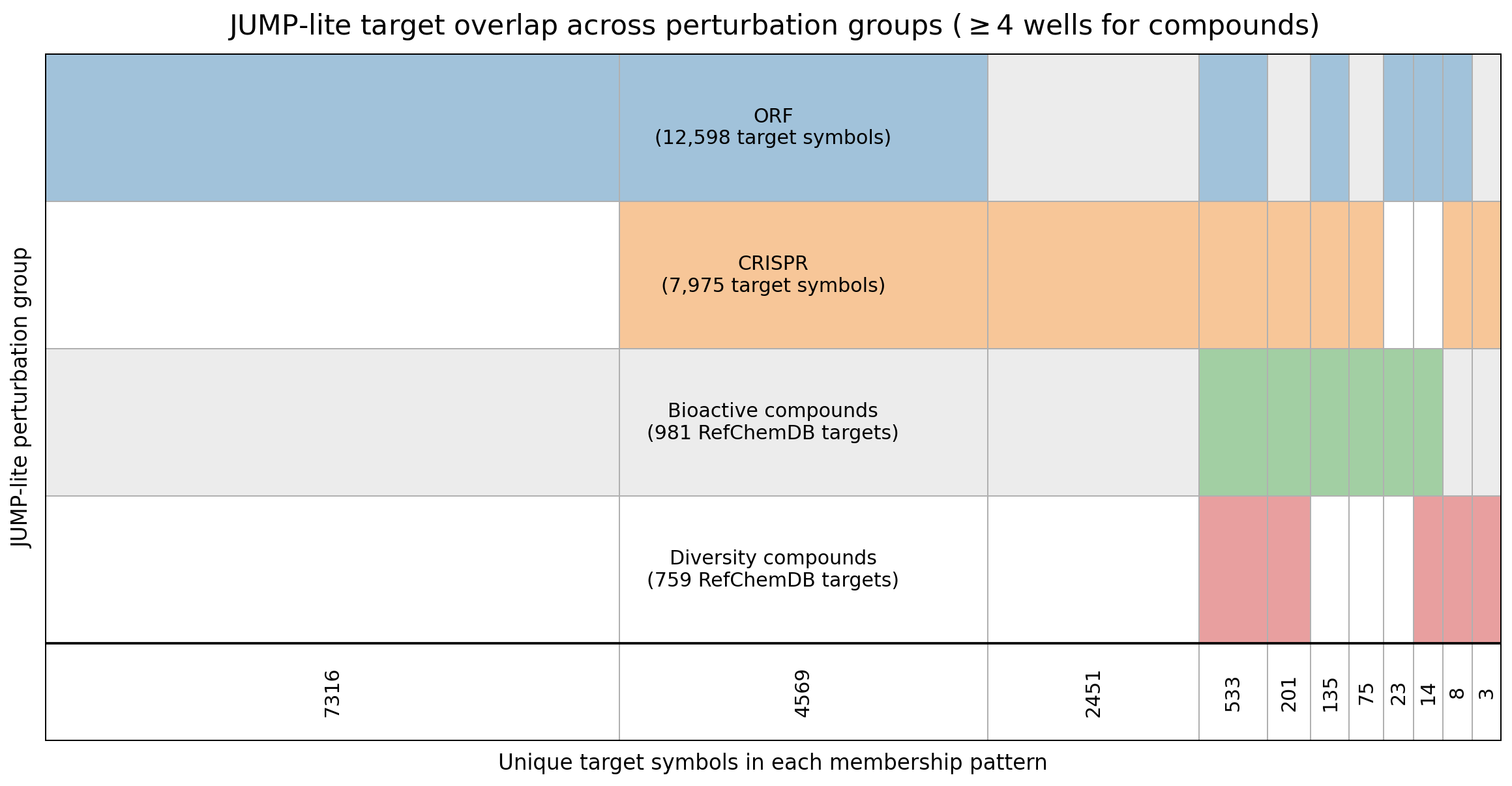}
    \caption{\rev{\textbf{Target overlap across JUMP-lite perturbation groups.} Each colored row is one target set, and each numbered vertical segment counts unique symbols with the indicated membership pattern. Compound sets use downstream RefChemDB Tier~1--3 targets linked to retained treatment compounds (759 diversity-source and 981 bioactive-source targets); RefChemDB was not an inclusion filter. CRISPR and ORF sets use retained genetic symbols after excluding negative controls and missing-value placeholders (7,975 and 12,598, respectively); modality-specific positive controls are retained. Compounds can map to multiple targets, and set totals overlap; these target-level counts therefore are not perturbation counts and should not be summed.}\addressedbacklink{perturbations}{A8}}
    \label{fig:overlap_per_source}
\end{figure*}

\rev{The planned protocol included available CRISPR and ORF perturbations and formed the compound candidate pool from JUMP identifiers whose InChIKey connectivity layer mapped into the union of MOTIVE's compound--compound or compound--gene graphs. RefChemDB was reserved for eligible target-based evaluation. We removed source~9, Target-2 plates, plates with less than 25\% selected-well occupancy, established quality-control/redlist plates, and negative-control-only plates; intersected wells with the available profile table; retained compound identifiers with at least four wells separately within the diversity-source and bioactive-source groups; and added the corresponding modality-specific negative controls. The frozen well manifest contains the resulting 163,776 analysis/release wells. Downstream RefChemDB target overlap across the compound and genetic groups is shown in Figure~\ref{fig:overlap_per_source}; it describes annotation coverage rather than a selection criterion.}

JUMP was generated at 12 partner sources using different instruments \cite{chandrasekaranJUMPCellPainting2023}. Source 7 profiled the bioactive library at 0.625 $\mu$M; three other sources profiled diversity compounds at 10~$\mu$M (Table \ref{tab:jump-lite-summary}).

Compound group, dose, source, and annotation coverage are confounded: 52\% of bioactive compounds occur at no other source, and 64\% have evaluation annotations versus 3--11\% at the large diversity sources. Group differences therefore do not isolate compound type.

\subsubsection{Image compression}
\label{sec:compression}
Each five-channel site was stored as a 3-D array in a \rev{directory specific to each compression setting}, with metadata-preserving names traceable to the Cell Painting Gallery \cite{weisbartCellPaintingGallery2024}. \rev{The full-release mapping is Raw/Zstd at compression level 9 (4.8 TB), HQ/\texttt{jpegxl\_lossy\_hq} at JPEG XL distance 1.0 with default effort (237.7 GB), MQ/\texttt{jpegxl\_lossy\_mq} at distance 3.0 with default effort (92.0 GB), and D20/\texttt{jpegxl\_lossy\_d20} at distance 20.0 with default effort (16.2 GB). Table~\ref{tab:zarr_benchmark} maps every Target-2 pilot label and reports pilot-store sizes and per-array timings. The pilot D20 store used effort 2, unlike the full-release D20 store.}

\begin{figure}[ht]
    \centering
    \includegraphics[width=\columnwidth]{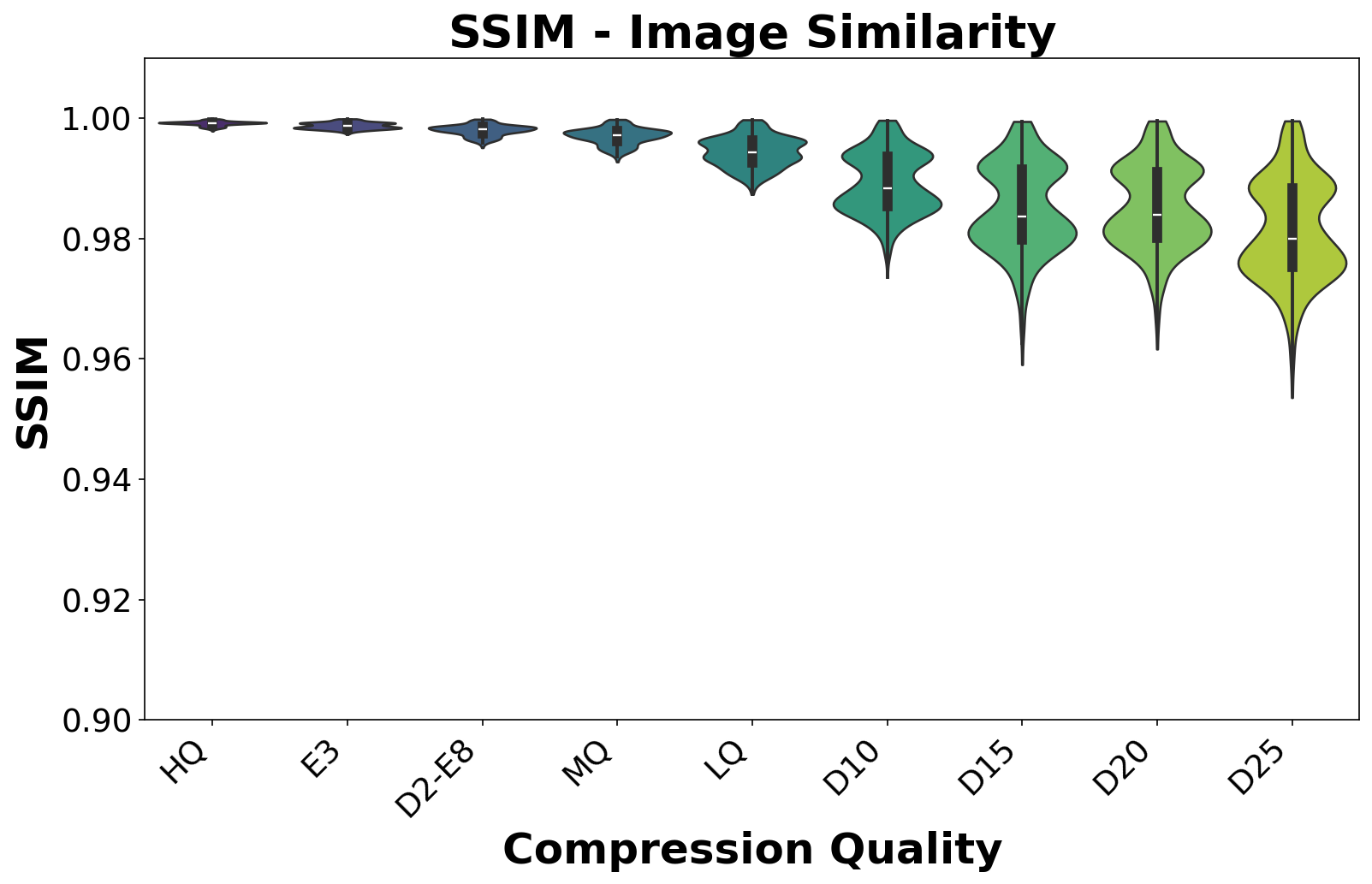}
    \caption{Image quality and similarity measure using Structural Similarity Index Measure (SSIM).}
    \label{fig:compression_impact_plot}
\end{figure}

SSIM between compressed and Raw images exceeded 0.96 at every level (Figure \ref{fig:compression_impact_plot}).

% \subsubsection{\rev{Compression reduces detected object counts}}
%
% \rev{To examine segmentation sensitivity to signal loss, we compared paired Cellpose object counts from MQ and lossless images and found a larger reduction for cells than nuclei (Figure~\ref{fig:paired-cp-measure-mq-lossless}).}

% Structure should be:
% Data Collection->Methods (curation, compression, featurization, parameter exploration) ->Benchmarks
\subsubsection{Image featurization pipeline}
ALIBY orchestrated two pipelines \cite{munozPhenotypingSingleCells2023}: segmentation followed by per-cell features, and tiled-image preprocessing followed by deep-learning embeddings. Nahual provided isolated model environments. Model-specific tiling, channels, and scaling followed their documentation (Table \ref{table:model_overview}); OpenPhenom additionally used 8-bit conversion \cite{recursionpharmaOpenPhenomModelCard2026}, and SubCell used min--max scaling \cite{gupta2025subcell}.

% Code: https://github.com/afermg/JUMP_core/blob/main/analysis/aliby_featurize.py#L31-L51
% Org mode to generate table: https://github.com/afermg/JUMP_lite/blob/main/draft/tables.org#L5-L12

\subsubsection{Model deployment: \textit{Nahual}}
\label{sec:methods_nahual}
Nahual exchanges parameter dictionaries and NumPy image arrays between ALIBY and model-specific processes. Each model and Cellpose ran in an isolated Nix Flake to avoid incompatible TensorFlow/PyTorch dependencies \cite{devresseNixBasedFully2015,dolstraNixSafePolicyFree2004}. Nahual was kept separate from ALIBY so model environments could remain fixed independently of pipeline development.

\subsubsection{Profile processing}
\label{sec:profile-processing}
~\\

We expanded an existing standardized post-processing pipeline developed by Arevalo et al. \cite{arevaloMOTIVEDrugTargetInteraction2024}. For this profile-processing pipeline we: (1) \rev{dropped features whose portion of missing or non-finite values exceeds 30\%, then dropped rows containing a missing or non-finite value in any retained feature}; (2) removed low-variance features using frequency and uniqueness cutoffs, \rev{dropping features whose unique values represent less than 1\% of the total or whose most common value occurs more than 20 times as often as the second most common}; (3) per-plate standardization via robust-MAD, calculating the median and MAD of the DMSO negative controls to center and rescale features relative to these controls; (4) drop outlier features with z-score exceeding 100; (5) for CellProfiler features, we enforced Gaussian distributions through a rank-based inverse normal transformation; (6) \rev{for CellProfiler, removed redundant features based on pairwise correlation (Pearson \(|r| > \text{threshold}\)) using a greedy independent-set algorithm fitted without held-out treatments}; and (7) batch correction via PCA-TVN (Typical Variation Normalization using CORAL-based covariance alignment on PCA transformed embeddings) \cite{celikBuildingBenchmarkingExploring2024} to reduce plate-to-plate technical variation. These steps collectively remove noise, confounding technical signal, and redundant features.

\rev{We used Hydra to search the profile-processing settings. For CellProfiler, we swept two normalization methods, two fitting cohorts for plate adjustment, two correlation thresholds, five PCA-TVN regularization values, and seven component counts. CellProfiler includes correlation pruning because its hand-engineered intensity, texture, and shape measurements contain many redundant features. The Cartesian product gives $2\times2\times2\times5\times7=280$ distinct CellProfiler pipelines. Figure~\ref{fig:performance_delta_raw} shows one point per pipeline.}

\rev{For Target-2, we used a smaller grid because the dataset was smaller. We varied inverse-normal transformation, outlier clipping, and low-variance filtering in Target-2, and we fixed those choices in the JUMP-lite sweep. For learned embeddings, the nominal grid used robust-MAD, standard z-scoring, or no rescaling, two fitting cohorts, two values of \texttt{use\_prune\_correlated}, five regularization values, and seven component counts, giving up to 420 directory configurations. Removing redundant fitting-cohort choices when no rescaling was used left 350 directory aliases per learned family. The resolved learned pipeline has no correlation-pruning step, so the two values of \texttt{use\_prune\_correlated} produce the same processing steps and collapse to 175 effective pipelines per family. The learned pipeline omits separate correlation pruning because its axes are learned jointly and PCA-TVN already models their covariance. The strict rerun preserves the historical pipeline and does not test whether adding correlation pruning would improve learned representations. Across five learned families, CellProfiler, and five Cell Count pipelines, the strict inventory contains 2,035 aliases and 1,160 effective pipelines. We also omitted inverse-normal transformation for learned embeddings because their output distributions are approximately normal.}

When assessing the impact of lossy compression on downstream tasks (Figure \ref{fig:model_comparison_4plate})  we swept all parameter settings for all combinations of model and compression level separately. Every configuration was evaluated using phenotypic activity and consistency metrics. The post-processed profiles from the best-performing configuration were then used for downstream analyses. We selected the settings with the highest PA-PC product for each model across compression levels. For configuration selection, we compute this product after min--max rescaling each metric to $[0, 1]$ over the swept configurations. This single scalar enables ranking across configurations that may trade off one metric against the other.

\subsubsection{\rev{Audit of models trained on JUMP}}
\label{sec:model-audit}
\rev{MorphEM and OpenPhenom were loaded from Hugging Face. Matching collection, source, batch, plate, filename stem, and channel in the current CHAMMI-75 inventory named by MorphEM's card identified 138,945 channels from 27,789 sites, 19,157 wells, and 261 plates (4.24\%, 11.70\%, and 47.37\% of JUMP-lite) \cite{caicedolabMorphEmModelCard2026}. OpenPhenom documents did not release details on the plates or sources used \cite{recursionpharmaOpenPhenomModelCard2026}.}

\rev{To evaluate sensitivity to potential JUMP exposure during MorphEM pretraining, we repeated the full split, Raw recipe selection, refitting, and held-out evaluation across five seeds after two exclusions. Direct matching removed 12,132--12,341 held-out wells per split and retained 107,616--109,790 wells; excluding every held-out well on a matched plate retained 72,204--74,137 wells. Figure~\ref{fig:pretraining-overlap-sensitivity} reports mean absolute Raw PA$\times$PC with two-sided 95\% Student-t intervals across seeds and marks MorphEM's same-seed full-held-out interval. MorphEM remained highest in both subsets. This analysis uses acquisition identity as an overlap proxy and does not establish a causal pretraining effect or leakage-free evaluation. The sites used to train OpenPhenom were not explicitly documented.}

\subsubsection{\rev{Rank sensitivities}}

\rev{To test whether the JUMP-lite tiers shown in Figure~\ref{fig:performance_ranking} depend on individual sources or target groups, we reaggregated the frozen results after informative source deletions and under paired target-grouped resampling.}

\subsubsection{CellProfiler feature comparability} 
For the full JUMP-lite evaluation, we used precomputed CellProfiler features from the Cell Painting Gallery \rev{only at Raw}. They were derived from six to nine imaging sites per well, compared with \rev{up to four sites} per well for the learned models. This site-count and sampling asymmetry \rev{prevents a controlled cross-representation comparison}; using the precomputed features nevertheless \rev{supplies a useful descriptive classical reference} while keeping the full-scale analysis tractable.

% \paragraph{\rev{Interim paired MQ/lossless cp\_measure sensitivity.}}
% \label{sec:cp-measure-paired-sensitivity}
% \rev{To assess CellProfiler feature preservation under compression despite the archived site-count mismatch, we processed paired Cellpose and cp\_measure outputs under MQ and lossless inputs. Figure~\ref{fig:paired-cp-measure-mq-lossless} expands the object-count component to all 632,672 paired sites in the frozen manifest (158,168 wells across 551 plates).}

\subsubsection{\rev{Pooled and compound profile-space clustering and annotation records}}
\label{sec:compound-clustering-methods}
\rev{To contextualize the annotation-focused compound selection relative to the entire JUMP dataset, we overlaid JUMP-lite perturbations on the JUMP UMAP and compared compound occupancy in label-blind $K=64$ and $K=128$ partitions (Figure~\ref{fig:cluster-selection-coverage}).}

\subsubsection{Statistical analyses}
\textbf{Phenotypic activity and consistency.} Using \textit{copairs} \cite{kalininVersatileInformationRetrieval2025}, PA measures retrieval of each perturbation's replicates against same-plate controls, while PC measures retrieval of compounds sharing RefChemDB targets against compounds with different targets \cite{judson2018workflow}. Statistical significance was assessed by per-perturbation mAP permutation tests with Benjamini--Hochberg correction. PA and PC were computed independently over all perturbations; PC was not restricted to phenotypically active compounds.

\textbf{Normalized Average Precision (NAP).} NAP adjusts mAP for its expected random baseline so random performance maps to zero, enabling comparisons across settings with different positive/negative counts. \rev{For PA, average precision is computed per replicate; for PC, it is computed per perturbation. Each value is normalized against its expected value under random ranking before averaging. The reported PA-PC product is $\mathrm{PA}\times\mathrm{PC}$; only configuration selection uses PA and PC min--max rescaled over the swept configurations.}

\subsection{Benchmark Annotations}
\label{sec:annotations}

\subsubsection{MOTIVE annotations}
\rev{MOTIVE evaluates whether profiles retrieve known partners in three relationship graphs. Compound--gene (CG) tests whether compound profiles retrieve ORF or CRISPR profiles of annotated targets. Compound--compound (CC) tests whether compound profiles retrieve related compounds, while gene--gene (GG) tests whether genetic profiles retrieve related genes within ORF or CRISPR.}

\rev{We started from the published MOTIVE tables, which contain 408,787 CG edges across 36 relationship types, 9.99 million CC edges across 16 types, and 2.98 million GG edges across 16 types. We mapped compounds from InChIKeys and genes from HGNC symbols to available JCP2022 perturbations. After mapping and deduplication, the full set contains 301,233 CG, 1,407,910 CC, and 2,459,220 GG edges. The full set documents annotation coverage, but it is not used to score the retrieval results in this paper. The published CG and CC records, before we constructed the strict set, were used earlier to define the annotation-linked compound candidate pool.}

\rev{Counts refer to directed evaluation rows rather than unique undirected biological relationships. CC and GG pairs are stored in both directions. GG connects perturbations within the same genetic modality, while one CG annotation can contribute separate rows for available ORF and CRISPR reagents.}

\rev{All five reported MOTIVE Recall@1\% tasks use a stricter annotation set. For CG, we kept target, binding, inhibition, activation, agonist, antagonist, and enzyme relationships. For GG, we kept protein interaction, gene binding, generic interaction, and post-translational modification relationships. The exact source labels are defined in \texttt{scripts/curate\_motive.py}.}

\rev{We rebuilt strict CC instead of filtering the full CC graph. Most raw CC records describe structural resemblance, drug interactions, or synergy, which do not necessarily imply similar morphology. We therefore linked two compounds when they shared at least one target retained in strict CG.}

\rev{The strict set contains 75,955 CG edges, 997,484 derived CC edges, and 1,134,062 GG edges, for 2,207,501 edges in total. These edges define the positive pairs used in every reported MOTIVE result.}

\subsubsection{RefChem annotations}
\rev{To prevent selection--evaluation leakage, we selected compounds through the MOTIVE union and reserved supported RefChemDB targets for downstream PC evaluation (Figure~\ref{fig:overview}).}

\subsection{Data and Infrastructure}

\subsubsection{Code and data}
\label{sec:artifact-availability}
Public repositories are available for JUMP-lite (\url{https://github.com/afermg/JUMP_lite}) and Nahual (\url{https://github.com/afermg/nahual}). JUMP-lite v1.0 images, profiles, annotations, and segmentation artifacts have been deposited in the Cell Painting Gallery and await gallery promotion. Release indices and related tables are archived on Zenodo (\url{https://doi.org/10.5281/zenodo.18705140}). We provide a machine-readable MLCommons Croissant 1.0 file for the five release metadata tables and release manifest. A Responsible AI artifact is not claimed.

\subsubsection{Hardware}
Experiments ran on a Gigabyte G493-ZB1 server with four NVIDIA H100 GPUs, 1.5 TB RAM, and an AMD EPYC 9684X 96-core CPU, \rev{running NixOS 25.11 as the Operative System}.

\subsection{Supplementary Tables}

% Supplementary table declarations are ordered by first explicit reference in
% the main/SI manuscript text. Tables not explicitly referenced in the paper
% follow the SI section they support.

\begin{table*}[!ht]
\centering
\caption{\textbf{\rev{Feature-extraction rates.}} \rev{Learned-model rates were calculated from the 9,216-site pilot. The Cellpose plus cp\_measure rate was measured with the newest cp\_measure version by summing the MQ and lossless queues.}\addressedbacklink{terminology}{A15}}
\small
\begin{tabular}{@{}p{0.48\textwidth}p{0.30\textwidth}@{}}
\hline
\rev{Pipeline/model} & \rev{Reported rate}\\ \hline
\rev{DINOv2} & \rev{17,076 sites/hour}\\
\rev{ViT-rand} & \rev{17,078 sites/hour}\\
\rev{MorphEM} & \rev{13,104 sites/hour}\\
\rev{OpenPhenom} & \rev{22,752 sites/hour}\\
\rev{SubCell} & \rev{16,248 sites/hour}\\
\rev{Cellpose plus cp\_measure} & \rev{332 sites/hour}\\
\hline
\end{tabular}
\label{table:speed_benchmark}
\end{table*}

\begin{table*}[htbp]
\centering
\tiny
\setlength{\tabcolsep}{2.5pt}
\caption{\rev{Mean percentage performance change from the Raw (lossless Zstd) baseline ($\pm$ standard deviation across normalization configurations). RefChemDB reports NAP; MOTIVE reports recall@1\%.}}
\label{tab:codec_delta_pct_combined}
\resizebox{\textwidth}{!}{%
\begin{tabular}{ll NNNN NN NN N NN N}
\toprule
 & & \multicolumn{4}{c}{} & \multicolumn{2}{c}{\textbf{\rev{RefChemDB}}} & \multicolumn{5}{c}{\textbf{MOTIVE}} & \multicolumn{1}{c}{\textbf{Mean}} \\
\cmidrule(lr){7-8} \cmidrule(lr){9-13} \cmidrule(lr){14-14}
 & & \multicolumn{6}{c}{\% NAP $\Delta$} & \multicolumn{5}{c}{\% Recall@1 $\Delta$} & \\
\cmidrule(lr){3-8} \cmidrule(lr){9-13}
 & & \multicolumn{4}{c}{PA} & \multicolumn{2}{c}{PC} & \multicolumn{2}{c}{CC} & \multicolumn{1}{c}{GG} & \multicolumn{2}{c}{CG $\to$ CRISPR} & \\
\cmidrule(lr){3-6} \cmidrule(lr){7-8} \cmidrule(lr){9-10} \cmidrule(lr){11-11} \cmidrule(lr){12-13}
\textbf{\rev{Compression setting}} & & {CRISPR} & {ORF} & {Divs} & {Bioact} & {Divs} & {Bioact} & {Divs} & {Bioact} & {CRISPR} & {Divs} & {Bioact} & {All} \\
\midrule
  & DINOv2 & {-5.0 $\pm$ 3.5} & {+0.6 $\pm$ 3.3} & {-3.3 $\pm$ 1.2} & {+1.1 $\pm$ 2.3} & {-3.7 $\pm$ 17.3} & {-11.3 $\pm$ 4.1} & {+1.3 $\pm$ 5.0} & {-4.3 $\pm$ 2.8} & {-1.0 $\pm$ 4.1} & {-2.0 $\pm$ 12.9} & {-7.1 $\pm$ 10.0} & {-3.1} \\
  & MorphEm & {-3.6 $\pm$ 3.4} & {-0.6 $\pm$ 1.7} & {-0.9 $\pm$ 0.9} & {+0.9 $\pm$ 3.3} & {+7.7 $\pm$ 13.5} & {-4.0 $\pm$ 3.3} & {-1.1 $\pm$ 4.0} & {-2.0 $\pm$ 2.2} & {+1.2 $\pm$ 2.7} & {+7.0 $\pm$ 16.0} & {-3.8 $\pm$ 10.0} & {+0.1} \\
 \rev{HQ} & OpenPhenom & {-8.0 $\pm$ 14.6} & {+4.7 $\pm$ 26.3} & {-7.8 $\pm$ 4.1} & {+1.0 $\pm$ 2.7} & {-2.0 $\pm$ 15.3} & {-3.4 $\pm$ 6.5} & {-5.2 $\pm$ 4.4} & {-4.2 $\pm$ 3.4} & {-0.7 $\pm$ 3.4} & {-0.5 $\pm$ 13.4} & {+2.0 $\pm$ 9.6} & {-2.2} \\
  & SubCell & {+0.4 $\pm$ 1.4} & {-0.1 $\pm$ 0.5} & {+1.1 $\pm$ 0.6} & {+2.6 $\pm$ 1.3} & {+3.1 $\pm$ 14.6} & {-5.1 $\pm$ 3.7} & {-0.6 $\pm$ 4.6} & {-0.8 $\pm$ 2.6} & {+0.9 $\pm$ 2.8} & {-0.1 $\pm$ 12.6} & {+1.8 $\pm$ 7.8} & {+0.3} \\
 & \textit{\rev{Mean}} & -4.1 & +1.1 & -2.7 & +1.4 & +1.3 & -6.0 & -1.4 & -2.8 & +0.1 & +1.1 & -1.8 & -1.2 \\
\midrule
  & DINOv2 & {-16.4 $\pm$ 9.3} & {-9.2 $\pm$ 12.9} & {-8.7 $\pm$ 2.1} & {-8.6 $\pm$ 4.0} & {-5.3 $\pm$ 18.0} & {-24.5 $\pm$ 3.7} & {-5.1 $\pm$ 4.8} & {-9.9 $\pm$ 2.9} & {-4.0 $\pm$ 3.7} & {-4.0 $\pm$ 13.4} & {-13.4 $\pm$ 9.5} & {-9.9} \\
  & MorphEm & {-14.1 $\pm$ 6.3} & {-4.7 $\pm$ 2.1} & {-5.7 $\pm$ 1.2} & {-7.5 $\pm$ 2.8} & {-0.1 $\pm$ 12.5} & {-10.3 $\pm$ 4.3} & {-4.1 $\pm$ 3.7} & {-7.5 $\pm$ 2.5} & {-0.6 $\pm$ 3.1} & {+2.2 $\pm$ 13.3} & {-11.0 $\pm$ 9.2} & {-5.8} \\
 \rev{MQ} & OpenPhenom & {-23.6 $\pm$ 19.5} & {-9.0 $\pm$ 19.9} & {-15.8 $\pm$ 6.6} & {-13.3 $\pm$ 11.3} & {-10.5 $\pm$ 13.4} & {-17.1 $\pm$ 6.9} & {-9.8 $\pm$ 5.2} & {-3.8 $\pm$ 3.0} & {-4.1 $\pm$ 3.5} & {+7.0 $\pm$ 20.5} & {-7.1 $\pm$ 11.5} & {-9.7} \\
  & SubCell & {-2.6 $\pm$ 1.1} & {-0.1 $\pm$ 0.6} & {+4.0 $\pm$ 1.2} & {+0.2 $\pm$ 1.3} & {+12.7 $\pm$ 13.8} & {-9.4 $\pm$ 3.2} & {-1.8 $\pm$ 5.3} & {-2.7 $\pm$ 2.9} & {-3.3 $\pm$ 3.3} & {-3.5 $\pm$ 11.8} & {-1.8 $\pm$ 9.7} & {-0.8} \\
 & \textit{\rev{Mean}} & -14.2 & -5.8 & -6.5 & -7.3 & -0.8 & -15.4 & -5.2 & -6.0 & -3.0 & +0.4 & -8.3 & -6.5 \\
\midrule
  & DINOv2 & {-37.0 $\pm$ 17.5} & {-30.9 $\pm$ 23.9} & {-22.3 $\pm$ 3.0} & {-30.9 $\pm$ 12.2} & {-27.4 $\pm$ 15.1} & {-42.7 $\pm$ 6.0} & {-13.9 $\pm$ 5.2} & {-23.3 $\pm$ 2.4} & {-11.5 $\pm$ 4.2} & {-10.0 $\pm$ 14.5} & {-20.2 $\pm$ 10.3} & {-24.5} \\
  & MorphEm & {-35.1 $\pm$ 9.4} & {-22.5 $\pm$ 5.9} & {-21.9 $\pm$ 2.6} & {-26.2 $\pm$ 6.7} & {-19.4 $\pm$ 9.5} & {-30.1 $\pm$ 3.1} & {-17.5 $\pm$ 4.5} & {-20.0 $\pm$ 2.2} & {-8.6 $\pm$ 6.2} & {-5.3 $\pm$ 12.4} & {-37.2 $\pm$ 10.6} & {-22.2} \\
 \rev{D20} & OpenPhenom & {-43.7 $\pm$ 20.4} & {-30.9 $\pm$ 22.8} & {-28.9 $\pm$ 6.5} & {-35.7 $\pm$ 22.0} & {-21.8 $\pm$ 13.2} & {-35.2 $\pm$ 4.3} & {-13.9 $\pm$ 5.5} & {-17.1 $\pm$ 3.8} & {-5.0 $\pm$ 6.8} & {+3.5 $\pm$ 17.2} & {-12.9 $\pm$ 10.9} & {-22.0} \\
  & SubCell & {-20.9 $\pm$ 6.3} & {-7.0 $\pm$ 4.1} & {-5.5 $\pm$ 1.0} & {-10.2 $\pm$ 4.7} & {+8.9 $\pm$ 24.8} & {-15.2 $\pm$ 9.2} & {-4.9 $\pm$ 5.5} & {-11.0 $\pm$ 2.5} & {-7.3 $\pm$ 3.2} & {-6.6 $\pm$ 11.2} & {-15.4 $\pm$ 8.1} & {-8.7} \\
 & \textit{\rev{Mean}} & -34.2 & -22.8 & -19.6 & -25.7 & -14.9 & -30.8 & -12.6 & -17.8 & -8.1 & -4.6 & -21.4 & -19.3 \\
\bottomrule
\end{tabular}
}
\end{table*}

% Generated by rebuttal/cellpose_count_s12_v1/render_manuscript_table.py
\begin{table*}[p]
\begin{revblock}
\centering
\caption{\rev{\textbf{Raw and MQ finite-subsample scores.}
Scores are reported as mean $\pm$ standard deviation (SD) across 20 repeated 80\% subsamples. ``All'' is the per-repeat average of the 11 task scores after normalizing each task by that repeat's best displayed representation. Raw includes CellProfiler\textsuperscript{*} and legacy Cell Count\textsuperscript{\textasciicircum}; MQ omits these Raw-only references and adds measured Cellpose Counts\textsuperscript{\ensuremath{\dagger}} from four MQ sites per well. These are ranking-sensitivity summaries, not common-well estimates of compression effects. Bold marks the highest displayed mean; SD is descriptive.}\addressedbacklink{uncertainty}{A4}}
\label{tab:rank-subsampling-scores}
\scriptsize
\setlength{\tabcolsep}{2.4pt}
\renewcommand{\arraystretch}{0.96}
\textbf{(a) Detailed Raw scores}\par\vspace{0.15em}
\resizebox{\textwidth}{!}{%
\begin{tabular}{lccccccc}
\toprule
Task & MorphEM & DINOv2 & SubCell & OpenPhenom & ViT-rand & Cell Count\textsuperscript{\textasciicircum} & CellProfiler\textsuperscript{*} \\
\midrule
PA--CRISPR & $0.7777\pm0.0008$ & $0.6448\pm0.0012$ & $0.5264\pm0.0015$ & $0.5226\pm0.0016$ & $0.0265\pm0.0006$ & $0.5461\pm0.0016$ & $\mathbf{0.8153\pm0.0013}$ \\
PA--ORF & $0.4636\pm0.0012$ & $0.2772\pm0.0011$ & $0.2815\pm0.0012$ & $0.3984\pm0.0007$ & $0.1613\pm0.0008$ & $0.1396\pm0.0006$ & $\mathbf{0.5252\pm0.0012}$ \\
PA--Divs & $0.5154\pm0.0049$ & $0.4087\pm0.0053$ & $0.4151\pm0.0055$ & $0.3688\pm0.0045$ & $0.0804\pm0.0018$ & $0.2991\pm0.0053$ & $\mathbf{0.5953\pm0.0039}$ \\
PA--Bioact & $0.3457\pm0.0035$ & $0.2766\pm0.0031$ & $0.2558\pm0.0027$ & $0.2899\pm0.0031$ & $0.0301\pm0.0007$ & $0.1440\pm0.0027$ & $\mathbf{0.3592\pm0.0031}$ \\
\addlinespace[0.15em]
PC--Divs & $0.0413\pm0.0118$ & $0.0288\pm0.0082$ & $0.0327\pm0.0066$ & $0.0322\pm0.0087$ & $0.0100\pm0.0040$ & $0.0114\pm0.0049$ & $\mathbf{0.0498\pm0.0092}$ \\
PC--Bioact & $\mathbf{0.0436\pm0.0043}$ & $0.0382\pm0.0048$ & $0.0334\pm0.0042$ & $0.0278\pm0.0042$ & $0.0035\pm0.0020$ & $0.0056\pm0.0015$ & $0.0432\pm0.0045$ \\
\addlinespace[0.15em]
CC--Divs & $0.0425\pm0.0027$ & $0.0327\pm0.0025$ & $0.0371\pm0.0025$ & $0.0330\pm0.0028$ & $0.0134\pm0.0011$ & $0.0093\pm0.0021$ & $\mathbf{0.0485\pm0.0032}$ \\
CC--Bioact & $\mathbf{0.0381\pm0.0010}$ & $0.0315\pm0.0011$ & $0.0283\pm0.0008$ & $0.0275\pm0.0010$ & $0.0128\pm0.0005$ & $0.0116\pm0.0032$ & $0.0375\pm0.0013$ \\
\addlinespace[0.15em]
GG--CRISPR & $0.0209\pm0.0008$ & $\mathbf{0.0212\pm0.0006}$ & $0.0212\pm0.0007$ & $0.0210\pm0.0005$ & $0.0133\pm0.0003$ & $0.0101\pm0.0019$ & $0.0197\pm0.0007$ \\
\addlinespace[0.15em]
CG--Divs & $\mathbf{0.0309\pm0.0038}$ & $0.0185\pm0.0026$ & $0.0224\pm0.0029$ & $0.0176\pm0.0037$ & $0.0128\pm0.0014$ & $0.0105\pm0.0033$ & $0.0171\pm0.0020$ \\
CG--Bioact & $\mathbf{0.0312\pm0.0029}$ & $0.0190\pm0.0019$ & $0.0181\pm0.0016$ & $0.0147\pm0.0018$ & $0.0105\pm0.0012$ & $0.0099\pm0.0028$ & $0.0264\pm0.0025$ \\
\midrule
All (normalized) & $\mathbf{0.9373\pm0.0115}$ & $0.7190\pm0.0092$ & $0.7098\pm0.0150$ & $0.6832\pm0.0141$ & $0.2574\pm0.0131$ & $0.3480\pm0.0214$ & $0.9341\pm0.0072$ \\
\bottomrule
\end{tabular}%
}\par
\vspace{0.20em}
\textbf{(b) Detailed MQ scores}\par\vspace{0.15em}
\resizebox{\textwidth}{!}{%
\begin{tabular}{lcccccc}
\toprule
Task & MorphEM & DINOv2 & SubCell & OpenPhenom & ViT-rand & Cellpose Counts\textsuperscript{\ensuremath{\dagger}} \\
\midrule
PA--CRISPR & $\mathbf{0.7242\pm0.0010}$ & $0.6034\pm0.0016$ & $0.5002\pm0.0014$ & $0.4605\pm0.0015$ & $0.0381\pm0.0007$ & $0.3379\pm0.0021$ \\
PA--ORF & $\mathbf{0.4520\pm0.0013}$ & $0.2602\pm0.0011$ & $0.2816\pm0.0012$ & $0.3898\pm0.0011$ & $0.1744\pm0.0007$ & $0.1445\pm0.0008$ \\
PA--Divs & $\mathbf{0.4970\pm0.0042}$ & $0.3666\pm0.0052$ & $0.4204\pm0.0054$ & $0.3371\pm0.0042$ & $0.1065\pm0.0027$ & $0.2267\pm0.0043$ \\
PA--Bioact & $\mathbf{0.3255\pm0.0037}$ & $0.2616\pm0.0032$ & $0.2638\pm0.0026$ & $0.2765\pm0.0030$ & $0.0386\pm0.0007$ & $0.1768\pm0.0017$ \\
\addlinespace[0.15em]
PC--Divs & $\mathbf{0.0471\pm0.0098}$ & $0.0426\pm0.0125$ & $0.0336\pm0.0067$ & $0.0380\pm0.0063$ & $0.0057\pm0.0046$ & $0.0089\pm0.0030$ \\
PC--Bioact & $\mathbf{0.0394\pm0.0043}$ & $0.0293\pm0.0043$ & $0.0308\pm0.0039$ & $0.0270\pm0.0034$ & $0.0027\pm0.0012$ & $0.0055\pm0.0010$ \\
\addlinespace[0.15em]
CC--Divs & $\mathbf{0.0411\pm0.0022}$ & $0.0338\pm0.0029$ & $0.0339\pm0.0023$ & $0.0302\pm0.0027$ & $0.0146\pm0.0014$ & $0.0112\pm0.0017$ \\
CC--Bioact & $\mathbf{0.0350\pm0.0010}$ & $0.0286\pm0.0007$ & $0.0282\pm0.0008$ & $0.0265\pm0.0009$ & $0.0141\pm0.0006$ & $0.0107\pm0.0003$ \\
\addlinespace[0.15em]
GG--CRISPR & $0.0202\pm0.0007$ & $\mathbf{0.0213\pm0.0008}$ & $0.0206\pm0.0007$ & $0.0204\pm0.0006$ & $0.0158\pm0.0004$ & $0.0097\pm0.0003$ \\
\addlinespace[0.15em]
CG--Divs & $0.0228\pm0.0030$ & $\mathbf{0.0239\pm0.0029}$ & $0.0224\pm0.0029$ & $0.0216\pm0.0031$ & $0.0166\pm0.0030$ & $0.0125\pm0.0015$ \\
CG--Bioact & $\mathbf{0.0235\pm0.0025}$ & $0.0156\pm0.0013$ & $0.0174\pm0.0013$ & $0.0136\pm0.0015$ & $0.0138\pm0.0012$ & $0.0150\pm0.0017$ \\
\midrule
All (normalized) & $\mathbf{0.9847\pm0.0088}$ & $0.8028\pm0.0159$ & $0.7902\pm0.0164$ & $0.7640\pm0.0178$ & $0.3389\pm0.0128$ & $0.3921\pm0.0164$ \\
\bottomrule
\end{tabular}%
}\par
\vspace{0.25em}
\textbf{(c) Task-family summaries, Raw}\par\vspace{0.15em}
\resizebox{0.90\textwidth}{!}{%
\begin{tabular}{lcccccc}
\toprule
Representation & PA NAP & PC NAP & CC R@1\% & GG R@1\% & CG R@1\% & All (normalized) \\
\midrule
MorphEM & $0.5256\pm0.0011$ & $0.0425\pm0.0056$ & $0.0403\pm0.0016$ & $0.0209\pm0.0008$ & $\mathbf{0.0310\pm0.0029}$ & $\mathbf{0.9373\pm0.0115}$ \\
DINOv2 & $0.4018\pm0.0011$ & $0.0335\pm0.0040$ & $0.0321\pm0.0016$ & $\mathbf{0.0212\pm0.0006}$ & $0.0188\pm0.0020$ & $0.7190\pm0.0092$ \\
SubCell & $0.3697\pm0.0012$ & $0.0330\pm0.0038$ & $0.0327\pm0.0013$ & $0.0212\pm0.0007$ & $0.0202\pm0.0019$ & $0.7098\pm0.0150$ \\
OpenPhenom & $0.3949\pm0.0010$ & $0.0300\pm0.0043$ & $0.0302\pm0.0016$ & $0.0210\pm0.0005$ & $0.0161\pm0.0023$ & $0.6832\pm0.0141$ \\
ViT-rand & $0.0746\pm0.0005$ & $0.0067\pm0.0026$ & $0.0131\pm0.0006$ & $0.0133\pm0.0003$ & $0.0117\pm0.0008$ & $0.2574\pm0.0131$ \\
Cell Count\textsuperscript{\textasciicircum} & $0.2822\pm0.0011$ & $0.0085\pm0.0026$ & $0.0104\pm0.0022$ & $0.0101\pm0.0019$ & $0.0102\pm0.0028$ & $0.3480\pm0.0214$ \\
CellProfiler\textsuperscript{*} & $\mathbf{0.5737\pm0.0012}$ & $\mathbf{0.0465\pm0.0048}$ & $\mathbf{0.0430\pm0.0021}$ & $0.0197\pm0.0007$ & $0.0217\pm0.0019$ & $0.9341\pm0.0072$ \\
\bottomrule
\end{tabular}%
}\par
\vspace{0.20em}
\textbf{(d) Task-family summaries, MQ}\par\vspace{0.15em}
\resizebox{0.90\textwidth}{!}{%
\begin{tabular}{lcccccc}
\toprule
Representation & PA NAP & PC NAP & CC R@1\% & GG R@1\% & CG R@1\% & All (normalized) \\
\midrule
MorphEM & $\mathbf{0.4997\pm0.0010}$ & $\mathbf{0.0433\pm0.0052}$ & $\mathbf{0.0381\pm0.0013}$ & $0.0202\pm0.0007$ & $\mathbf{0.0231\pm0.0024}$ & $\mathbf{0.9847\pm0.0088}$ \\
DINOv2 & $0.3730\pm0.0011$ & $0.0359\pm0.0062$ & $0.0312\pm0.0015$ & $\mathbf{0.0213\pm0.0008}$ & $0.0198\pm0.0017$ & $0.8028\pm0.0159$ \\
SubCell & $0.3665\pm0.0012$ & $0.0322\pm0.0040$ & $0.0310\pm0.0013$ & $0.0206\pm0.0007$ & $0.0199\pm0.0018$ & $0.7902\pm0.0164$ \\
OpenPhenom & $0.3660\pm0.0010$ & $0.0325\pm0.0032$ & $0.0284\pm0.0015$ & $0.0204\pm0.0006$ & $0.0176\pm0.0020$ & $0.7640\pm0.0178$ \\
ViT-rand & $0.0894\pm0.0007$ & $0.0042\pm0.0022$ & $0.0143\pm0.0008$ & $0.0158\pm0.0004$ & $0.0152\pm0.0017$ & $0.3389\pm0.0128$ \\
Cellpose Counts\textsuperscript{\ensuremath{\dagger}} & $0.2215\pm0.0011$ & $0.0072\pm0.0017$ & $0.0110\pm0.0008$ & $0.0097\pm0.0003$ & $0.0137\pm0.0012$ & $0.3921\pm0.0164$ \\
\bottomrule
\end{tabular}%
}\par
\end{revblock}
\end{table*}

\begin{table*}[htbp]
\centering
 \caption{\rev{Composition of JUMP-lite benchmark datasets grouped by modality. JCP stands for JUMP Cell Painting. Multiple JCP IDs may refer to the same biological target. For example, ORF and CRISPR perturbations may independently target the same gene.}\addressedbacklink{perturbations}{A8}\addressedbacklink{accounting}{A9}}
\label{tab:jump-lite-summary}
\begin{tabular}{lrrrr}
\toprule
Perturbation Type & \rev{JCP identifiers} & Sources & Plates & Wells \\
\midrule
CRISPR & 7,977 & 1 & 142 & 48,881 \\
Compound-Diversity & 1,187 & 3 & 84 & 9,155 \\
Compound-Bioactive & 3,415 & 1 & 100 & 24,077 \\
ORF & 15,131 & 1 & 225 & 81,663 \\
\midrule
\textbf{Total} & \textbf{27,710} & \textbf{6} & \textbf{551} & \textbf{163,776} \\
\bottomrule
\end{tabular}
\end{table*}

\begin{table*}[!ht]
\centering
\caption{\rev{\textbf{Target-2 pilot compression mapping, storage, and speed benchmark.} $D$ and $E$ denote JPEG XL distance and effort; ``default'' means effort was not explicitly set. Size is the total pilot-store size. Times are mean $\pm$ SD across 200 five-channel site arrays per level.}\addressedbacklink{codec-names}{A10}}
\label{tab:zarr_benchmark}
\begin{tabular}{llrrr}
\toprule
Level & Stored setting & Size & Comp. Time (s/array) & Decomp. Time (s/array) \\
\midrule
Raw & Zstd, level 9 & 55.00G & $0.001 \pm 0.001$ & $0.019 \pm 0.005$ \\
HQ & JPEG XL $D=1$, default $E$ & 2.90G & $0.824 \pm 0.077$ & $0.091 \pm 0.015$ \\
E3 & JPEG XL $D=1$, $E=3$ & 2.60G & $0.085 \pm 0.016$ & $0.088 \pm 0.016$ \\
D2-E8 & JPEG XL $D=2$, $E=8$ & 1.60G & $7.018 \pm 0.672$ & $0.083 \pm 0.015$ \\
MQ & JPEG XL $D=3$, default $E$ & 1.20G & $0.785 \pm 0.081$ & $0.099 \pm 0.019$ \\
LQ & JPEG XL $D=5$, default $E$ & 0.76G & $0.761 \pm 0.081$ & $0.171 \pm 0.020$ \\
D10 & JPEG XL $D=10$, default $E$ & 0.44G & $0.820 \pm 0.120$ & $0.178 \pm 0.029$ \\
D15 & JPEG XL $D=15$, default $E$ & 0.34G & $0.794 \pm 0.117$ & $0.168 \pm 0.026$ \\
D20 & JPEG XL $D=20$, $E=2$ & 0.33G & $0.519 \pm 0.051$ & $0.111 \pm 0.023$ \\
D30 & JPEG XL $D=30$, default $E$ & 0.26G & $0.681 \pm 0.067$ & $0.110 \pm 0.017$ \\
\bottomrule
\end{tabular}

\end{table*}

\begin{table*}[htbp]
\centering
\caption{\textbf{Model inputs and preprocessing.} CP denotes Nuclei, AGP, Mito, RNA, and ER; ``standard'' is z-scoring, and ``clip $X\%$'' clips both tails before rescaling.\addressedbacklink{pretraining}{A6}\addressedbacklink{model-scope}{A12}}
\label{table:model_overview}
%    \resizebox{\textwidth}{!}{
\begin{tabular}{lccr}
\hline
Model & Size & Channels & Preprocessing\\
\hline
cp\_measure &  & CP & minmax 0-1\\
ViT\_random & 224 & Nuclei,AGP,Mito & standard\\
DINOv2 & 224 & Nuclei,AGP,Mito & standard\\
MorphEM & 224 & CP & standard\\
OpenPhenom & 256 & CP & clip 0.5\%, 8 bit, standard\\
SubCell & 448 & Nuclei,AGP,Mito,RNA & clip 0.5\%, minmax 0-1\\
\hline
\end{tabular}
\end{table*}

\subsection{Supplementary Figures}

% Supplementary figure declarations are ordered by first explicit reference
% in the main/SI manuscript text. Figure environments are kept as complete
% blocks.

% 1. fig:overlap_per_source

% 2. fig:compression_impact_plot

% 3. fig:segmentation_example_sup
\begin{figure}[ht]
    \centering
    \begin{minipage}[t]{0.48\columnwidth}\centering
        \centerline{\small \rev{Raw image}}
        \includegraphics[width=\columnwidth]{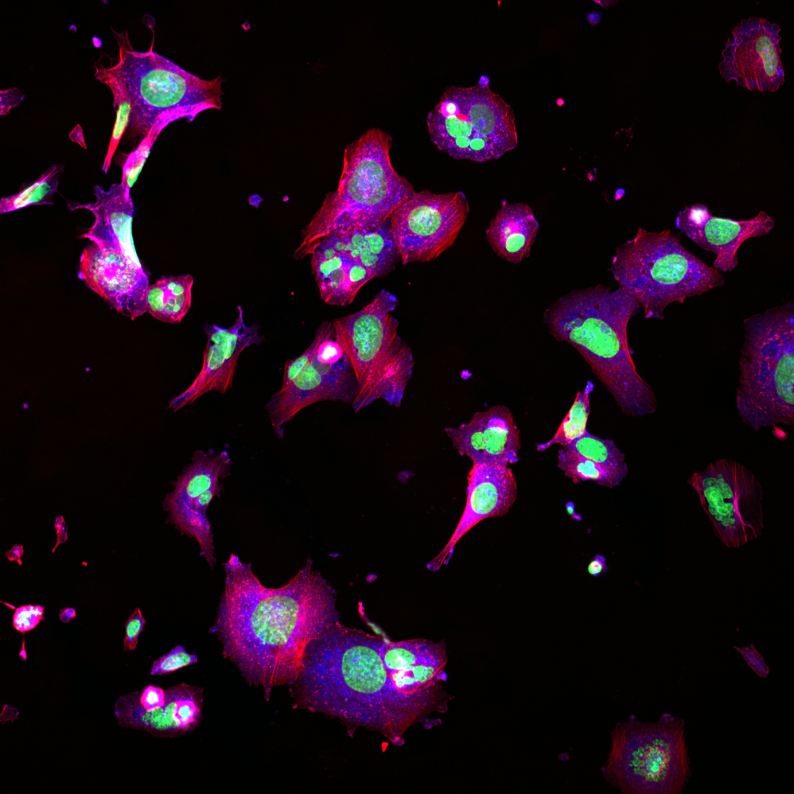}
    \end{minipage}%
    \begin{minipage}[t]{0.48\columnwidth}\centering
        \centerline{\small Segmentation}
        \includegraphics[width=\columnwidth]{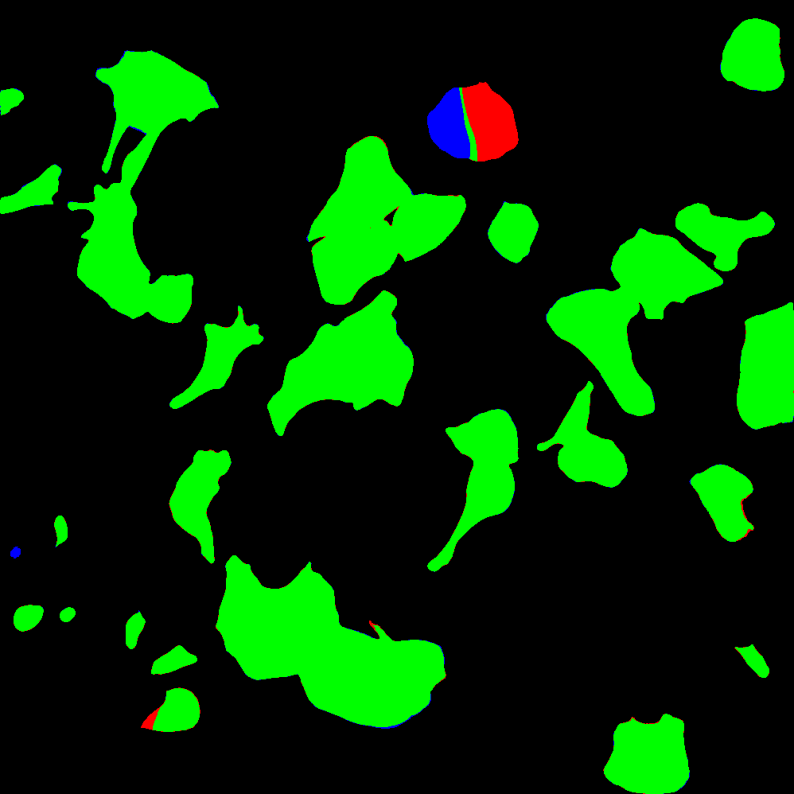}
    \end{minipage}
    \caption{\textbf{Example segmentation comparison.} Segmentation mask disagreement between the uncompressed image and \rev{E3} compression.}
    \label{fig:segmentation_example_sup}
\end{figure}

% 4. fig:sup_seg_f1
\begin{figure*}[ht]
    \centering
    \includegraphics[width=\textwidth]{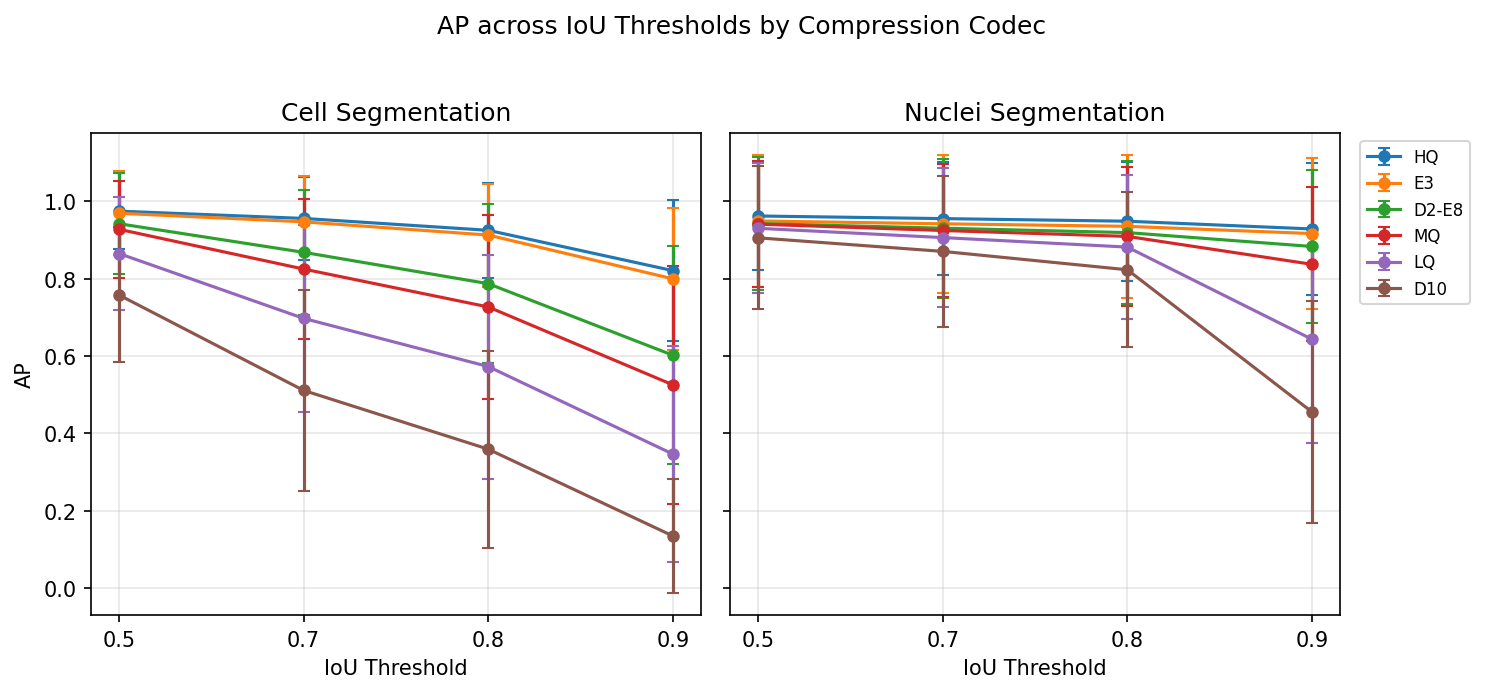}
    \caption{\textbf{\rev{Segmentation-mask agreement at IoU.}} \rev{Segmentation average precision (AP) at IoU=$x$ for the Target-2 plate subset, measured by comparing masks from compressed images against Raw-derived masks. The Raw-derived masks are a consistency reference, not manual ground truth. Cell-level AP counts cells with IoU above $x$ as matched across compression levels, following the evaluation setup used in the 2018 Data Science Bowl segmentation challenge \cite{caicedoNucleusSegmentationAcross2019}.}\addressedbacklink{segmentation}{A13}
    }
    \label{fig:sup_seg_f1}
\end{figure*}

% 5. fig:sup_feature_correlation
\begin{figure*}[ht]
    \centering
    \includegraphics[width=\columnwidth]{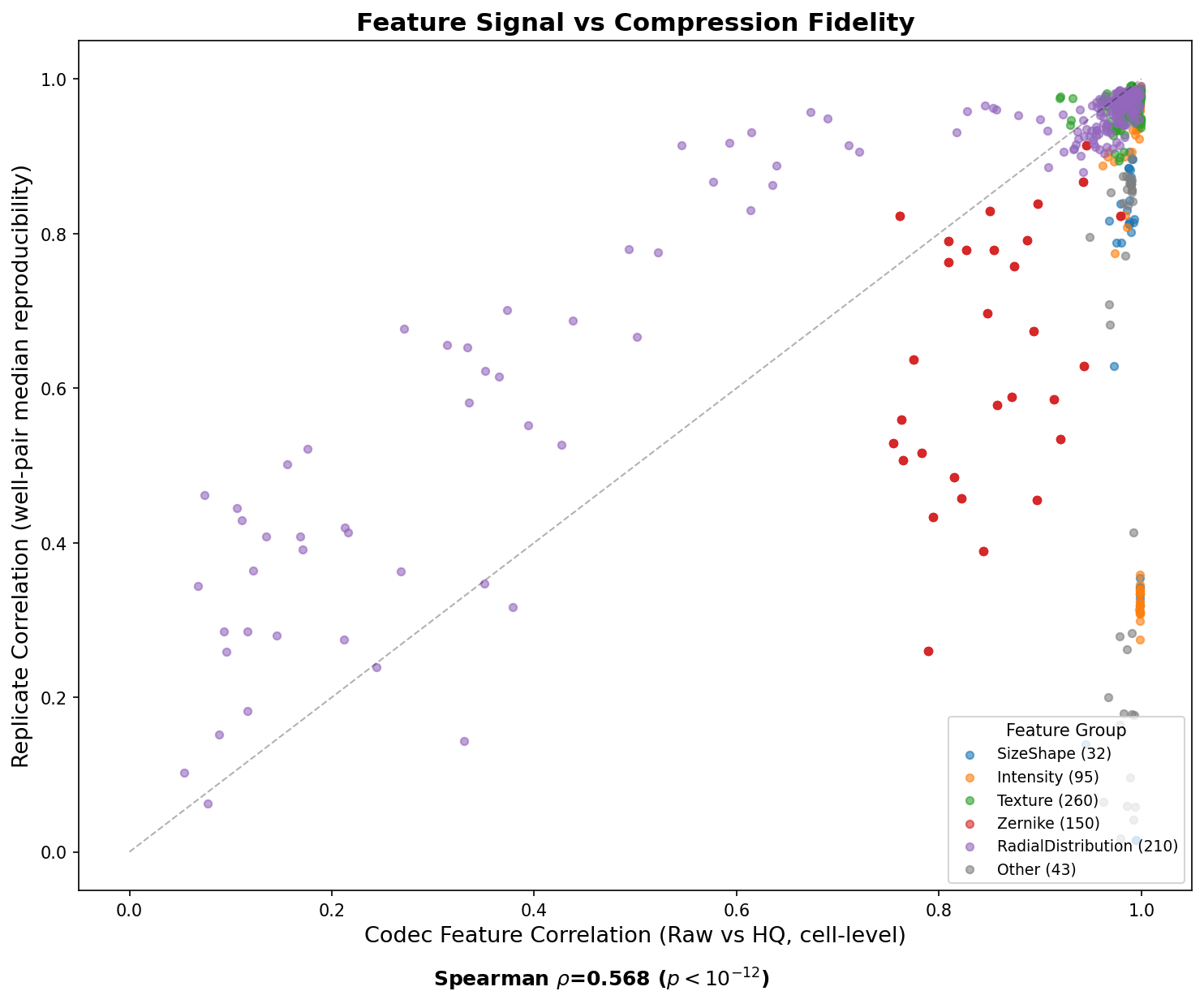}
    \caption{\textbf{Feature signal vs.\ compression fidelity.} Each point represents one CellProfiler feature, colored by feature group. The $x$-axis shows Pearson correlation between features from uncompressed and high-quality compressed images (\rev{compression fidelity}); the $y$-axis shows median Pearson correlation between replicate wells of the same treatment (biological reproducibility). Top-right, features that are both biologically reproducible and preserved by compression (desirable); top-left, features reproducible across replicates but degraded by compression (signal destroyed); bottom-right, features stable across compression levels but not across replicates, likely reflecting systematic technical artifacts rather than biological signal; bottom-left, features unstable in both dimensions (noise). The overall positive association (Spearman $\rho = 0.568$) indicates that most biologically informative features are also robust to compression, supporting the use of lossy JPEG~XL for downstream profiling tasks.
    }
    \label{fig:sup_feature_correlation}
\end{figure*}

% 6. fig:mq-d2e8-explanation
\begin{figure}[ht]
    \centering
    \includegraphics[width=\columnwidth]{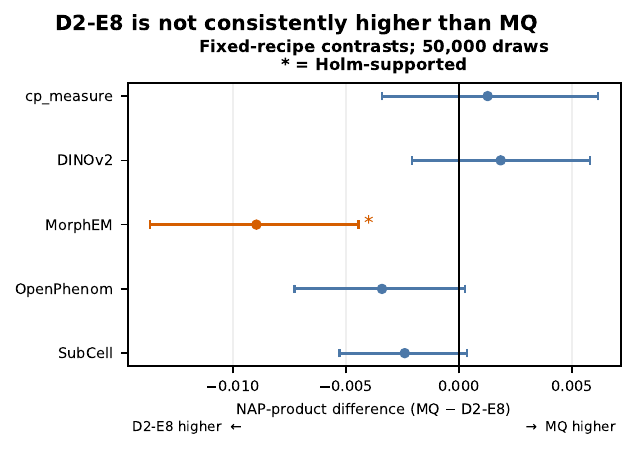}
    \caption{\rev{\textbf{D2-E8 is not consistently higher than MQ.} Comparison of the difference between D2-E8 and MQ using bootstrap intervals. Only MorphEM favors D2-E8 after Holm-Bonferroni correction; the other four contrasts are unresolved. Whiskers are 95\% bootstrap intervals.}\addressedbacklink{nonmonotonic}{A7}}
    \label{fig:mq-d2e8-explanation}
\end{figure}

% 7. fig:segmentation_iou
\begin{figure}[ht]
    \centering
     \includegraphics[width=\columnwidth]{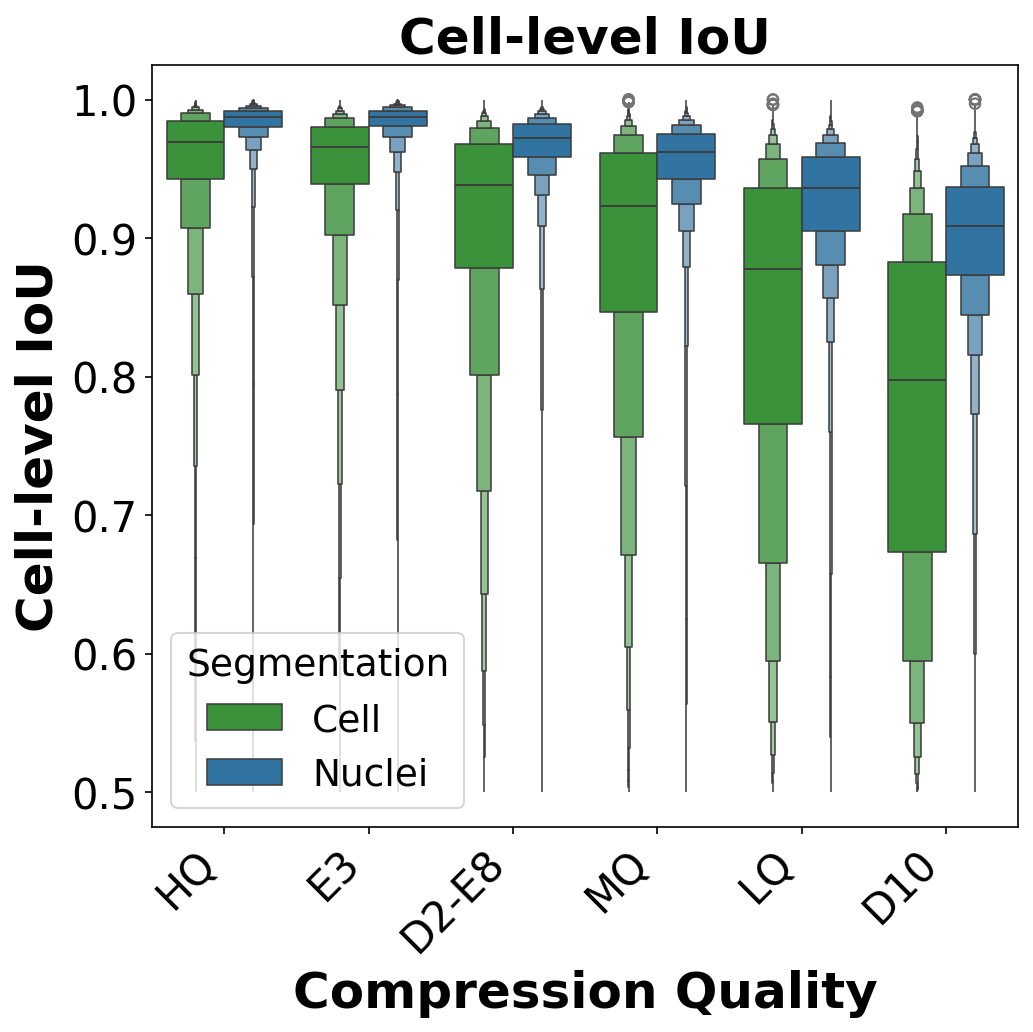}
    \caption{\textbf{Matched-cell segmentation IoU across compression levels.} Per-cell intersection-over-union (IoU) for matched cells (those with IoU above 0.5) between segmentation masks from uncompressed and compressed images. IoU remains high across all compression levels, complementing the cell-level average precision in Figure \ref{fig:compression_iou_feature_correlation}a.}
    \label{fig:segmentation_iou}
\end{figure}

% 8. fig:main-results-combined-access-format
\begin{figure*}[!ht]
        \centering
        \begin{minipage}{0.42\textwidth}
        \includegraphics[width=\textwidth]{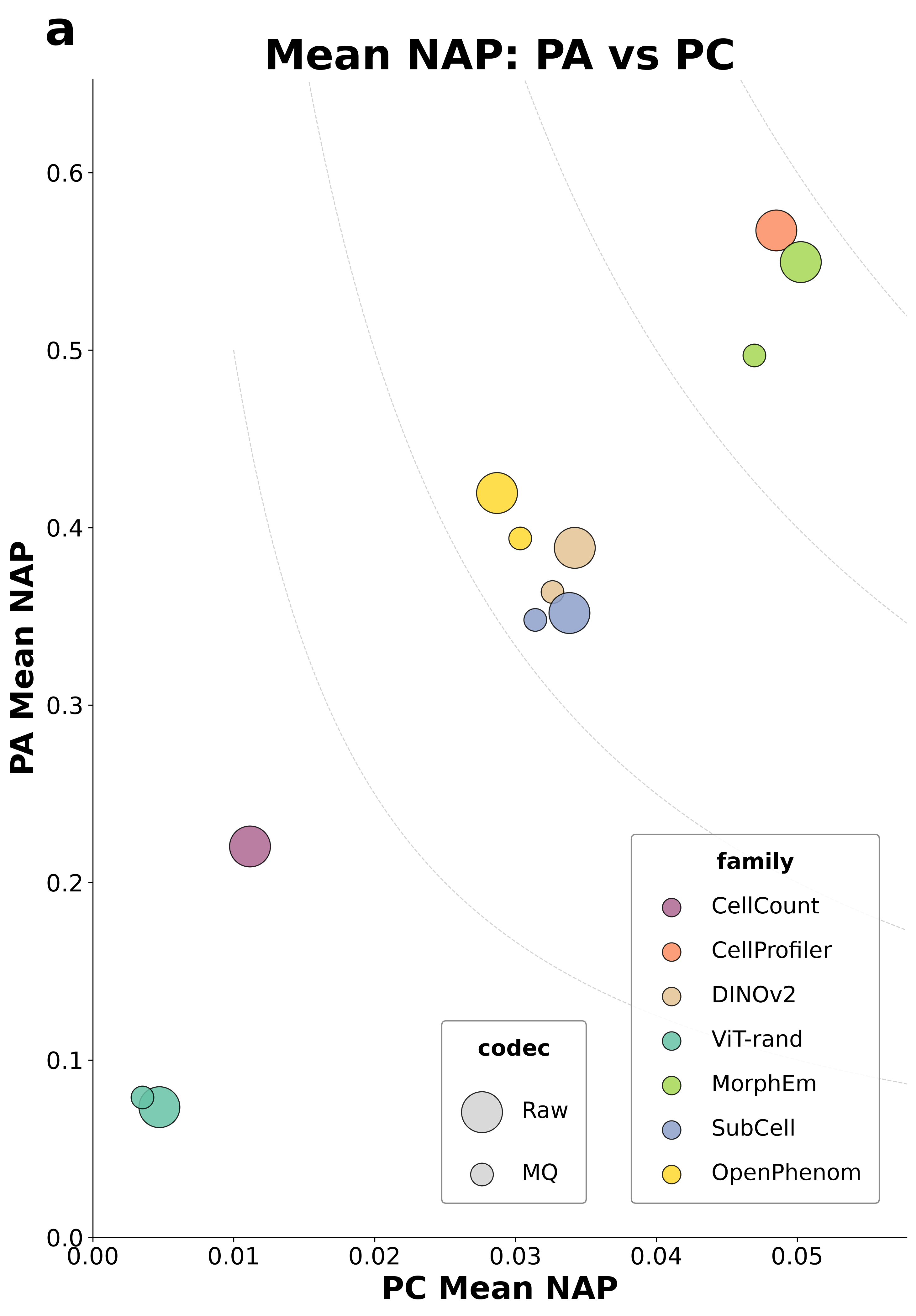}
            % \centerline{\small \textbf{(a)} PA vs. PC.}
        \end{minipage}
        \centering
        \begin{minipage}{0.42\textwidth}
        \includegraphics[width=\textwidth]{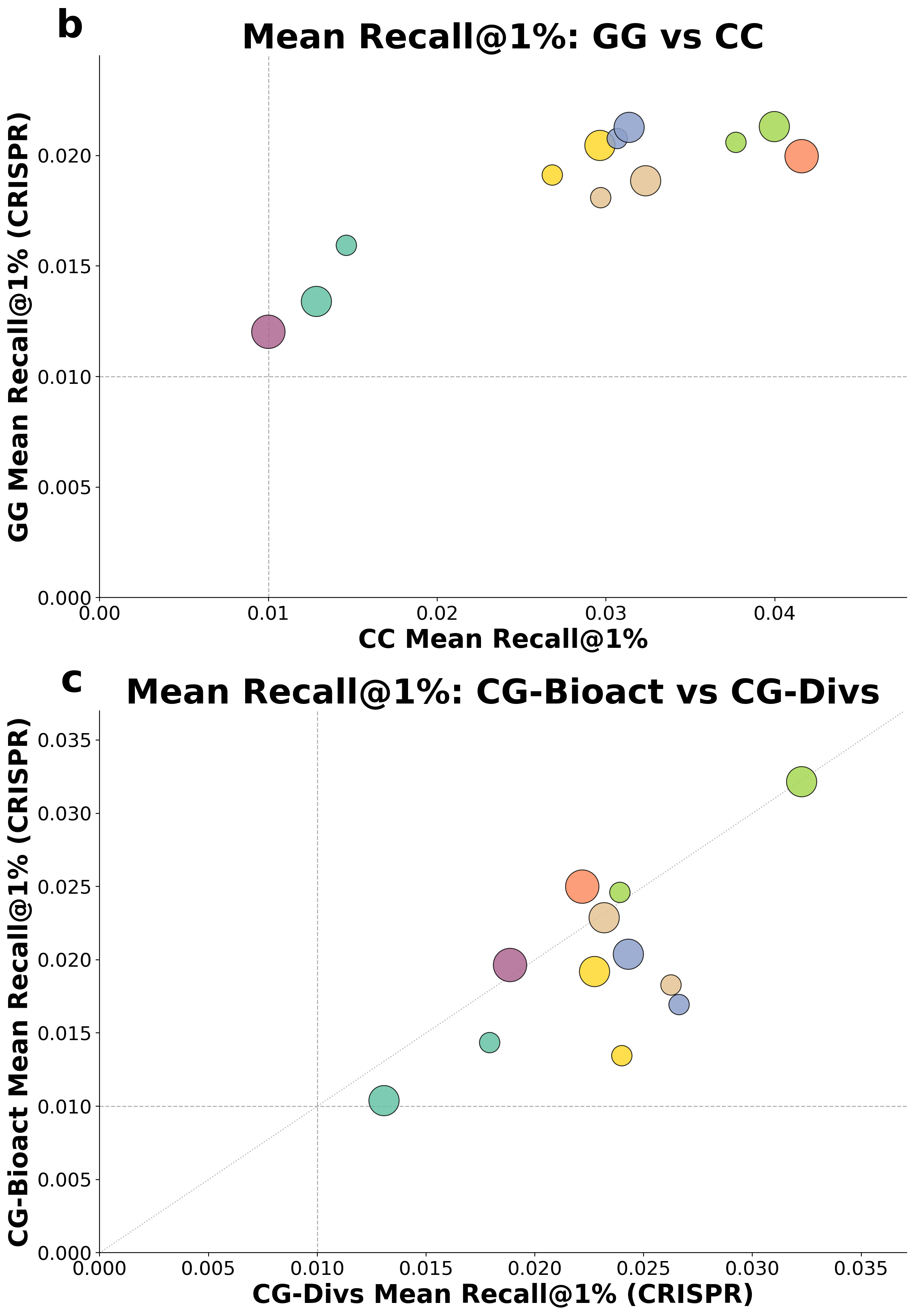}
            % \centerline{\small \textbf{(b) (c)} MOTIVE.}
        \end{minipage}
        \caption{\textbf{Benchmarking model performance across tasks} \textbf{(a)} \rev{RefChemDB} phenotypic metrics show the difference between representations. x-axis shows phenotypic consistency and y-axis phenotypic activity. \textbf{(b)} MOTIVE dataset's Gene-Gene (GG) matching recall at 1\% on the x-axis and Compound-Compound (CC) recall at 1\% on the y-axis. \textbf{(c)} MOTIVE annotations recall at 1\%. Compound-Gene (CG) matching for the Bioactivity d (dataset) and on the Diversity dataset (y-axis).}
        \label{fig:main-results-combined-access-format}
\end{figure*}

% 9. fig:sub_motive_results_full
\begin{figure*}[ht]
    \centering
    \includegraphics[width=\textwidth]{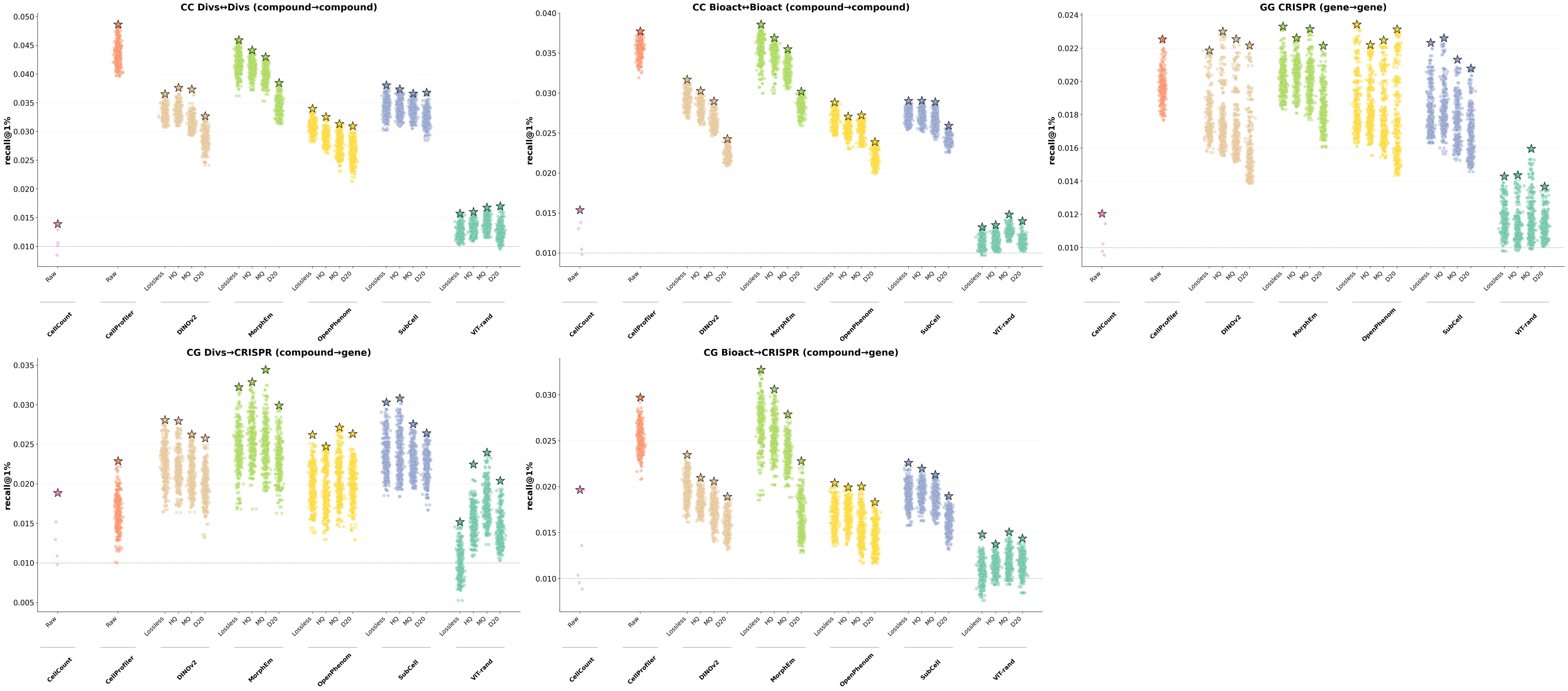}
    \caption{\textbf{MOTIVE strict set performance}
    Calculating recall \@ 1 for each of the combinations of CRISPR x Bioactive, Diverse. As well as within group compound/gene grouping.
    Showing recall enrichment across the board.}
    \label{fig:sub_motive_results_full}
\end{figure*}

% 10. fig:sub_motive_delta_full
\begin{figure*}[ht]
    \centering
    \includegraphics[width=\textwidth]{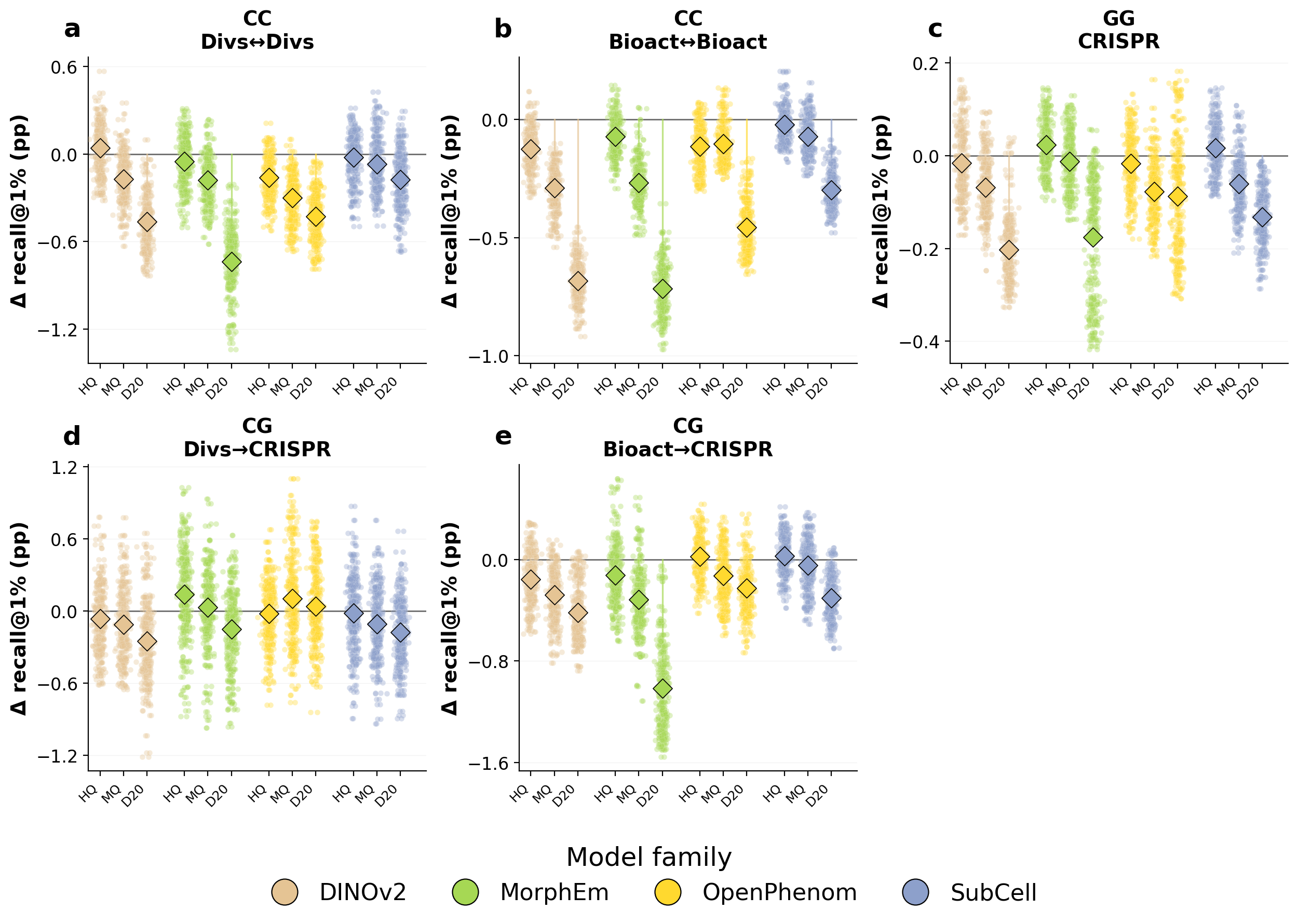}
    \caption{\textbf{\rev{Difference between MOTIVE tasks of compressed vs original \rev{Raw} images}}
    Calculating recall \@ 1 for each of the combinations of CRISPR x Bioactive,Diverse. As well as within group compound/gene grouping.
    \rev{Relative performance difference between original and compressed images, showing an increasing performance drop with stronger compression.} }
    \label{fig:sub_motive_delta_full}
\end{figure*}

% 11. fig:sup_rank_stability
\begin{figure*}[ht]
    \centering
    \includegraphics[width=\textwidth]{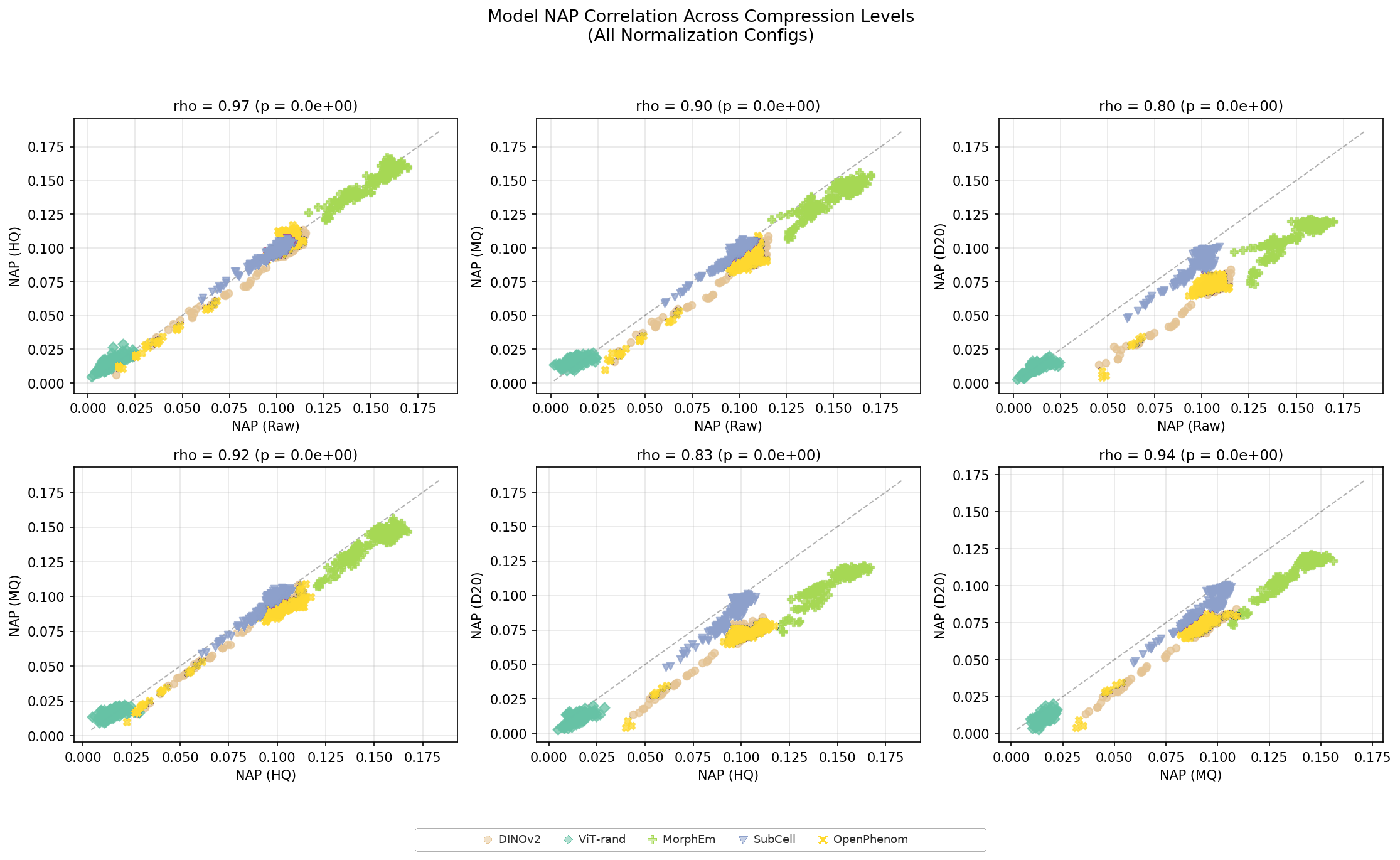}
    \caption{\rev{\textbf{Rank association of learned-model NAP across compression levels.} Pairwise scatter plots of per-configuration NAP for the four learned representations plus the ViT-rand baseline at each pair of levels (Raw, HQ, MQ, D20), with one point per normalization configuration and color indicating the model. CellProfiler and Cell Count are excluded because they have no compression sweep. Spearman $\rho$ is shown above each panel and generally decreases as compression levels become more separated. The strong association for Raw/HQ/MQ supports persistence of the model tiers in the main text.}
    }
    \label{fig:sup_rank_stability}
\end{figure*}

% 12. fig:performance_ranking_mq
\begin{figure*}[!ht]
    \centering
    \includegraphics[width=\textwidth]{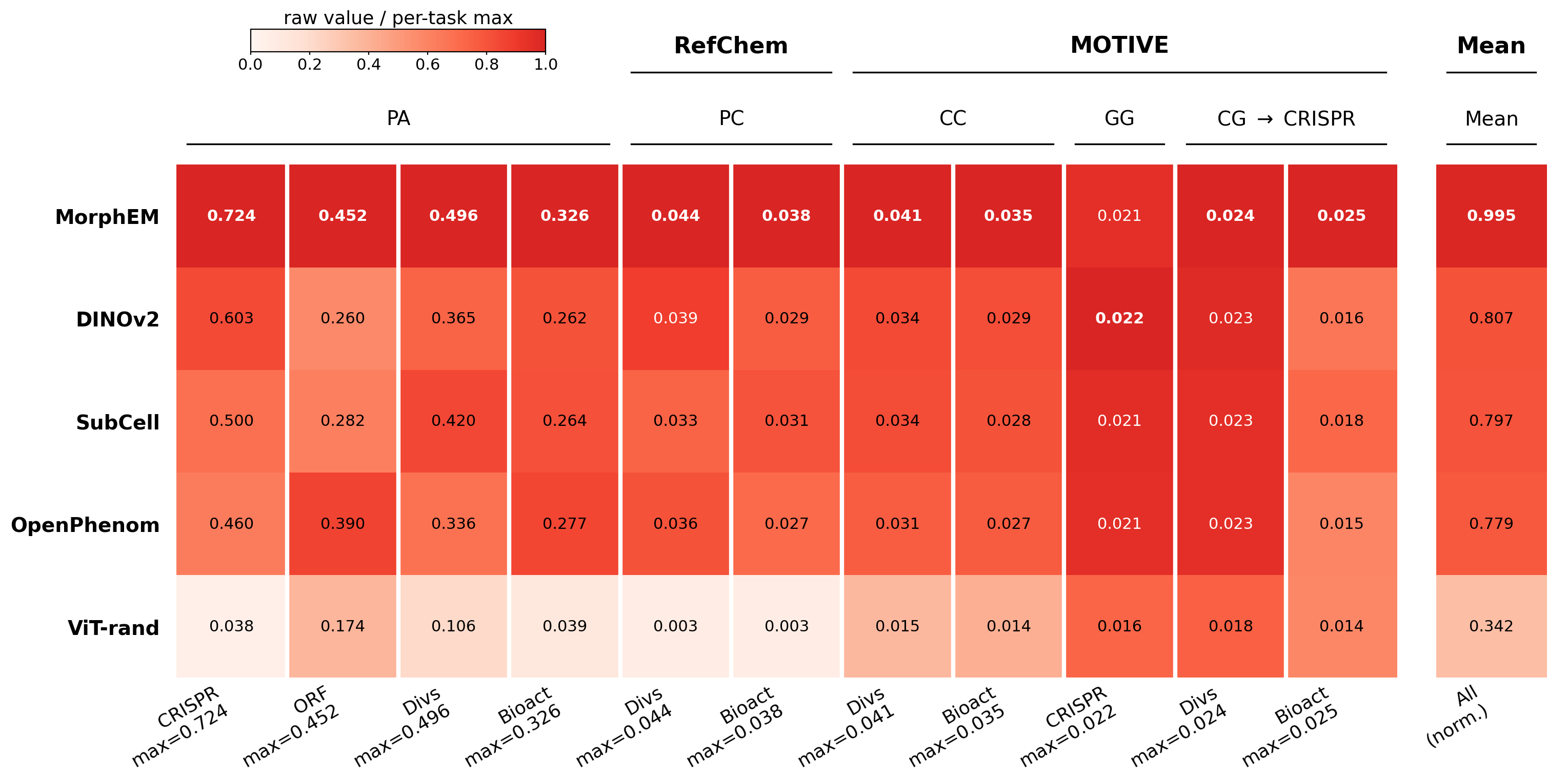}
    \caption{\rev{\textbf{MQ model performance per task, normalized over representations with measured MQ profiles.} Each cell shows a representation's performance on one task: Normalized Average Precision (NAP) for the six RefChemDB tasks (PA across CRISPR, ORF, Diverse (Divs), and Bioact; PC across Divs and Bioact) and Recall@1\% for the five MOTIVE tasks (CC and CG$\rightarrow$CRISPR across Divs and Bioact; GG on CRISPR). For each data source and representation, one post-processing configuration was selected by its mean score across compression settings and used for this MQ display. Cells are colored by the raw score divided by the maximum among the five displayed representations for that task; absolute task maxima are printed under the column labels. Rows are ordered by the mean of their normalized task scores in the rightmost column. Raw-only CellProfiler and Cell Count are excluded before normalization, winner identification, mean calculation, and ordering. Bold values identify the highest-scoring displayed representation for each task and for the mean.}\addressedbacklink{cellprofiler}{A5}}
    \label{fig:performance_ranking_mq}
\end{figure*}

% 13. fig:rank-subsampling
\begin{figure*}[!t]
    \centering
    \includegraphics[width=0.98\textwidth,trim=0 0 0 24bp,clip]{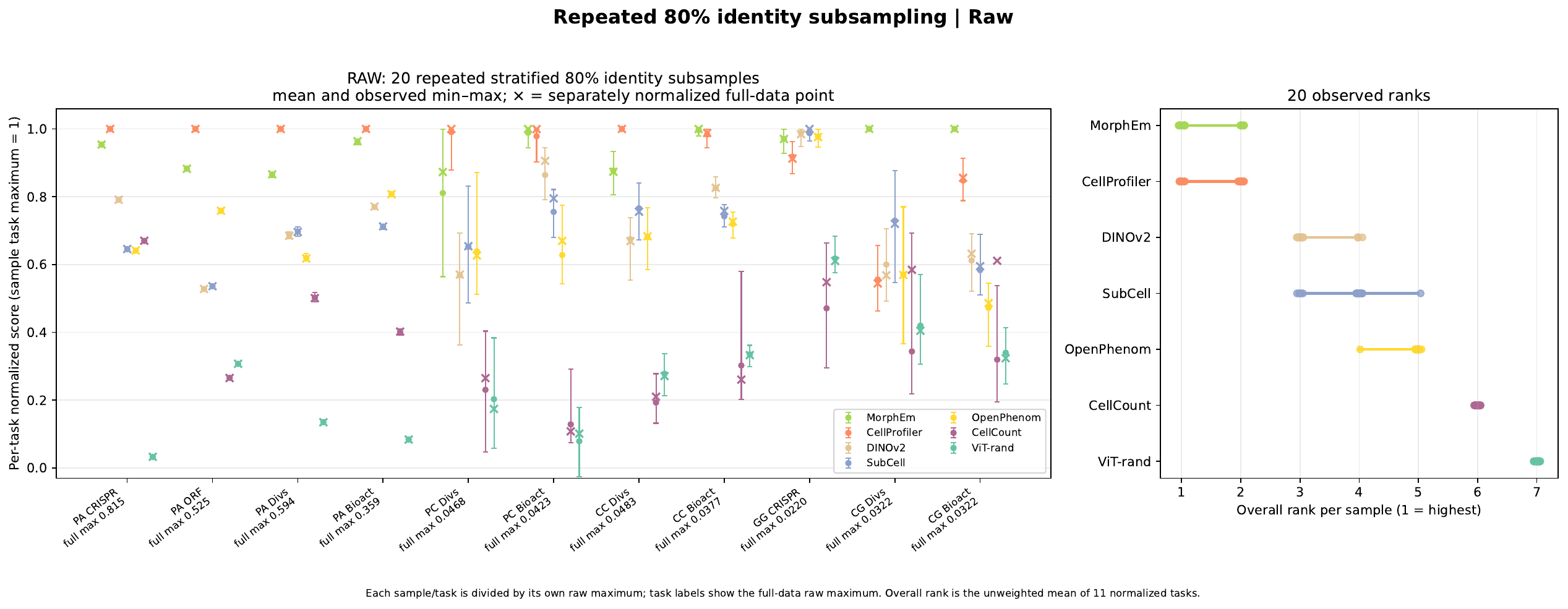}
    \par\vspace{-0.5em}
    \includegraphics[width=0.98\textwidth,trim=0 0 0 24bp,clip]{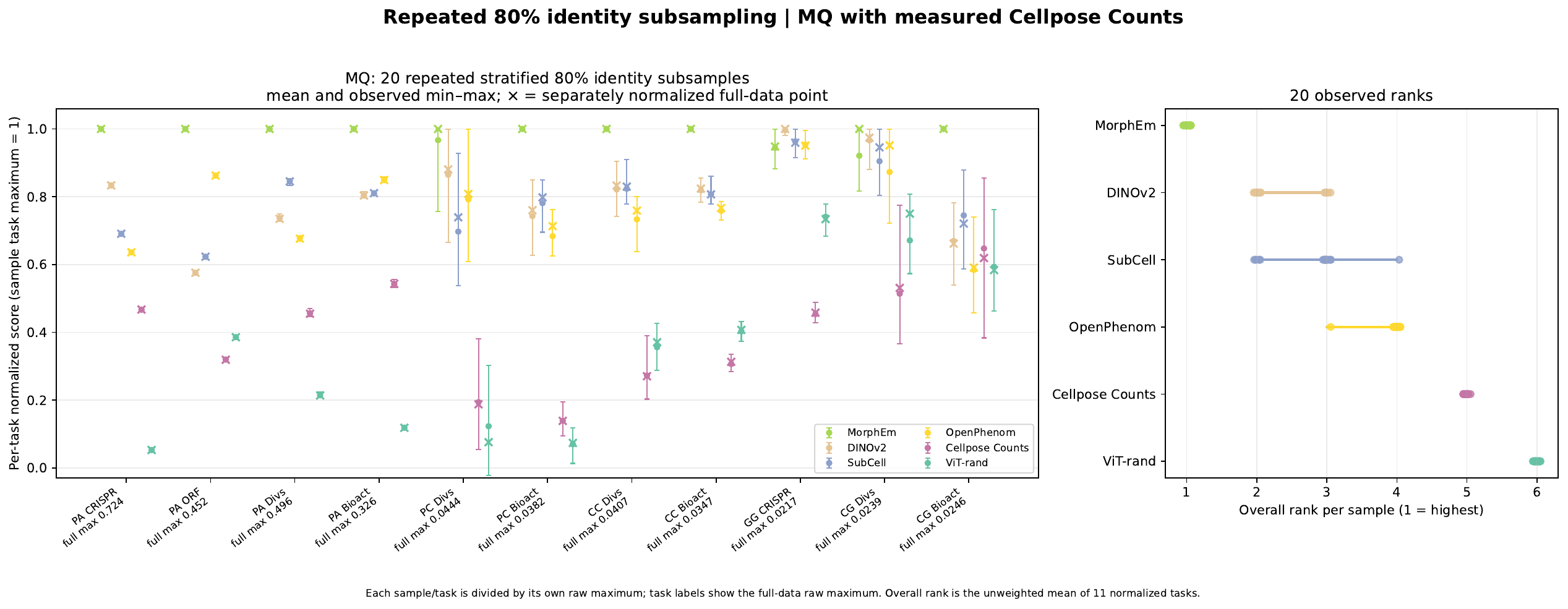}
    \caption{\rev{\textbf{The model tiers are sustained upon repeated sub-sampling.}
    \textbf{Top: Raw. Bottom: MQ.}
    Left panels show normalized per-task means and observed ranges; $\times$ marks the separately normalized full-data value. Each subsample/task is divided by its own raw maximum, and the task labels report the full-data raw maximum. Right panels show the 20 overall ranks, computed as the unweighted mean of the 11 normalized tasks.
    Raw retains the original CellProfiler and legacy Cell Count references. MQ omits both Raw-only references and instead includes Cellpose Counts, comprising independently labeled cell and nuclei mask counts measured from four MQ sites per well. Because profile coverage and the available representations differ between panels, their normalized heights are not direct estimates of compression effects.}\addressedbacklink{uncertainty}{A4}}
    \label{fig:rank-subsampling}
\end{figure*}

% 14. fig:sub_pa_pc_results_full
\begin{figure*}[ht]
    \centering
    \includegraphics[width=\textwidth]{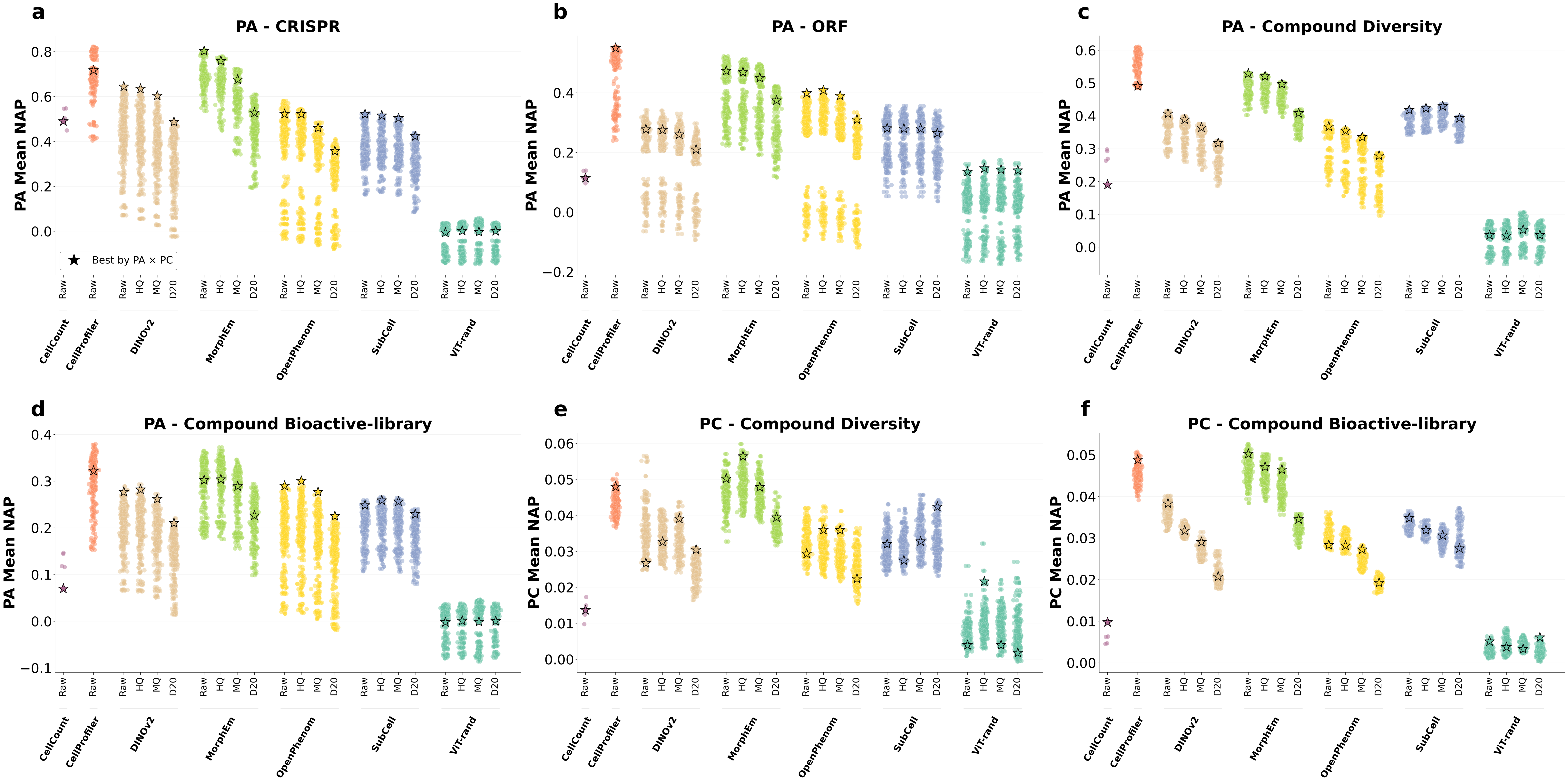}
    \caption{\textbf{PA and PC \rev{RefChemDB} performance}
    Calculating PA and PC for each of the of CRISPR, ORF, compound Bioactive, compound Diverse.}
    \label{fig:sub_pa_pc_results_full}
\end{figure*}

% 15. fig:cluster-selection-coverage
\begin{figure*}[!t]
    \centering
    \includegraphics[width=0.98\textwidth]{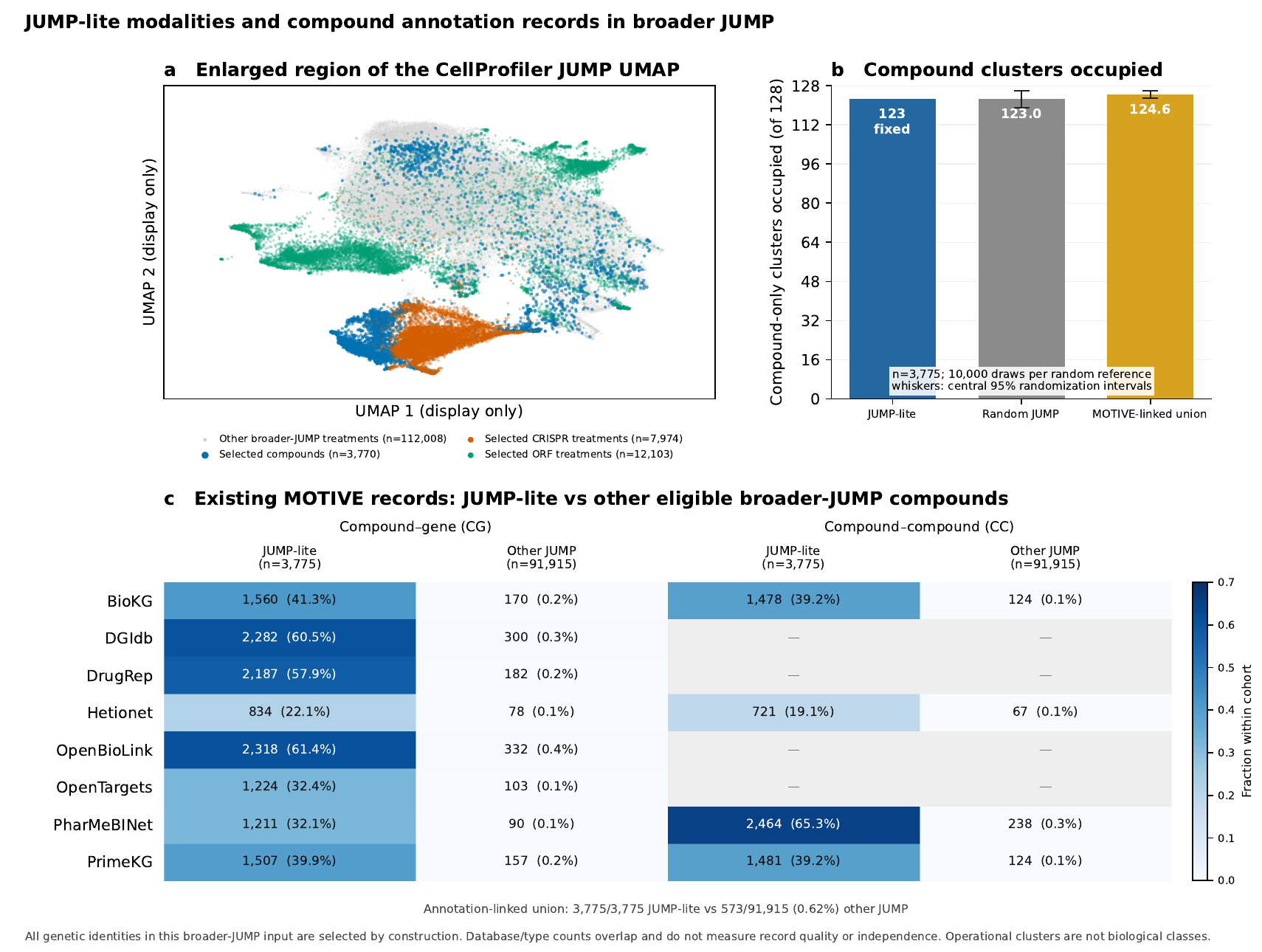}
    \caption{\rev{\textbf{JUMP-lite emphasizes treatments linked to external annotation instead of capturing JUMP phenotypic diversity.} \textbf{(a)} The UMAP uses frozen treatment coordinates generated from the assembled JUMP ALL/v1.0b CellProfiler Harmony profiles. The view focuses on the dense region. Gray denotes other broader-JUMP treatments; blue, orange, and green denote selected compounds, CRISPR treatments, and ORF treatments, respectively. A minuscule fraction of treatments (503 of 136,358; 0.37\%) lies outside the displayed region. UMAP is used only for display. \textbf{(b)} JUMP-lite compounds occupy 123 of 128 operational clusters. \textbf{(c)} Existing MOTIVE records cover all 3,775 JUMP-lite compounds and 573 of 91,915 other eligible compounds.}\addressedbacklink{diversity}{A1}}
    \label{fig:cluster-selection-coverage}
\end{figure*}

% 16. fig:heldout_fixed_recipe_codec
\begin{figure*}[!t]
    \centering
    \includegraphics[width=0.98\textwidth]{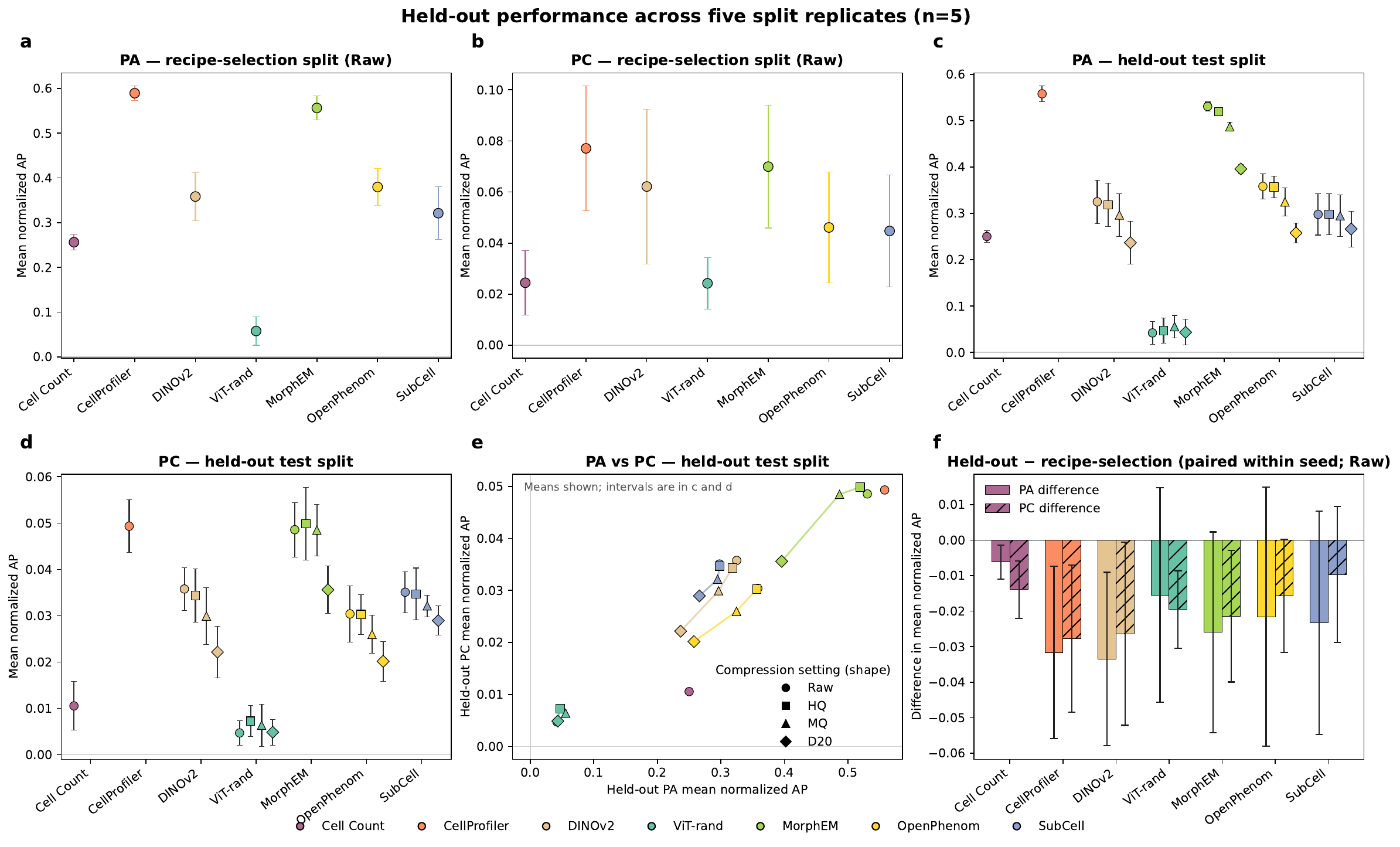}
    \caption{\rev{\textbf{Strict held-out performance across five split replicates.} \textbf{(a--b)} Raw PA and PC on the recipe-selection split. \textbf{(c--d)} Held-out PA and PC after applying the Raw-selected configurations to Raw, HQ, MQ, and D20; horizontal offsets separate compression settings. \textbf{(e)} Held-out PA versus PC means, with lines connecting compression settings within each representation. Intervals are shown separately in panels~(c--d). \textbf{(f)} Paired held-out-minus-selection differences for Raw profiles. Shapes distinguish compression settings, and whiskers in panels~(a--d, f) show 95\% Student-t intervals.}\addressedbacklink{postprocessing}{A3}\addressedbacklink{uncertainty}{A4}\addressedbacklink{codec-names}{A10}\addressedbacklink{operating-points}{A11}}
    \label{fig:heldout_fixed_recipe_codec}
\end{figure*}

% 17. fig:nonlite-pa-pc-sampled
\subsection{\rev{Model tiers in currently usable non-JUMP-lite treatments}}
\label{sec:nonlite-pa-pc-sampled}

\begin{figure*}[!t]
    \centering
    \includegraphics[width=0.98\textwidth]{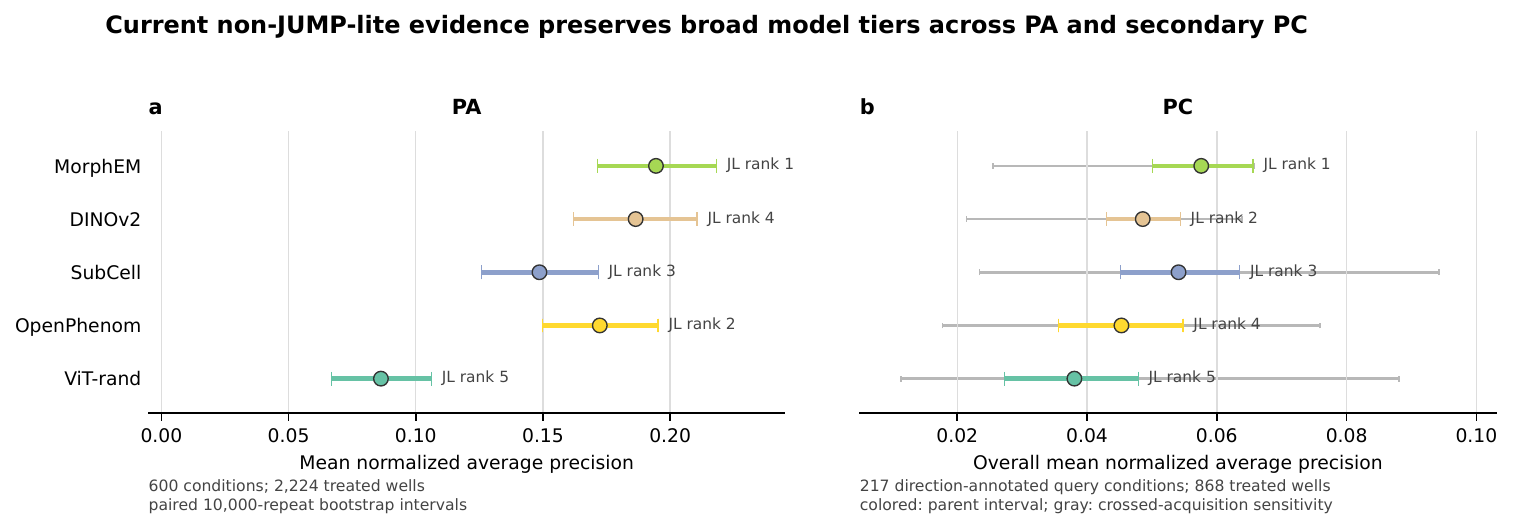}
    \caption{\rev{\textbf{The overall ranking is broadly sustained outside JUMP-lite.} \textbf{(a)} Mean phenotypic activity (PA) for the currently usable non-JUMP-lite cohort, evaluated with five models at fixed checkpoints: 600 treatment conditions, 2,224 treated wells, and all 26,725 DMSO controls on 453 plates. Bars show 95\% intervals from 10,000 paired condition-bootstrap repeats. \textbf{(b)} Secondary direction-stratified phenotypic consistency (PC) for 217 qualified direction-annotated query conditions, representing 191 unique compounds and 868 treated wells, within the fixed 512-condition annotated non-JUMP-lite profile pool. Colored bars show 95\% intervals from the parent-multiplier procedure; thin gray bars show the crossed-acquisition sensitivity analysis. ``JL rank'' labels are Raw JUMP-lite ranks. The PC sample is smaller because direction annotations are limited outside JUMP-lite. The distance between MorphEM and the other models, as well as the overall mean normalized AP is lower, indicating that the overall signal of these samples is lower.} \addressedbacklink{broader-ranking}{A2}}
    \label{fig:nonlite-pa-pc-sampled}
\end{figure*}

\begin{figure*}[ht!]
    \centering
    \includegraphics[width=\textwidth]{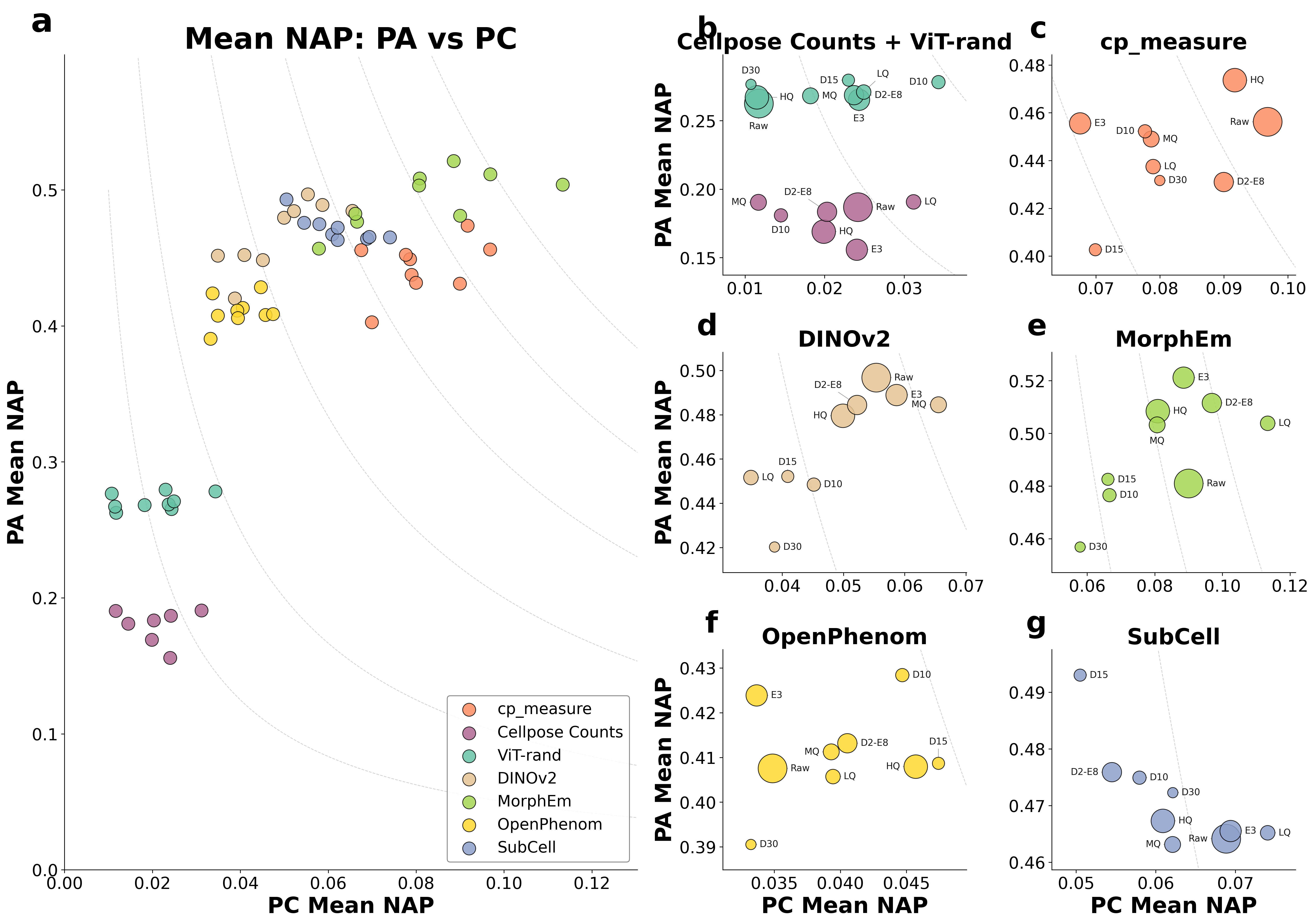}
    \caption{\rev{\textbf{Model performance on the Target-2 subset across compression levels.} Each point represents the best-performing normalization configuration for one model--compression combination, scored by the rescaled PA-PC product; lines show contours of equal product. \textbf{(a)} Across all models, model choice drives performance differences more than compression level. \textbf{(b--g)} Per-model results, with Cellpose Counts and ViT-rand grouped as baselines. Only MorphEM \textbf{(e)} and cp\_measure \textbf{(c)} show consistent degradation when the uncompressed and high-fidelity settings (Raw and HQ) are compared with aggressive compression (D10, D15, and D30). We define ``appreciable degradation'' as a relative difference greater than 10\% in PA or PC. Within this four-plate Target-2 comparison, cp\_measure lies within the PA and PC ranges of the learned representations. The small assay limits generalization, and DINOv2, OpenPhenom, and SubCell show non-monotonic ordering across compression settings in these results.}\addressedbacklink{cellprofiler}{A5}\addressedbacklink{nonmonotonic}{A7}}
    
    \label{fig:model_comparison_4plate}
\end{figure*}

\begin{figure*}[!t]
    \centering
    \includegraphics[width=0.93\textwidth]{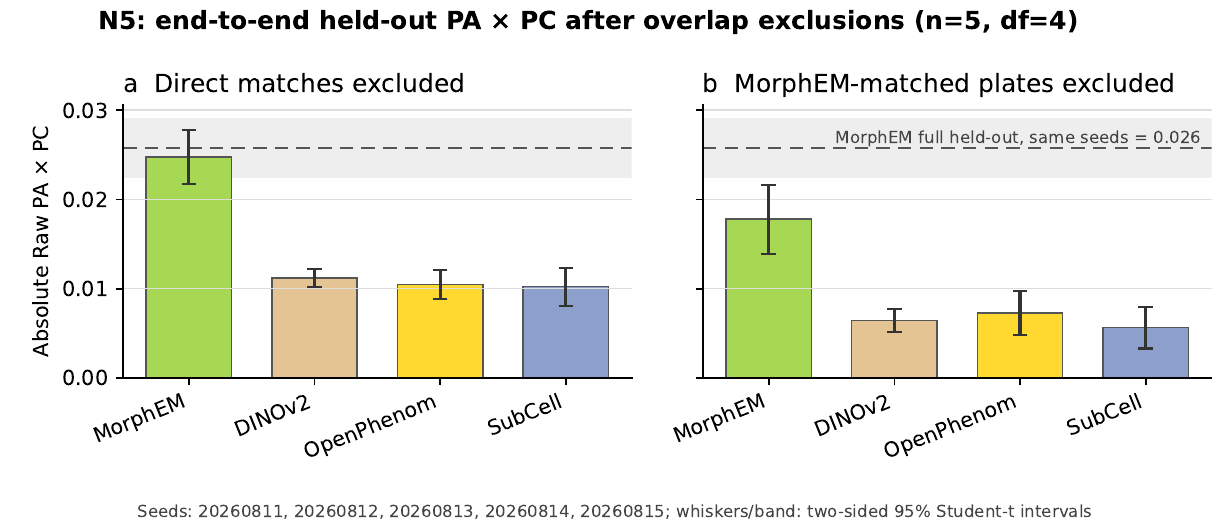}
    \caption{\rev{\textbf{MorphEM remains higher than other models even if we discard the images and plates in its training dataset.} Raw PA, PC, and their within-seed product were recomputed for MorphEM, DINOv2, OpenPhenom, and SubCell after independently repeating the strict split, Raw recipe selection, refitting, and held-out evaluation for five seeds. Bars show the mean absolute Raw PA$\times$PC product, and whiskers show two-sided 95\% Student-t intervals across seeds ($n=5$, $df=4$). \textbf{(a)} Held-out wells directly matching the frozen CHAMMI-75 physical-acquisition inventory were excluded. \textbf{(b)} Every held-out well on a MorphEM-matched plate was excluded. The dashed line and shaded band show MorphEM's mean and same-seed interval on the complete held-out sets. MorphEM remained highest in both subsets.}\addressedbacklink{pretraining}{A6}}
    \label{fig:pretraining-overlap-sensitivity}
\end{figure*}

\rev{To test learned-model rankings outside JUMP-lite, we evaluated PA using the five models' currently usable embeddings saved at fixed checkpoints. We limited the analysis to physical non-JUMP-lite plates with enough exact DMSO controls and compared the result with the available secondary PC estimate (Figure~\ref{fig:nonlite-pa-pc-sampled}).}

% 18. fig:paired-cp-measure-mq-lossless
% \begin{figure*}[!t]
%     \centering
%     \includegraphics[width=0.98\textwidth]{main/figures/rebuttal/paired_cp_measure_mq_lossless_all_sites.pdf}
%     \caption{\rev{\textbf{MQ compression reduces object counts on average across all paired sites.} Both panels combine nuclei (blue) and cells (orange) from the same 632,672 sites. \textbf{(a)} Absolute MQ and lossless counts. The dashed line marks equal counts. Mean MQ minus lossless differences were $-5.35$ nuclei masks/site and $-17.82$ cell masks/site. \textbf{(b)} The count fraction at each site, calculated as the MQ count divided by the lossless count for each object type. Mean fractions were 0.985 for nuclei and 0.947 for cells after omitting 423 nuclei and 94 cell records with zero lossless counts. Counts were independently labeled under each compression setting, so the fractions do not track individual masks. Panel~(a) is limited to 0 to 700 masks/site, and panel~(b) is limited to fractions from 0 to 1.2. The percentage within each limit is printed in the panel.}}
%     \label{fig:paired-cp-measure-mq-lossless}
% \end{figure*}

% 19. fig:model_comparison_4plate

% 20. fig:pretraining-overlap-sensitivity

\end{document}